\documentclass[aps, prd, amssymb, amsmath, amsfonts, eqsecnum, showpacs, nofootinbib, floatfix, twocolumn, twoside, a4paper, superscriptaddress, longbibliography]{revtex4-2}

\usepackage{mathrsfs}
\usepackage{siunitx}
\usepackage{comment}
\usepackage{soul}
\usepackage[T3,T1]{fontenc}
\DeclareMathAlphabet{\mathpzc}{OT1}{pzc}{m}{it} 
\usepackage{mathrsfs} 
\usepackage{bm} 
\usepackage{microtype} 
\def\MT@register@subst@font{
  \MT@exp@one@n\MT@in@clist\font@name\MT@font@list
  \ifMT@inlist@\else\xdef\MT@font@list{\MT@font@list\font@name,}\fi}
\makeatother

\usepackage{orcidlink}
\usepackage{multirow}
\usepackage{booktabs} 
\usepackage{array} 
\usepackage{tabularx}

\newcolumntype{C}{>{\centering\arraybackslash}X}

\usepackage{xcolor}
\usepackage{hyperref} 
\definecolor{dodgerblue}{HTML}{1E90FF}
\definecolor{DARK_ORANGE}{HTML}{FF5B00}
\definecolor{DARK_GREEN}{HTML}{02590F}
\definecolor{DARK_RED}{HTML}{B22222}
\definecolor{LIGHTER_DARK_RED}{HTML}{A52A2A}
\hypersetup{
    colorlinks=true,    
    citecolor=blue,     
    linkcolor=red,      
    urlcolor=blue,     
}

\usepackage{graphicx} %
\graphicspath{{./figures/}}
\usepackage{xspace}

\newcommand{\phX}{\texttt{IMRPhenomX}\xspace}

\newcommand{\phXP}{\texttt{IMRPhenomXP}\xspace}

\newcommand{\seobnr}{\texttt{SEOBNRv2}\xspace}
\newcommand{\phPvT}{\texttt{IMRPhenomPv2\_NRTidalv2}\xspace}

\newcommand{\pycbc}{\textsc{PyCBC}\xspace}
\newcommand{\bilby}{\textsc{Bilby}\xspace}
\newcommand{\bilbytgr}{\mbox{\textsc{Bilby TGR}}\xspace}
\newcommand{\pesummary}{\mbox{\textsc{PESummary}}\xspace}
\newcommand{\lalsuite}{\textsc{LALSuite}\xspace}

\newcommand{\lalsimulation}{\textsc{LALSimulation}\xspace}
\newcommand{\dynesty}{\textsc{Dynesty}\xspace}
\newcommand{\numpy}{\textsc{NumPy}\xspace}
\newcommand{\scipy}{\textsc{SciPy}\xspace}
\newcommand{\matplotlib}{\textsc{Matplotlib}\xspace}

\newcommand{\QGR}{\ensuremath{\mathcal{Q}_{\mathrm{GR}}}\xspace}
\newcommand{\mchirp}{\ensuremath{\mathcal{M}_c}\xspace}

\newcommand{\chieff}{\ensuremath{\chi_{ \ensuremath{\mathrm{eff}}}}\xspace}
\newcommand{\chip}{\ensuremath{\chi_{\mathrm{p}}}\xspace}
\newcommand{\dphi}{\ensuremath{\delta\hat{\varphi}}\xspace}
\newcommand{\ff}{\ensuremath{\mathcal{FF}}\xspace}
\newcommand{\match}{\ensuremath{{\mathcal{M}}}\xspace}
\newcommand{\ovpnet}{\ensuremath{\mathcal{O}_{\rm{net}}}\xspace}
\newcommand{\rhot}{\ensuremath{\tilde{\rho}}\xspace}
\newcommand{\hinj}{\ensuremath{h_{\mathrm{inj}}}\xspace}
\newcommand{\hrec}{\ensuremath{h_{\mathrm{rec}}}\xspace}

\newcommand{\tbank}{\ensuremath{\mathcal{T}_B}\xspace}

\newcommand{\Hz}{\ensuremath{\,\mathrm{Hz}}\xspace}

\newcommand{\flow}{\ensuremath{f_{\mathrm{low}}}\xspace}

\newcommand{\fhigh}{\ensuremath{f_{\mathrm{high}}}\xspace}

\newcommand{\dd}{\mathrm{d}\xspace}

\newcommand{\msun}{\ensuremath{\,M_\odot}\xspace}

\newcommand{\nth}{\textsuperscript{th}}

\newcommand\abs[1]{\ensuremath{\lvert#1\rvert}}

\newcommand\inp[2]{\langle #1 \,|\, #2 \rangle}

\def \densityUnit {\,\si{\gram\per\cubic\centi\meter}}
\DeclareSIUnit\Mpc{Mpc}

\newcommand{\Nikhef}{Nikhef -- National Institute for Subatomic Physics, Science Park 105, 1098 XG Amsterdam, The Netherlands}
\newcommand{\UU}{Institute for Gravitational and Subatomic Physics (GRASP), \mbox{Utrecht University}, Princetonplein 1, 3584 CC Utrecht, The Netherlands}
\newcommand{\UCLouvain}{Centre for Cosmology, Particle Physics and Phenomenology - CP3, Université Catholique de Louvain, Louvain-La-Neuve, B-1348, Belgium}
\newcommand{\ROB}{Royal Observatory of Belgium, Avenue Circulaire, 3, 1180 Uccle, Belgium}
\newcommand{\IFAE}{Institut de F\'isica d’Altes Energies (IFAE), The Barcelona Institute of Science and Technology,\\ Campus UAB, 08193 Bellaterra (Barcelona), Spain}
\newcommand{\GRAPPA}{Gravitation Astroparticle Physics Amsterdam (GRAPPA),\\ University of Amsterdam, 1098 XH Amsterdam, The Netherlands}

\begin{document}

\title{Compact Binary Coalescences in Dense Gaseous Environments Can Pose as ones in Vacuum}

\author{Soumen Roy\:\orcidlink{0000-0003-2147-5411}}
\email{soumen.roy@uclouvain.be}
\affiliation{\UCLouvain}
\affiliation{\ROB}
\affiliation{\Nikhef}
\affiliation{\UU}

\author{Rodrigo Vicente\:\orcidlink{0000-0002-3150-9315}}
\email{r.l.lourencovicente@uva.nl}
\affiliation{\IFAE}
\affiliation{\GRAPPA}

\date{\today}

\begin{abstract}
The gravitational-wave events observed by the LIGO-Virgo-KAGRA collaboration are attributed to compact binary coalescences happening in vacuum. However, several studies suggest that gaseous environments may play a significant role in the formation and evolution of compact binaries. Why have we not seen environmental effects in LVK signals?
While matched-filtering remains the most effective technique for gravitational-wave searches, it comes with a burden: we might only observe signals that align with our (vacuum) expectations, potentially missing unexpected or unknown phenomena. Even more concerning is the possibility that environmental effects could mimic vacuum waveforms, introducing biases in parameter estimation and impacting population studies. Here, we use numerical relativity simulations of binary black hole mergers inside stellar envelopes to show that: (i) a \texttt{GW150914}-like event would be detected (with a false alarm rate smaller than $10^{-4}\, \mathrm{yr^{-1}}$) using a template bank of vacuum waveforms, even when immersed in a stellar envelope of density larger than $10^{7}\,\mathrm{g/cm^3}$; (ii) environmental effects can pass routinely performed tests of vacuum General Relativity, while leading to considerable biases in parameter estimation; but (iii) phenomenological environment waveforms are effectual in detecting environmental effects and can resolve systematics.
\end{abstract}
\maketitle

\section{Introduction} 

One century after its prediction by Einstein~\cite{Einstein:1916cc, Einstein:1918btx}, gravitational waves (GWs) were finally observed in 2015 by LIGO~\cite{LIGOScientific:2016aoc}. 
Since then, many more detections have been recorded by the Advanced LIGO~\cite{LIGOScientific:2014pky} and Advanced Virgo~\cite{VIRGO:2014yos} detectors~\cite{LIGOScientific:2018mvr, LIGOScientific:2020ibl, LIGOScientific:2021usb, KAGRA:2021vkt, KAGRA:2013rdx}, which have been used (among others) to test the theory of General Relativity (GR)~\cite{LIGOScientific:2016lio, LIGOScientific:2018dkp, LIGOScientific:2019fpa, LIGOScientific:2020tif}, measure the rate of expansion of the Universe~\cite{LIGOScientific:2017adf, LIGOScientific:2019zcs, LIGOScientific:2021aug}, and probe the interior structure of neutron stars~\cite{LIGOScientific:2018cki}. Their potential of discovery for (astro)physics is vast~\cite{Barack:2018yly}. Planned detectors for the next decade will give us the means to fully realise it~\cite{Maggiore:2019uih, Evans:2021gyd, Colpi:2024xhw}. However, their exquisite sensitivity demands an extremely accurate modelling to completely handle systematics.

Astrophysical environments---from common envelopes and circumbinary disks to active galactic nuclei (AGN) disks---may play an important role on the formation and evolution of compact binaries, and are often invoked as a potential channel (complementary to isolated evolution and dynamical formation~\cite{Mandel:2021smh, Mandel:2018hfr}) to explain the observed merger rates~\cite{KAGRA:2021duu}. Additionally, in some non-standard theoretical scenarios binary black holes (BBH) could coalesce in a dense stellar envelope~\cite{Woosley:2016nnw, Fedrow:2017dpk}. However, the effect of gaseous environments has been continuously neglected in the analysis of the GW transients observed by the LIGO-Virgo-KAGRA (LVK) collaboration. Recently, we have performed the first search for environmental effects in the GTWC-1 and in selected events from the GWTC-2~\cite{CanevaSantoro:2023aol}. While we could not find evidence for environmental effects in the data, we were able to derive relevant constraints on the density of putative gaseous environments.

While matched-filtering remains the most effective technique for GW searches, it leads to observational bias: can we detect compact binaries coalescences (CBCs) in dense gaseous environments by using a template bank containing only vacuum waveforms? A positive answer to such question might be even more troubling than a negative one, as it could signify that CBCs in gaseous environments are able to disguise as in vacuum. That would strongly impact the estimation of binary parameters and, thus, bias astrophysical population studies.

Most numerical relativity (NR) simulations of CBCs in environments do not study the environmental effects on GW waveforms, focusing exclusively on their electromagnetic counterpart (e.g., \cite{Farris:2011vx, Farris:2012ux, Paschalidis:2021ntt, Ruiz:2023hit}). An exception is the NR simulations of Ref.~\cite{Fedrow:2017dpk} (see also~\cite{Choudhary:2020pxy, Zhang:2022rex, Aurrekoetxea:2023jwk}), which consider the GW waveforms of binary black hole (BBH) mergers within dense stellar envelopes. Such configuration is motivated by the theoretical possibility of BBH formation through the dynamical fragmentation of a rapidly spinning massive star core undergoing gravitational collapse~\cite{Loeb:2016fzn, Reisswig:2013sqa}. For example,~Ref.~\cite{Woosley:2016nnw} showed that the progenitor of \texttt{GW150914} could be a star with mass~$\gtrsim150M_\odot$ going through chemically homogeneous evolution with no mass loss, which could then form two BHs at its centre due to bifurcation of angular momentum in the disk during core collapse. The BBH inspiral would happen then within a stellar envelope of density~$\rho\gtrsim 10^6\densityUnit $~\cite{Woosley:2016nnw, Dai:2016ejj}. Here, we use NR simulations as a testbed to show that: (i) a \texttt{GW150914}-like event would be detected (with a false alarm rate smaller than~$10^{-4}\, \mathrm{yr^{-1}}$) using a template bank of vacuum waveforms, even when immersed on a stellar density larger than~$10^{7}\densityUnit$; (ii) routinely performed tests of vacuum GR might not flag (unmodelled) environmental effects, even when they lead to considerable biases in parameter estimation (PE). Additionally, we demonstrate, for the first time, that phenomenological models can be highly effectual in detecting environmental effects and resolving the systematic biases. We use a geometrized unit system with $G = c = 1$.

\section{How do environments affect GW waveforms?}

One natural way to tackle such question is to perform NR simulations of CBCs in environments, focusing on the effects on compact binary dynamics and associated GW waveforms. Such approach involves the fewer assumptions. But, for the densities of most astrophysical environments, the effects on GW waveforms are expected to be small and need to accumulate over several cycles, which can make a full NR approach unfeasible. Additionally, it can be very ineffective to explore the large parameter space and plethora of astrophysical environments using \emph{only} numerical simulations. Phenomenological (semi)analytic models of the leading environmental effects (e.g., accretion or dynamical friction) for a given system constitute an alternative simpler approach. But, as these are based on many assumptions and approximations, without (NR) simulations to compare with, it is hard to assess how accurate (or trustable) these environment waveforms are. 

In this section, we consider for the first time an hybrid approach to environmental effects on GWs, comparing phenomenological against NR environment waveforms, analysing the systematic biases incurred by not including environmental effects in the recovery model for the Bayesian PE, and demonstrating how phenomenological waveforms can resolve them. A study of systematic biases from environmental effects using NR waveforms was performed in Ref.~\cite{Leong:2023nuk} for a toy model consisting of a BBH inside a very dense scalar field bubble. 

\subsection{NR Waveforms}

We use publicly available waveforms from GR hydrodynamics simulations performed in Ref.~\cite{Fedrow:2017dpk}, which made use of the \texttt{Einstein Toolkit}~\cite{EinsteinToolkit:2024_05, Loffler:2011ay} to evolve the Einstein's equations in the BSSN formalism~\cite{Shibata:1995we, Baumgarte:1998te} with fourth-order finite differences and adaptative mesh refinement (for further details, we direct the reader to Ref.~\cite{Fedrow:2017dpk}). The constraint satisfying initial data describes quasi-circular orbits of two equal-mass non-spinning punctures at a separation of~$11.6 M$, with~$M$ the total BBH mass. 

The stellar material was modelled as a gas with a~$\Gamma$-law equation of state~$P=(\Gamma-1)\rho \epsilon$, with~$\Gamma=4/3$ and setting the initial specific internal energy~$\epsilon$ assuming a gas dominated by relativistic degenerate electrons~\cite{Shapiro:1983du}. The density~$\rho$ was smoothly tapered to an atmosphere value of~$10^{-16}M^{-2}$ outside of a radius of~$80 M$. Four different waveforms were generated corresponding to the initial (homogeneous) densities~$\rho_0 M^2=\{10^{-10},10^{-9},10^{-8},10^{-7}\}$.\footnote{A fiducial vacuum simulation was also performed for comparison.} In this work, we focus on a \texttt{GW150914}-like event with total mass~$M=60 M_\odot$, which scales the initial puncture separation to~$1030\,{\rm km}$ and initial densities to~$\rho_0\approx\{10^{4.2},10^{5.2},10^{6.2},10^{7.2}\}\densityUnit$. We note that the initial sound speed is~$c_s=\sqrt{\dd P/\dd \rho}\sim 10^{-4}[\rho_0/10^{7.2}\mathrm{g\,cm^{-3}}]^{1/6}$.

The environmental effects on the GW waveforms are remarkably simple. Even for the largest density considered, the environment waveform retains all general features of the vacuum waveform, acquiring only a considerable de-phasing with respect to it (cf. Fig.~3 of Ref.~\cite{Fedrow:2017dpk}). The interaction with the environment (e.g., via dynamical friction) accelerates the inspiral.

\subsection{Phenomenological Waveforms}\label{sec:phenom}

The initial velocity of the binary components in the simulations is~$v_a\gtrsim 0.1\gg c_s$, implying that the motion is highly supersonic.	
To construct an environment waveform, we start by parametrizing the change in each BH mass by the Hoyle-Littleton accretion rate~\cite{Hoyle1941, Edgar:2004mk}
\begin{equation}
	\dot{m}_a = \Big(\frac{1+q}{q^{2-a}}\Big)^3\frac{4 \pi m_a^2 \tilde{\rho}_{\rm accr}}{v^3}\,, \quad (a=1,2)
\end{equation}
and the total torque from dynamical friction as~\cite{CanevaSantoro:2023aol}
\begin{equation}
	\frac{\dot{L}_z^{\rm df}}{M}=-\Big(\frac{1-2 \eta}{\eta^2}\Big)\frac{4\pi M^2 \tilde{\rho}_{\rm df}}{v^4}\,,
\end{equation}
with the effective velocity~$v\equiv (M \Omega)^{1/3}$, where~$\Omega$ is the orbital angular frequency, and the mass-ratios~$q\equiv m_2/m_1\leq1$ and~$\eta\equiv q/ (1+q)^{2}\leq 1/4$. We defined the effective densities~$\tilde{\rho}_{\rm acc}\equiv \bar{\rho}\, \mathcal{I}_{\rm acc}$ and~$\tilde{\rho}_{\rm df}\equiv \bar{\rho}\, \mathcal{I}_{\rm df}$, where~$\bar{\rho}$ is the local average gas density, and~$\mathcal{I}_{\rm accr}$ and~$\mathcal{I}_{\rm df}$ are functions of~$v/c_s$.
In this work, we restrict to quasi-circular orbits of symmetric binaries ($q\approx 1$).

A word of caution is in order: the above parametrizations have been shown both analytically and numerically to be effectual for individual objects or widely separated binaries, but they are not guaranteed to provide a good description of symmetric binaries in the supersonic regime, since, for such systems, the Hoyle-Lyttleton radius of each component is~$R_{\rm HL}\equiv 2 m_a/v_a^2\sim 4 R$, i.e., four times larger than the orbital separation distance. So, here we treat~$\tilde{\rho}_{\rm acc}$ and~$\tilde{\rho}_{\rm df}$ as unknown constants, assuming our model can still capture accurately the dynamics of these systems. By comparison with the NR simulations, we will show (in Sec.~\ref{sec:PE}) that not only do our phenomenological waveforms match remarkably well the NR ones, but also that~$\tilde{\rho}_{\rm acc}$ and~$\tilde{\rho}_{\rm df}$ provide good estimators of the asymptotic density~$\rho_0$. 

From angular momentum conservation, one finds at Newtonian order that, in the early inspiral, the orbital frequency evolution is described by
\begin{equation}\label{eq:orbit_phasing}
	\begin{split}
		M^2 \dot{\Omega}&=\frac{96}{5}\frac{q}{(1+q)^2}v^{11}+12 \pi q\, \tilde{\rho}_{\rm df} M^2 \\
		&\quad+3 \Big[q \dot{m}_1+\frac{\dot{m}_2}{q}+\frac{2}{3}\dot{M}\Big]v^3\,,
	\end{split}
\end{equation}
where the first term in the right-hand side originates from the torque of (quadrupole) GW radiation reaction, the second term from dynamical friction, and the last one from the change in the binary's momentum of inertia. 
We construct environment waveforms using the stationary phase approximation (SPA)~\cite{Cutler:1994ys, Droz:1999qx}, where the inspiral waveform in frequency domain for the (leading) quadrupole mode is given by~$h(f)\propto (M^2 \dot{\Omega})^{-1/2} e^{i(2\pi f t(f)-\phi(f)-\pi/4)}$, with~$t$ defined as the instant at which~$\Omega(t)=\pi f$, i.e.,
\begin{equation}
	t(f)=t_{\rm c}-\int_f^{+\infty}\dd f' \frac{\pi}{\dot{\Omega}(f')}\,,
\end{equation}
and the waveform phase
\begin{equation}
	\phi(f)=\phi_{\rm c}-\int_f^{+\infty} \dd f' \frac{2 f'}{\dot{\Omega}(f')}\,,
\end{equation}
where~$\dot{\Omega}$ is found from Eq.~\eqref{eq:orbit_phasing} and~$v=(\pi M f)^{1/3}$. The amplitude and phase corrections to the inspiral waveform are implemented in the IMR models \phXP~\cite{Pratten:2020ceb} for BBH and \phPvT~\cite{Dietrich:2019kaq} for BNS systems, which incorporate spin-induced precession effects.\footnote{Refs.~\cite{Barausse:2014tra, Cardoso:2019rou, CanevaSantoro:2023aol} have considered the leading-order corrections to the inspiral phasing in the waveform.}

\subsection{Overlap and Systematic Biases}
\label{sec:PE}

We now compare the (injected) NR waveforms against the vacuum and environment models to show that: (i) our phenomenological environment waveforms have a large overlap with NR waveforms; (ii) environmental effects can significantly bias the BBH parameters when recovering them with a vacuum model; and (iii) our phenomenological waveforms can successfully resolve the systematic bias and accurately estimate the environment density.

We use the \lalsimulation package from the LIGO Algorithms Library (LAL) software suite to generate the waveforms~\cite{lalsuite}, and perform the PE analyses using the \bilby package~\cite{Ashton:2018jfp} with the \dynesty sampler~\cite{Speagle:2019ivv}.
The lower cutoff frequency is set to 26\Hz, as limited by the NR simulations. 
The injections were simulated assuming a zero noise realization, weighted by the \texttt{GW150914} event PSDs of Hanford (H1) and Livingston (L1) detectors, and a $\mathrm{SNR}{\,\approx\,} 24$.

\begin{figure}[t]
    \centering
    \includegraphics[width=\linewidth]{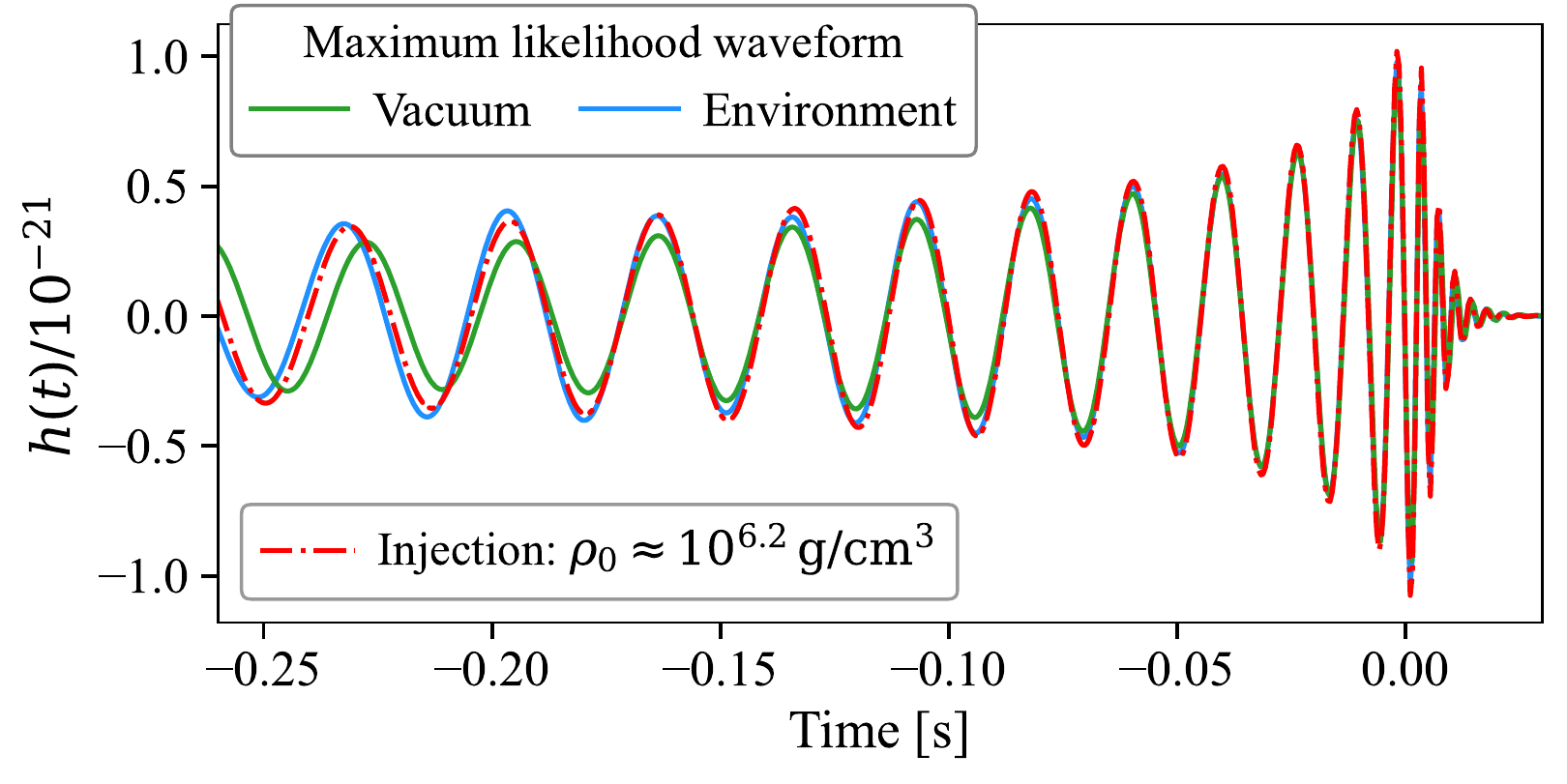}
    \caption{Injected NR waveform (red dash-dotted) of a non-spinning \mbox{$(30+30)\msun$} BBH coalescence in stellar material of density $\rho_0 \approx 10^{6.2}\, \densityUnit$, compared against the maximum likelihood waveforms obtained from Bayesian analyses with vacuum (green) and phenomenological environment (blue) models, shown for the H1 detector. Our environment model has a larger overlap with NR waveforms than vacuum.
    }
    \label{fig:compare_maxL_wf}
\end{figure}

To quantify the agreement between the recovered waveform (\hrec) and the injection waveform (\hinj) for a network of detectors, we compute the network overlap (\ovpnet) as given in Ref.~\cite{Becsy:2016ofp, Roy:2022teu},
\begin{equation}\label{eq:overlap}
\ovpnet \equiv \frac{ \sum_k \inp{\hinj^k}{\hrec^k} }{ \left(  \sum_k\inp{ \hinj^k }{\hinj^k} \cdot \sum_k\inp{\hrec^k}{\hrec^k} \right)^{1/2} }\, , 
\end{equation}
where $k$ represents the $k\nth$ detector. We consider the configuration of two detectors, H1 and L1. The angular brackets, $\inp{\cdot}{\cdot}$, represent the noise-weighted inner product, defined as
\begin{equation}
 \inp{a}{b} \equiv 4 \Re \int_{\flow}^{\fhigh}  df \, \frac{ \tilde{a}^{\ast}(f) \, \tilde{b}(f) }{S_n (f)} \,, 
\end{equation}
where $S_n(f)$ is the one-sided power spectral density (PSD) of the detector noise, and $ \tilde{a}(f) $ and $\tilde{b}(f) $ denote the Fourier transform of $a(t)$ and $b(t)$, respectively. 

Figure~\ref{fig:compare_maxL_wf} shows the comparison of an injected NR waveform for a BBH coalescence inside a star core against the maximum likelihood waveforms obtained from a Bayesian analysis using the vacuum and phenomenological environment waveforms for the H1 detector. The larger overlap of our phenomenological environment model with NR, as compared against vacuum waveforms, is evident. Quantitatively, in terms of what a network of the detectors H1 and L1 can ``see'', we find a network overlap with the NR waveform of~$\ovpnet=0.974$ for the vacuum and~$\ovpnet=0.994$ for the environment models.

\begin{figure}
    \centering
    \includegraphics[width=\linewidth]{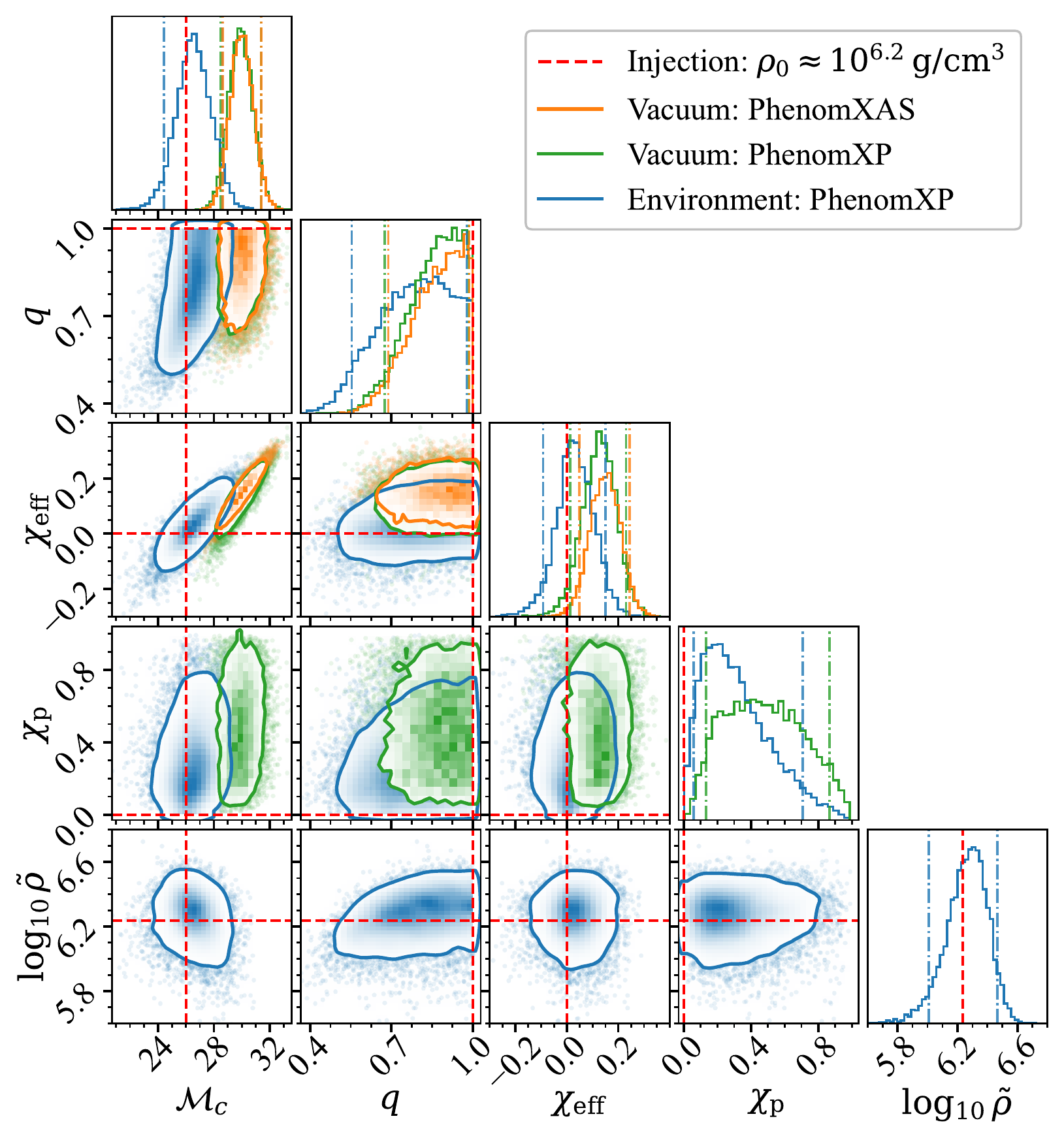}
    \caption{Marginalized posterior distributions for an injection of a NR waveform of a non-spinning \mbox{$(30+30)\msun$} BBH coalescence in stellar material of density $\rho_0\approx 10^{6.2}\densityUnit$. The corner plot compares the recovered posteriors with vacuum (orange for aligned-spin, green for precessing-spin) and phenomenological environment (blue) models. The red dashed lines indicate the injected values. The vertical dashed-dot lines in 1D histograms represent the 90\% credible interval. The non-inclusion of environmental effects leads to biases in the BBH parameters.}
    \label{fig:contour_G3sim}
\end{figure}

Figure\:\ref{fig:contour_G3sim} shows a corner plot comparing the recovered marginalized posterior distributions with vacuum and environment models from an injected NR waveform of a BBH coalescence inside a star core (the true parameters are indicated by red dashed lines). This figure clearly demonstrates that the non-inclusion of environmental effects on the recovery model can lead to significant systematic biases in the estimation of the BBH parameters (note, e.g., the 90\% credible regions on the \mchirp--\chieff, \mchirp--\chip, and \chieff--\chip planes, which are significantly away from the true values).\footnote{As customary in GW astronomy, instead of the individual component spins $\vec{\chi}_{1,2}$ (which are not very well determined), we use the effective spin \chieff along the Newtonian orbital angular momentum and the effective precession spin \chip~\cite{Hannam:2013oca, Schmidt:2014iyl}. The latter depends on the in-plane spin components and drives orbital precession.}

It is also seen that the phenomenological environment waveforms can accurately estimate the order of magnitude of the density as shown in the bottom panel of Fig.~\ref{fig:contour_G3sim}; we used~$\rhot \equiv \frac{1}{2}(\rhot_{\rm df}+\frac{1}{2}\rhot_{\rm acc})$, since $\rhot_{\rm df}$ and $\frac{1}{2}\rhot_{\rm acc}$ are completely degenerate for~$q=1$. Nevertheless, we note that one can measure~$\rhot_{\rm df}$ and~$\rhot_{\rm acc}$ individually, even for~$q\approx1$, though with much larger uncertainty than~$\tilde{\rho}$. For asymmetric binaries, $\rhot_{\rm df}$ and~$\rhot_{\rm acc}$ are expected to be measured with higher precision.  We also note that \rhot is correlated with $\mathcal{M}_c$ and $q$, such that a higher $\mathcal{M}_c$ with a lower \rhot can be compensated by a lower $\mathcal{M}_c$ with a higher \rhot, and similarly, a lower $q$ with a lower \rhot can offset a higher $q$ with a higher \rhot. Our phenomenological waveforms can be used to derive evidence for the presence of environmental effects against the vacuum hypothesis, for the NR simulation with $\rho_0\approx 10^{6.2} \densityUnit$, we find $\log_{10} \mathcal{B}_{\mathrm{vac}}^{\mathrm{env}}=1.3$.

\section{Can we detect signals with environmental effects using a vacuum template bank?}
\label{sec:search}

We have just demonstrated that the non-inclusion of environmental effects in the recovery model for PE can lead to substantial systematic biases: but is this a problem? After all, it could well be possible that we miss any waveform with large environmental effects, when doing a matched-filter search using a vacuum template bank. We show here that a \texttt{GW150914}-like event would be detected with a false alarm rate smaller than~$10^{-4}\, \mathrm{yr^{-1}}$, using a template bank of vacuum waveforms as in the first observing run (O1) of Advanced LIGO, even when immersed on a stellar core with density larger than~$10^{7}\,\densityUnit$. For such large densities, the biases in the BBH parameters are even stronger than the ones shown in Fig.~\ref{fig:contour_G3sim} (see Fig.~\ref{fig:contour_G4sim}).

To assess the detectability of the environment waveforms using a vacuum template bank as in the O1 run of advanced LIGO, we first calculate the fitting factor of the NR environment waveform against the vacuum template bank. Then, we perform a software injection study to determine the sensitivity range for the detection of the NR environment waveform. To reduce the computational cost of these analyses, we truncate the O1 template bank~\cite{LIGOScientific:2016vbw, Capano:2016dsf} by restricting the total mass ($M$) to be more than 30\msun and the mass ratio ($q$) to be greater than 0.1. Thus, the truncated template bank ($\tbank$)  spans the mass parameter ranges of $M \in  [30, 100]\msun$ and $q \in [0.1, 1]$, and the individual dimensionless spin magnitude $\abs{\chi_{1,2}} \in[0.0, 0.99]$. In this detectability study, we meticulously retain the O1 search configuration as used in \pycbc pipeline~\cite{pycbc-config}. We employ the \seobnr waveform model~\cite{Taracchini:2013rva} and carry out the analyses using \pycbc framework~\cite{Usman:2015kfa, Allen:2005fk, Allen:2004gu, Nitz:2017svb, DalCanton:2014hxh} over a frequency band spanning $[30,1024]\Hz$.

\begin{table}[t]
    \centering
    \begin{tabularx}{0.40\textwidth}{X c c c}
        \toprule[1pt]
        \toprule[1pt]
        $\rho_0\: [\densityUnit]$  & $10^{7.2}$ & $10^{6.2}$  & Vacuum  \\
        \midrule[0.5pt]
        Fitting Factor  & 0.915  & 0.967 & 0.987  \\
        \bottomrule[1pt]
        \bottomrule[1pt]
    \end{tabularx}
    \caption{Fitting factor of the NR environment waveform with the vacuum template bank used in O1 searches at LIGO.}
    \label{tab:FittingFactor}
\end{table}

The fitting factor (\ff) between an arbitrary signal ($h_a$) and the template bank (\tbank) is defined as the maximum match (\match) value across all templates in the bank~\cite{Apostolatos-1995}, 
\begin{equation}
\displaystyle{\ff(h_a) = \max_{\lambda \in \tbank} \match \left(h_a, h(\lambda) \right)}.
\end{equation}
The match is defined as the noise power spectral density (PSD) weighted inner product between two normalized waveforms, maximized over reference time and phase. We employ the harmonic PSD, calculated as the harmonic mean of the PSDs from LIGO Hanford (H1) and Livingston (L1). The same PSD is utilized to construct the template bank. The quantity $1-\ff(h_a)$ indicates the minimum fractional loss in the recovered matched-filter SNR over the template bank. 
Table~\ref{tab:FittingFactor} lists the fitting factors for three different injections of NR environment waveforms. It indicates a loss in SNR of only $8\%$, even in extremely dense environments $(\rho_0\sim10^{7.2}\densityUnit)$.

\begin{figure}[t]
    \centering
    \includegraphics[width=0.98\linewidth]{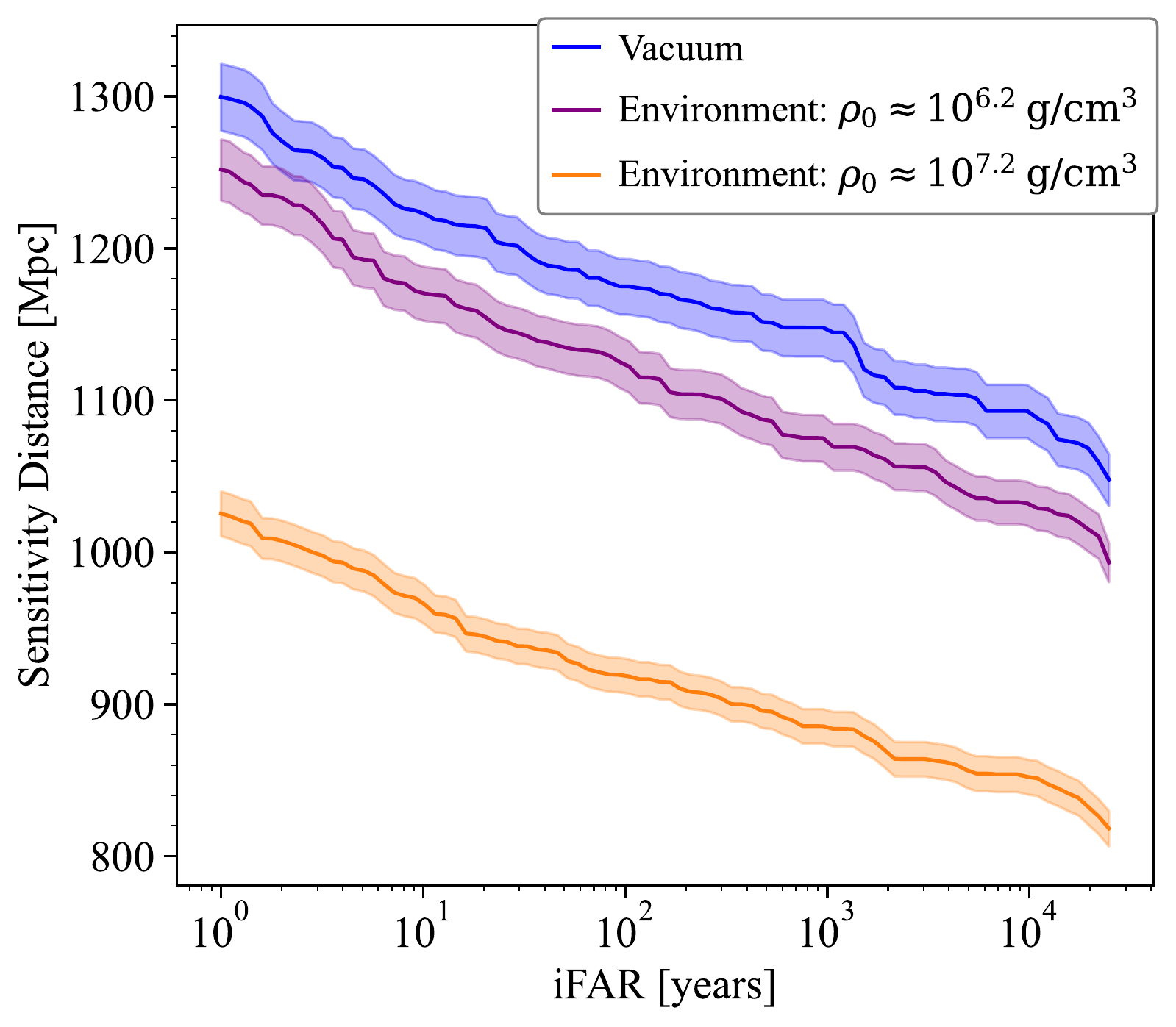}
    \caption{Sensitivity distance to an NR waveform of a non-spinning \mbox{$(30+30)\msun$} BBH coalescence in stellar material of density $\rho_0$, shown as a function of inverse false-alarm rate (iFAR) for a O1-like search with a vacuum template bank. Solid lines indicate the median of the distribution and shaded regions show the $\pm 1\sigma$ uncertainty regions. We include the sensitivity to a vacuum signal from the same BBH system. Even in a large density environment as~$\rho_0\sim 10^{7.2}\densityUnit$, a \texttt{GW150914}-like event would be detected.}
    \label{fig:sensitivity_dist}
\end{figure}

The ability to detect a GW signal depends not only on the waveform model faithfulness, but also on other factors, like the detector noise levels, the detector alignment and calibration, the source location and orientation, and the binary parameters. To quantify the sensitivity of a search pipeline, we generate a large number of CBC sources based on a prescribed astrophysical model. The sensitivity is commonly measured as the fraction of detected sources below a specified false-alarm rate (FAR) threshold, $\mathcal{F}$. We estimate the volume around a detector (or a network of detectors) within which sources are detectable by~\cite{Usman:2015kfa}
\begin{equation}
\displaystyle \mathcal{V}(\mathcal{F}) = \mathcal{V}_{\rm inj} \int  \epsilon(\mathcal{F};\vec{\lambda}) \, p_{\rm{pop}}(\vec{\lambda})  \, d\vec{\lambda},
\end{equation}
where $\epsilon$ is the probability of recovering a signal produced by a source with a FAR below~$\mathcal{F}$, whose source parameters are represented by $\vec{\lambda}$, respectively. The function $p_{\rm{pop}}$ describes the distribution of the source population, and~$\mathcal{V}_{\rm inj}$ the volume within which it is sampled from.

To evaluate the sensitivity of the search, we conducted a software injection study over ten days of LIGO’s O1 data. We generated a population of BBHs with intrinsic parameters fixed to a \texttt{GW150914}-like system evolving inside a star core of density~$\rho_0$. These sources were randomly distributed across the sky, with varying orientations and distances, and injected uniformly in time with a 256-second interval. Figure\:\ref{fig:sensitivity_dist} depicts the sensitivity distance to environment waveforms in a search with a vacuum template bank, for different matter densities (including vacuum). Even for a very dense environment with $\rho_0\sim 10^{7.2}\densityUnit$, the sensitivity distance is reduced only by $\sim25\%$ as compared to a source in vacuum: still much larger than \texttt{GW150914}'s luminosity distance ${\approx\,}\SI{440}{\Mpc}$. For reference, \texttt{GW150914} was detected with an $\rm{FAR}\lesssim 5\times 10^{-6}\mathrm{yr^{-1}}$~\cite{LIGOScientific:2016vbw}.

While this can lead to a loss of many events ($\sim 58\%$ as compared to vacuum), we conclude that existing searches with vacuum template banks would be able to detect a significant fraction of \texttt{GW150914}-like BBH systems in gaseous environments, even if the coalescence occurs within an environment as dense as $\rho_0\sim 10^{7.2}\densityUnit$. Finally, we remark that the recently developed beyond-GR search framework with post-Newtonian (PN) deviation terms~\cite{Narola:2022aob, Sharma:2023djw} are expected to be more efficient in detecting signals with environmental effects.

\begin{figure*}
    \centering
    \includegraphics[width=0.9\linewidth]{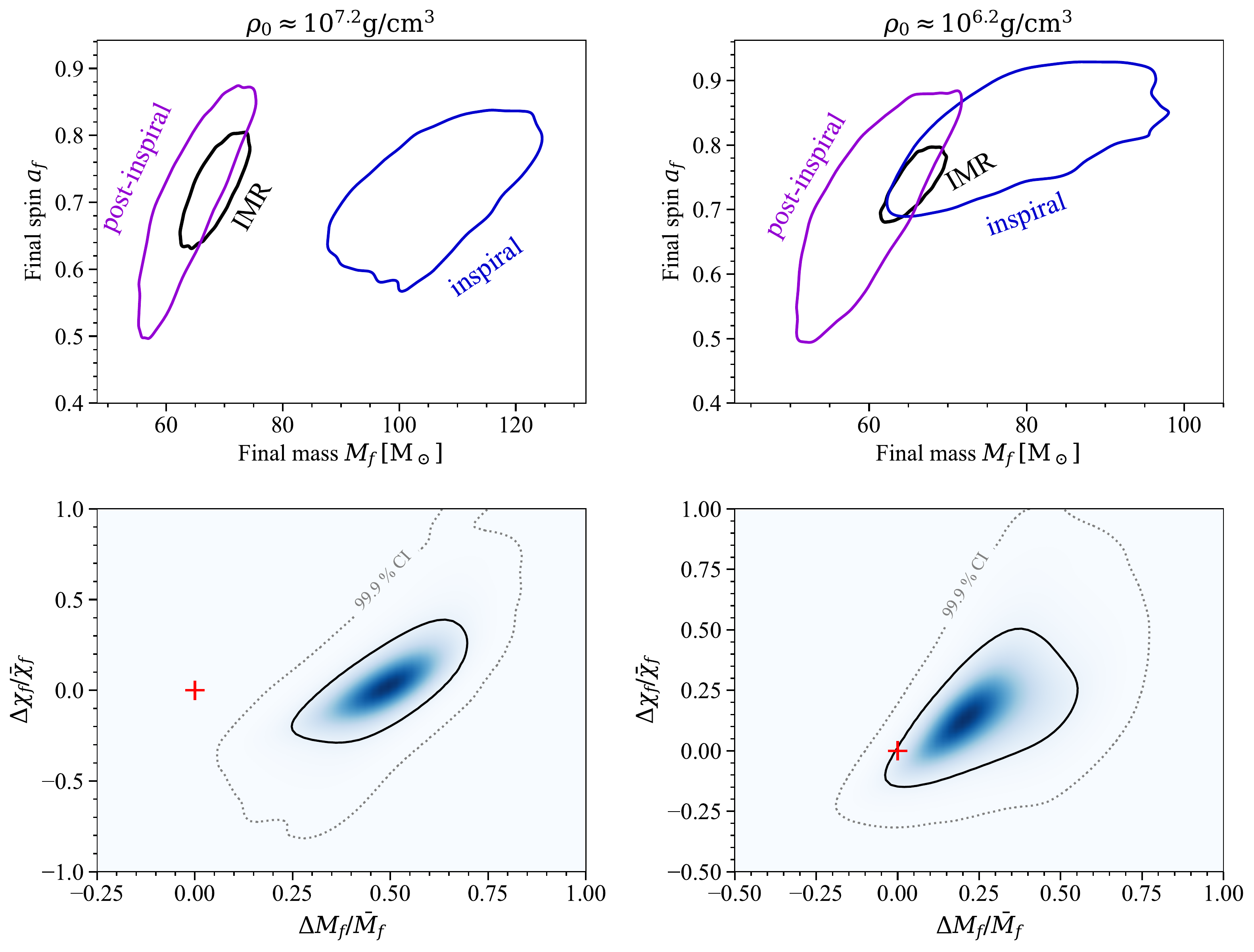}
    \caption{Results from an IMR consistency test. {\sc Top panel:}
    solid contours represent the 90\% credible region of the joint probability distribution for the mass and spin of the final object, as obtained from the inspiral and post-inspiral parts of the signal, and from the full IMR signal. 
    {\sc Bottom panel:} joint probability distribution of the fractional deviation parameters $\Delta M_f/\bar{M}_f$ and $\Delta a_f/\bar{a}_f$. The injected waveforms were generated using the NR simulations of a $(30+30)\msun$ BBH coalescence inside a star core of density~$\rho_0$, using the \texttt{GW150914}-event PSD of H1 and L1, and an injected $ \mathrm{SNR\,}{\approx\,}24$.}
    \label{fig:imr_consistency}
\end{figure*}

\section{Can signals with environmental effects pass routine vacuum GR tests?}
\label{sec:tgr}
We have shown that GW signals with large environmental effects can be detected with current searches using vacuum templates, and that their non inclusion in the recovery model might lead to substantial systematic biases in PE. But would the routinely performed vacuum GR tests be able to flag the presence of environmental effects? In this section, we show that, for a range of environment densities (which depends on the particular binary system), GW signals can pass vacuum GR tests even when leading to systematic biases.

We consider the routinely performed Inspiral-Merger-Ringdown (IMR) consistency test~\cite{Ghosh:2016qgn, Ghosh:2017gfp}, which examines the consistency between low- and high-frequency components of the signal, and the parameterized deviations test~\cite{Li:2011cg, Agathos:2013upa, Meidam:2017dgf, roy:2024tiger}, that probes modifications to the waveform by varying PN and phenomenological coefficients. Here, we assess the effectiveness of such tests in spotting environmental effects in GW signals.

\subsection{IMR Consistency Test}
The IMR consistency test is based on the comparison of the remnant’s mass and spin estimates obtained from the low- and high-frequency segments of the waveform. These estimates rely on the relationship between the binary component parameters---specifically, their (redshifted) masses and dimensionless spins---and the remnant's mass and spin, as provided by NR calibrated fits. We divide the signal between inspiral and post-inspiral regimes using a cutoff frequency~$f_c^{\mathrm{IMR}}$, corresponding to the dominant GW frequency of the innermost stable circular orbit of the remnant (Kerr) BH, determined from the analysis of the full IMR signal. 
The posterior distributions of the remnant's mass $(M_f)$ and dimensionless spin $(\chi_f)$ are obtained from the inspiral and from the post-inspiral parts of the signal. If the vacuum GR hypothesis is true, the joint posterior of $(M_f, \chi_f)$ obtained from the two separate analyses are expected to have a significant overlap. 

The top left (right) panel of Fig.\:\ref{fig:imr_consistency} shows the 90\% credible region of the 2D posteriors of $M_f$ and $\chi_f$ for injected NR waveforms of a non-spinning \mbox{$(30+30)\msun$} BBH coalescence in a star core of density $\sim 10^{7.2}\,\, ( 10^{6.2}) \densityUnit$. We also provide the 90\% credible region using the full IMR signal. For the larger density~($10^{7.2} \densityUnit$), the inspiral contour is well separated from the post-inspiral one, which indicates a deviation from vacuum GR; thus, the environmental effects are spotted. However, as shown in the top-right panel, for a smaller density ($ 10^{6.2} \densityUnit$), the IMR test cannot exclude vacuum GR, as the 90\% contours have significant overlap and the IMR contour is within their intersection.

\begin{figure*}[ht]
    \centering
    \includegraphics[width=0.9\linewidth]{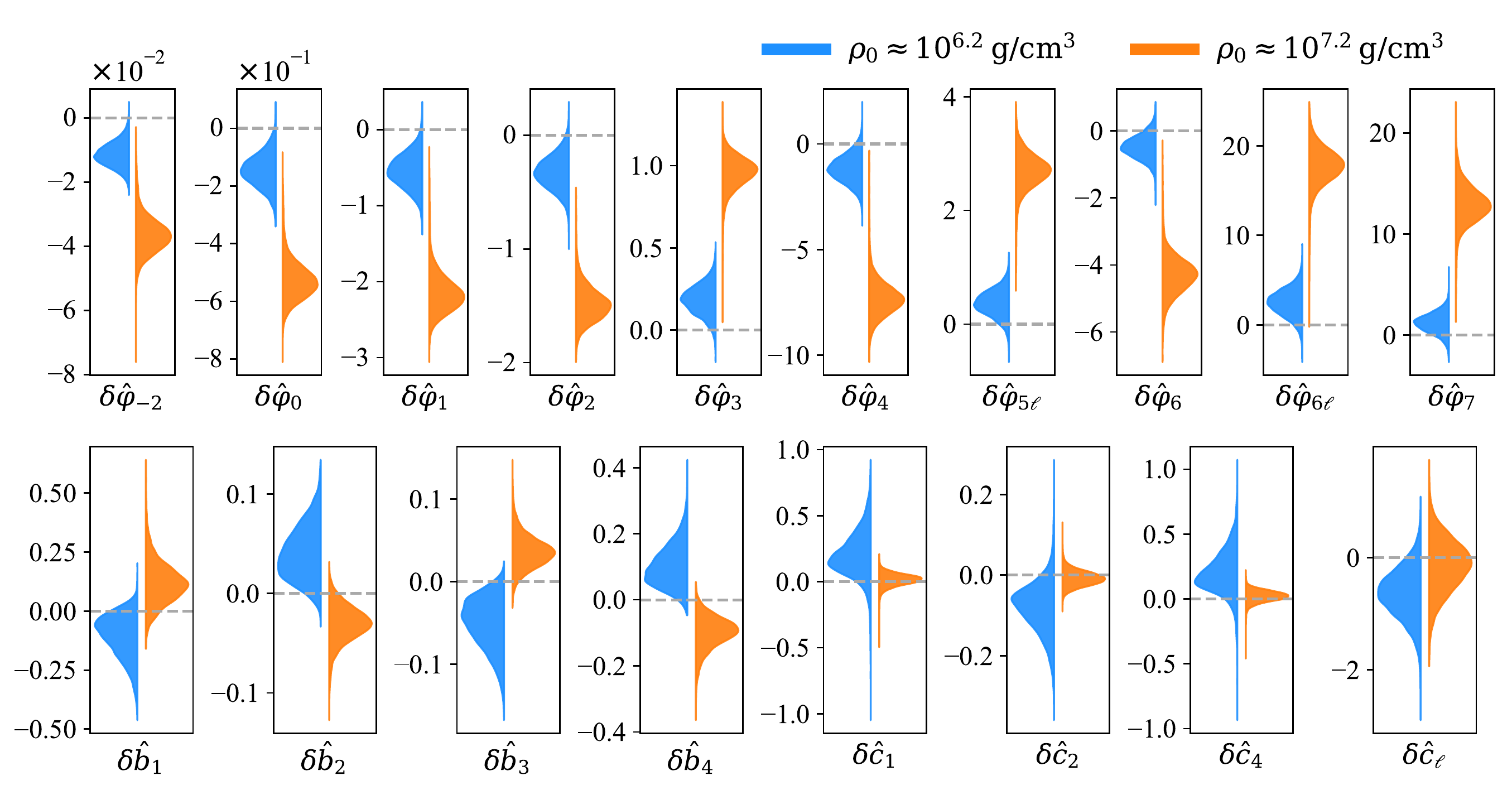}
    \caption{Results from a parametrized test of vacuum GR, showing the marginalized posterior distributions of deviation parameters. The injected waveforms were generated using the NR simulations of a $(30+30)\msun$ BBH coalescence inside a star core of density~$\rho_0$, using the \texttt{GW150914}-event PSD of H1 and L1, and an injected $\mathrm{SNR\,} {\approx\,}24$.}
    \label{fig:violin_tiger}
\end{figure*}

To assess the significance of our findings, we introduce the fractional deviations~$\Delta M_f/\bar{M}_f$ and $\Delta a_f/\bar{a}_f$,
\begin{equation}
\displaystyle \frac{\Delta M_f}{\bar{M}_f} \equiv 2 \frac{M_f^{\mathrm{I}} - M_f^{\mathrm{PI}}}{M_f^{\mathrm{I}} + M_f^{\mathrm{PI}}}\,, \quad
\frac{\Delta \chi_f}{\bar{\chi}_f} \equiv 2 \frac{\chi_f^{\mathrm{I}} - \chi_f^{\mathrm{PI}}}{\chi_f^{\mathrm{I}} + \chi_f^{\mathrm{PI}}}\,,
\end{equation}
where $M_f^{\mathrm{I}}$ ($\chi_f^{\mathrm{I}}$) and $M_f^{\mathrm{PI}}$ ($\chi_f^{\mathrm{PI}}$) represent the final mass (spin) of the remnant BH obtained from the inspiral and post-inspiral portions of the signal, respectively. For a vacuum GR signal, the joint distribution of these fractional deviations should be consistent with $(0, 0)$.\footnote{We obtained the joint probability distribution of the fractional deviations using the \texttt{summarytgr} code in \pesummary~\cite{Hoy:2020vys}.}

The bottom panels of Fig.\:\ref{fig:imr_consistency} show the 90\% and 99.9\% credible regions of the 2D posterior distributions of the fractional deviation parameters. For a \texttt{GW150914}-like event in a star core of density~$\sim 10^{7.2} \densityUnit$, the vacuum GR hypothesis is ruled out at confidence level $\gg 99.9\%$. But for an environment density of~$\sim 10^{6.2} \densityUnit$, the vacuum GR hypothesis is consistent with the injected NR waveform at a confidence level of 88\% (vacuum GR cannot be ruled out, environmental effects might be missed).
 
The significant deviation seen for the largest density could be anticipated. In that case, due to the high accretion rates, the relative change in the BH masses during the coalescence is non-negligible: the (dominant) ringdown frequency is $265\Hz$, which should be compared with $296 \Hz$ for the same initial BBH in vacuum (for the simulation in~$\rho_0\sim 10^{6.2}\densityUnit$, this frequency is~$290 \Hz$).

\subsection{Parametrized Deviations Test}
The parametrized tests probe departures from vacuum GR by allowing deviations ($\dphi_k$) to the phasing coefficients $(\phi_k^{\mathrm{GR}})$ of frequency domain vacuum GR waveforms,
\begin{equation}
\phi_k^{\mathrm{GR}} \rightarrow (1+\dphi_k)\: \phi_k^{\mathrm{GR}}.
\end{equation}
LVK analyses so far have employed two implementations: the FTI approach~\cite{LIGOScientific:2018dkp, Mehta:2022pcn, Sanger:2024axs}, which can be applied to any aligned-spin frequency-domain waveform model, and the TIGER approach~\cite{Li:2011cg, Agathos:2013upa, Meidam:2017dgf}, which is based on a (frequency-domain) phenomenological waveform family. In this study, we perform the analyses using the TIGER framework, specifically implemented within the $\phX$ waveform family in \lalsuite~\cite{roy:2024tiger, lalsuite}, and use the \bilbytgr package~\cite{bilby-tgr} for the PE. For the dominant mode, the phasing coefficients in the $\phX$ waveform family are divided into three regimes~\cite{Pratten:2020fqn}:
\begin{equation*}
\big\{\underbrace{ \varphi_0,...,\varphi_7,  \varphi_{5\ell}, \varphi_{6\ell}}_{\text{\normalsize inspiral}}, \  \underbrace{ b_1, b_2, b_3, b_4}_{\text{\normalsize intermediate}}, \ \underbrace{c_1, c_2, c_4, c_L}_{\text{\normalsize merger-ringdown}}\big\},
\end{equation*}
where the inspiral terms are PN coefficients, and the intermediate and merger-ringdown terms are phenomenological coefficients. Several alternative theories of gravity involving scalar fields predict non-zero contributions at the $-1$PN and 0.5PN orders, which are absent in (vacuum) GR~\cite{LIGOScientific:2018dkp}. Routinely performed LVK analyses also include modifications at these orders, corresponding to changes in phase \mbox{$\Delta\Phi_k = \dphi_k \frac{3}{128\eta} v^{k-5}$}, where $k = -2, 1$ correspond to the $-1$PN and 0.5PN terms, respectively. Here, we perform the parametrized vacuum GR test on injected environment waveforms for each coefficient from $-1$PN to 3.5PN ($\dphi_k$), and for the intermediate ($\delta\hat{b}_k$) and merger-ringdown ($\delta\hat{c}_k$) phenomenological coefficients, using the \phXP model~\cite{Pratten:2020ceb}.

Figure\:\ref{fig:violin_tiger} shows in blue (orange) the posterior distribution of the deviation parameters for injected NR waveforms of a non-spinning \mbox{$(30+30)\msun$} BBH coalescence in a star core of density $\sim 10^{6.2}\,\, ( 10^{7.2}) \densityUnit$. For the denser case, all inspiral deviation parameters strongly rule out vacuum GR with a credible interval far beyond $3\sigma$ (99.7\%). There are slight deviations in the intermediate parameters, while the merger-ringdown ones remain consistent with vacuum GR. For a density~$\sim 10^{6.2}\densityUnit$, vacuum GR is outside the $2\sigma$ (95.5\%) credible interval, but within the $3\sigma$ interval. This suggests the presence of deviations, though they are not statistically significant enough to indicate a clear inconsistency with vacuum GR. 

To further quantify the significance of these deviations, we consider how often Gaussian noise would bring $\delta \hat{\varphi}_k=0$ (vacuum GR) within the~$2\sigma$ credible interval of the $\delta \hat{\varphi}_k$ posterior. So, we injected the NR waveform for~$\rho_0\sim 10^{6.2}\densityUnit$ with 20 different Gaussian noise realizations and performed the parameterized deviation tests for the $-1$PN and 0.5PN terms. For the $-1$PN (0.5PN) test, we find that in 5 (8) and 12 (15) out of 20 cases, vacuum GR lies within the $2\sigma$ and $3\sigma$ credible intervals, respectively. The vacuum GR is within the posterior boundaries for all the cases.

To assess the effectiveness of the different PN deviation coefficients at spotting environmental effects, we compute the minimum credible interval that encompasses the vacuum GR value, expressed as \mbox{$\QGR(\delta\hat{\varphi}_k=0) = \abs{2 P(\delta\hat{\varphi}_k<0) -1}$}, where $P(\delta\hat{\varphi}_k<0)$ represents the fraction of posterior samples below 0. Among the PN deviation parameters considered, $\QGR$ is highest for the $-1$PN term and lowest for the 3.5PN term, following a descending trend with increasing PN order. This is expected, as the leading-order correction (from both dynamical friction and Hoyle-Lyttleton accretion) enters at $-$5.5PN~\cite{Cardoso:2019rou, CanevaSantoro:2023aol}.

We remark that in the analysis of several real events from GWTC-1/2 (e.g., \texttt{GW150914} and \texttt{GW190814}) vacuum GR was also found at the tail of the posterior, with $\QGR>97\%$. These results were interpreted as systematic errors likely caused by factors such as inaccuracies in GR waveform modeling, calibration uncertainties, and noise artifacts. 

\begin{figure}[t]
    \centering
    \includegraphics[width=0.98\linewidth]{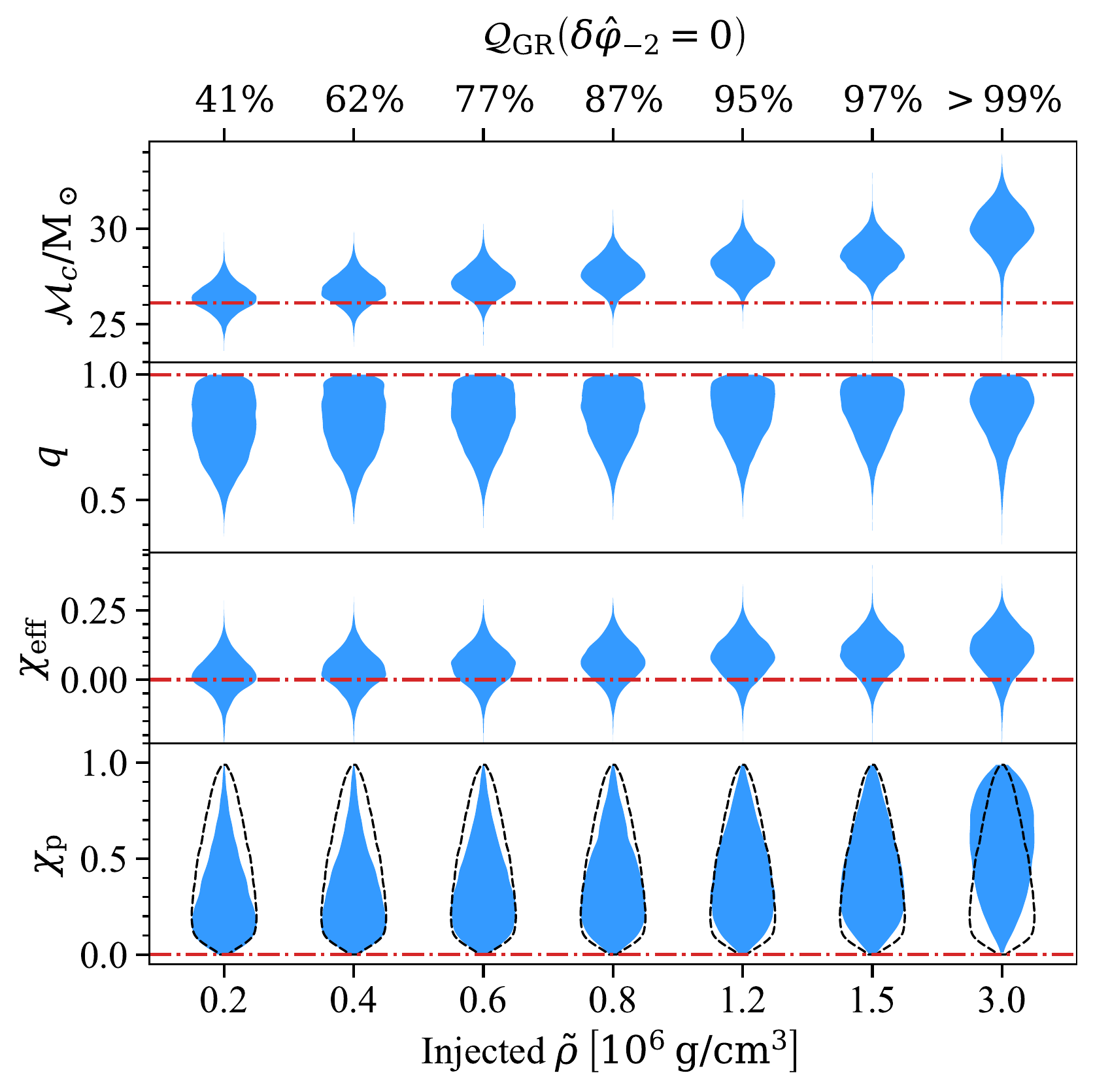}
    \caption{Systematic biases can be present in the posteriors of the chirp mass $\mathcal{M}_c$, mass-ratio $q$, effective spins $\chieff$ and $\chip$ recovered using the vacuum model, even if vacuum GR cannot be confidently excluded. At the top we show the minimum credible interval that encompasses the vacuum GR value $(\QGR)$ for the parametrized $-$1PN test. The black dashed line in the bottom panel represent the $\chip$ prior. We injected \texttt{GW150914}-like phenomenological environment waveforms for a $30+30\msun$ BBH in an environment with effective density~$\tilde{\rho}$, using the \texttt{GW150914}-event PSD of H1 and L1, and an injected $\mathrm{SNR\,}{\approx\,}24$.}
    \label{fig:bias_gw150914-like_inj}
\end{figure}

\subsection{Passing vacuum GR tests while leading to systematic biases}
\label{subsec:systematic}

To quantify the systematic biases for different values of~$\QGR(\delta \hat{\varphi}_2=0)$, which characterizes the effectiveness of a parameterized $-1$PN test in spotting environmental effects, we explore now a range of environment densities. As we have access to only three NR waveforms in the range $\rho_0\in [10^{5.2},10^{7.2}]\densityUnit$, we simulate a series of injections using our phenomenological environment waveforms (introduced in Sec.~\ref{sec:phenom} and validated against NR waveforms in Sec.~\ref{sec:PE}).
As before, we consider a \texttt{GW150914}-like system of a non-spinning $(30+30)\msun$ BBH coalescence in stellar environment with densities in the range $[0.2,3]\times10^6\densityUnit$.  These injections were simulated with a zero-noise realization, using the \texttt{GW150914}-event PSD of H1 and L1, and an injected $\mathrm{SNR} \approx 24$. 

Figure\:\ref{fig:bias_gw150914-like_inj} shows that for densities $\tilde{\rho} \lesssim 1.5\times10^6\densityUnit$, vacuum GR is within the 97\% credible interval of the $\delta\hat{\varphi}_{-2}$ posterior, though strong biases are present in the chirp mass, and moderate ones in the effective spin along the orbital angular momentum. Surprisingly, the biases shift both the~$\mchirp$ and $\chieff$ posteriors towards larger (positive) values as the density increases. Note that the bias in \mchirp towards larger values (which accelerates the coalescence) is so large that needs to be counterbalanced by a bias towards larger positive values of \chieff (slightly delaying it). However, the effect in $\chieff$ is reversed for very large densities~$\gtrsim 10^{7.2} \densityUnit$, when the signal consists mostly of a merger-ringdown stage, in which case the $\chieff$ posterior supports negative values (cf. Fig.\:\ref{fig:contour_G4sim} and Ref.~\cite{Fedrow:2017dpk}). 

A uniform prior on individual spin magnitudes and isotropic orientations results in a non-uniform prior on $\chip$, favouring values around $0.2$. The black dashed line in the bottom panel of Fig.\:\ref{fig:bias_gw150914-like_inj} represents such prior distribution. For non-spinning injections with a zero-noise realization and when environmental effects are small (as for $\tilde{\rho} = 2\times10^5\densityUnit$), the $\chip$ posterior tends to support the lower region of the prior. As the density increases, the $\chip$ posterior shifts toward higher values, retaining a non-zero support near zero (except for very large densities). 
If environmental effects are significant but not included in the PE, they can lead to biased conclusions on the binary's formation history~\cite{Gerosa:2018wbw}.

\section{Discussion and Conclusions}

Despite the general consensus that astrophysical environments might play an important role in the formation and evolution of compact binaries (e.g.,~\cite{Banerjee:2010, Chatterjee:2017, Ziosi:2014sra, Rodriguez:2015oxa, Mapelli:2021gyv, Tagawa:2019osr, Sedda:2023big, Rowan:2023, Ishibashi:2020zzy}), their effects on GW waveforms have only more recently started to be thoroughly investigated~\cite{Barausse:2007dy, Kocsis:2011dr, Yunes:2011ws, Barausse:2014tra, Tamanini:2019usx, Cardoso:2019rou, Toubiana:2020drf, Derdzinski:2020wlw, Zwick:2021dlg, Sberna:2022qbn, Zwick:2022dih, Vijaykumar:2023tjg} (also for dark matter environments~\cite{Annulli:2020lyc, Kavanagh:2020cfn, Baumann:2021fkf, Vicente:2022ivh, Traykova:2023qyv, Duque:2023seg, Cardoso:2022whc, Cole:2022yzw, Tomaselli:2023ysb, Rahman:2023sof, Karydas:2024fcn, Kavanagh:2024lgq}). All detectability studies, including the analysis of real data in Ref.~\cite{CanevaSantoro:2023aol}, have used phenomenological models of environmental effects, which have never been validated against NR simulations, with most studies focusing on future detectors (particularly, on LISA~\cite{Colpi:2024xhw}). The rationale for this has been that environmental effects are more relevant in the early inspiral, since they usually enter at negative PN orders~\cite{Barausse:2014tra}, and to be observable need to accumulate over many cycles during long inspirals (hence, the focus on future observations); this picture is also supported by the findings of Ref.~\cite{Spieksma:2024voy}, which indicate that the merger-ringdown may not be much sensitive to environmental effects. 

There are some astrophysical relevant scenarios where environments could be dense enough for their effects to be significant even for the few cycles observed in the LVK band. One such scenario is if close compact binaries can form from core fragmentation via a non-axisymmetric instability of rapidly rotating massive stars undergoing gravitational collapse~\cite{Loeb:2016fzn, Reisswig:2013sqa}. In these cases, the resultant CBCs could happen within (remnant) dense stellar material~\cite{Loeb:2016fzn, Fedrow:2017dpk}. While this is not a standard formation channel (for some difficulties, see~\cite{Woosley:2016nnw, Dai:2016ejj}), it has not been ruled out by theory. GW observations may soon have the final say~\cite{Dai:2016ejj, Fedrow:2017dpk, CanevaSantoro:2023aol}. These channel could additionally source a gamma-ray burst~\cite{Connaughton:2016umz, Loeb:2016fzn, Woosley:2016nnw}, providing an interesting opportunity for multi-messenger astronomy.

Motivated by current observations and the available NR simulations in environments, we focused here on CBCs within dense stellar material~\cite{Loeb:2016fzn, Reisswig:2013sqa}. We used these GR hydrodynamics simulations, together with Bayesian analyses, to explore the ability of CBCs in dense gaseous environments to disguise as coalescences in vacuum, and the subsequent systematic biases in PE.
We considered a \texttt{GW150914}-like system as a testbed to show that a large fraction of such systems could be detected using a template bank of vacuum waveforms, even when immersed in a stellar density larger than~$10^{7}\densityUnit$. 
Although large environmental effects are spotted by routine tests of vacuum GR (as also recently found for LISA massive BBHs in circumbinary disks~\cite{Garg:2024qxq}), we showed that, for a non-negligible range of environment densities, gaseous environmental effects can be large enough to impart substantial biases in PE, while still passing tests of vacuum GR. Finally, we demonstrated that phenomenological models can efficiently reproduce NR environment waveforms, be used to derive evidence for gaseous environments, and resolve systematic biases. 

In this work we focused on quasi-circular inspirals. Torques in gaseous environments can lead to a residual eccentricity in CBCs~\cite{Cardoso:2020iji}, as predicted in some AGN binary formation channels~\cite{Tagawa:2019osr, Sedda:2023big, Rowan:2023, Ishibashi:2020zzy}; previous studies indicate that \texttt{GW190521} could be such an event~\cite{Gayathri:2020coq, CalderonBustillo:2020xms}. Similarly to our findings, neglecting eccentricity in the PE can also bias the chirp mass estimation to larger values~\cite{Favata:2021vhw, Divyajyoti:2023rht}; for a \texttt{GW150914}-like event such biases are significant only for eccentricities~$e_{10\Hz}\gtrsim 0.1$ at a reference GW frequency of~$10\,\mathrm{Hz}$~\cite{Favata:2021vhw}. The degeneracy between eccentricity and gas density was studied in Ref.~\cite{Garg:2024oeu} for LISA.

While the results of our study depend on the particular type of binary system and environment (and also on the detector's frequency band), we expect some of our qualitative conclusions to scale towards smaller densities for longer inspirals (with more cycles). For instance, doing a similar type of injection study with phenomenological waveforms as in Sec.~\ref{subsec:systematic}, but for a low-mass \texttt{GW170817}-like BNS system~\cite{LIGOScientific:2017vwq, LIGOScientific:2018hze}, we found similar (qualitative) results as the ones shown in Fig.~\ref{fig:bias_gw150914-like_inj}. We obtained that for densities $\tilde{\rho}<1.2 \densityUnit$, vacuum GR falls (at least) within the 97\% credible interval of the $\delta\hat{\varphi}_{-2}$ posterior, though significant biases are present in chirp mass, mass-ratio, and the spin parameters. These biases persist for densities as low as $\tilde{\rho}\gtrsim 0.1\densityUnit$.\footnote{We considered a network of three detectors---H1, L1, and Virgo (V1)---assuming at their advanced design sensitivities: \texttt{aLIGOZeroDetHighPower} for H1 and L1~\cite{aLIGO_ZDHP}, and \texttt{AdvVirgo} for V1~\cite{2012arXiv1202.4031M}. With a starting frequency of 15\Hz, the injection optimal SNR is $85$. We considered densities $\tilde{\rho}$ ranging from $0.01$ to $10 \densityUnit$} For the near-future detectors Einstein Telescope~\cite{Punturo:2010zz} and Cosmic Explorer~\cite{Reitze:2019iox} similar conclusions are expected to hold for densities four orders of magnitude smaller~\cite{CanevaSantoro:2023aol}, possibly reaching the typical densities in residual circumbinary disks in a post-common envelope stage ($\lesssim 10^{-5} \densityUnit$)~\cite{Martin:2018iov, Tuna:2023jgw}. Note, however, that the environmental effects from circumbinary disks may not be well captured by the phenomenological model of this present work.

Systematic biases in spin from unmodelled environmental effects could affect our ability to infer the astrophysical origin of compact binaries (see, e.g., \cite{Gerosa:2018wbw}), while the ones in mass might hinder the exploitation of the narrow mass function of NSs to measure the Hubble constant~\cite{Taylor:2011fs, Taylor:2012db}. This underscores the need for robust waveform models beyond the vacuum GR paradigm. The construction of phenomenological environment waveforms tuned against numerical simulations may well play a key role in realising the full potential of GW astronomy.

\section*{Acknowledgements}
We are grateful to Harris M. K., Maria Haney, Justin Janquart, and Vitor Cardoso for useful comments on our draft.
We thank Gareth Cabourn Davies and Thomas Dent for their help with running the \pycbc pipeline on O1 data.
R.V. is supported by grant no. FJC2021-046551-I funded by MCIN/AEI/10.13039/501100011033 and by the European Union NextGenerationEU/PRTR. R.V. also acknowledges the support from the Departament de Recerca i Universitats from Generalitat de Catalunya to the Grup de Recerca 00649 (Codi: 2021 SGR 00649).
The authors are grateful for computational resources provided by the LIGO Laboratory and supported by National Science Foundation Grants PHY-0757058 and PHY-0823459. The authors are also grateful for the computational resources provided by Cardiff University supported by STFC grant ST/I006285/1.
This research has made use of data or software obtained from the Gravitational Wave Open Science Center (\href{https://www.gwosc.org}{https://www.gwosc.org}), a service of the LIGO Scientific Collaboration, the Virgo Collaboration, and KAGRA. This material is based upon work supported by NSF's LIGO Laboratory which is a major facility fully funded by the National Science Foundation, as well as the Science and Technology Facilities Council (STFC) of the United Kingdom, the Max-Planck-Society (MPS), and the State of Niedersachsen/Germany for support of the construction of Advanced LIGO and construction and operation of the GEO600 detector. Additional support for Advanced LIGO was provided by the Australian Research Council. Virgo is funded, through the European Gravitational Observatory (EGO), by the French Centre National de Recherche Scientifique (CNRS), the Italian Istituto Nazionale di Fisica Nucleare (INFN) and the Dutch Nikhef, with contributions by institutions from Belgium, Germany, Greece, Hungary, Ireland, Japan, Monaco, Poland, Portugal, Spain. KAGRA is supported by Ministry of Education, Culture, Sports, Science and Technology (MEXT), Japan Society for the Promotion of Science (JSPS) in Japan; National Research Foundation (NRF) and Ministry of Science and ICT (MSIT) in Korea; Academia Sinica (AS) and National Science and Technology Council (NSTC) in Taiwan. We have used \numpy~\cite{Harris:2020xlr} and \scipy~\cite{Virtanen:2019joe} for analyses, and \matplotlib~\cite{Hunter:2007ouj} for preparing the plots in this manuscript.

\section*{Data Availability}
The NR waveforms used in this work were obtained in  Ref.~\cite{Fedrow:2017dpk} and can be found at~\cite{BBHGAS_StellarCollapse}. The \texttt{GW150914} event PSDs used to perform the Bayesian analyses are available at~\cite{gw150914psds}. The designed sensitivity curves for Advanced LIGO and Virgo are available at \lalsuite software package~\cite{lalsuite}.

\appendix

\section{Systematic Biases for Large Environmental Effects}

\begin{figure}[t]
	\centering
	\includegraphics[width=\linewidth]{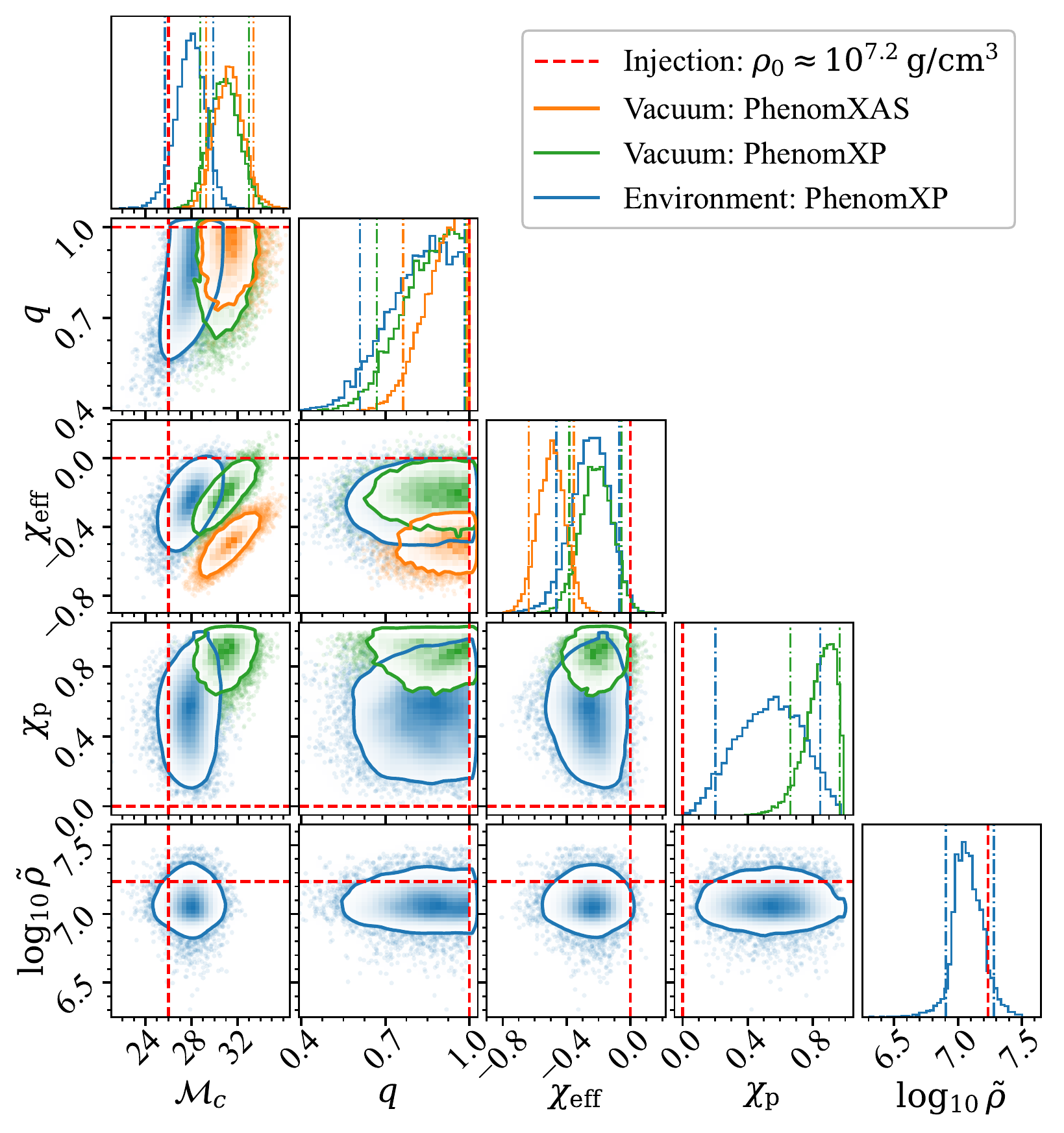}
	\caption{Same as Fig.\,\ref{fig:contour_G3sim} but for a (30+30)\msun BBH coalescence in stellar material of density $\rho_0\approx10^{7.2}\, \densityUnit$.}
	\label{fig:contour_G4sim}
\end{figure}

As discussed in Sec.~\ref{sec:PE}, we also perform an injection analysis for a NR waveform of a (30+30)\msun BBH merger in a very dense stellar material with $\rho_0 \approx 10^{7.2} \densityUnit$. Figure\:\ref{fig:contour_G4sim} shows a corner plot comparing the recovered marginalized posterior distributions with the vacuum and phenomenological environment models. It shows that the non inclusion of environmental effects on the recovery model leads to significant systematic biases in the estimation of the BBH parameters. Similarly to the case $\rho_0 \approx 10^{6.2} \densityUnit$ shown in Fig.~\ref{fig:contour_G3sim}, the 90\% credible regions on the \mchirp--\chieff, \mchirp--\chip, and \chieff--\chip planes are significantly offset from the true values.

The spin estimation with the vacuum model is strongly biased, suggesting a merger of a highly precessing BBH. A particularly notable result is that the effective spin \chieff posterior favours an anti-aligned configuration, which is in stark contrast to the findings for the $\rho_0 \approx 10^{6.2} \densityUnit$ case. This is likely due to the strong environmental contribution to the merger-ringdown part of the signal. Typically, environmental effects alter predominantly the early inspiral phase of binary evolution, but in an extremely dense environment the accretion rate can be large enough to lead to modifications in the merger-ringdown regime~\cite{Leong:2023nuk}. This also explains why such environmental effects were caught by the IMR consistency test (see Fig.~\ref{fig:imr_consistency}). 

We note that the parameters estimated with our environment model also shows biases, with some of the injected values at the edge of their 90\% credible intervals. This is not surprising, as our phenomenological model does not include environmental correction in the merger-ringdown phase, neither it accounts for the (non negligible) evolution of the binary mass components. Despite that, our model is effective in detecting environmental effects, as indicated by the log Bayes factor $\log_{10} \mathcal{B}_{\mathrm{vac}}^{\mathrm{env}}=3.7$.

\bibliography{reference}

\begin{thebibliography}{138}%
\makeatletter
\providecommand \@ifxundefined [1]{%
 \@ifx{#1\undefined}
}%
\providecommand \@ifnum [1]{%
 \ifnum #1\expandafter \@firstoftwo
 \else \expandafter \@secondoftwo
 \fi
}%
\providecommand \@ifx [1]{%
 \ifx #1\expandafter \@firstoftwo
 \else \expandafter \@secondoftwo
 \fi
}%
\providecommand \natexlab [1]{#1}%
\providecommand \enquote  [1]{``#1''}%
\providecommand \bibnamefont  [1]{#1}%
\providecommand \bibfnamefont [1]{#1}%
\providecommand \citenamefont [1]{#1}%
\providecommand \href@noop [0]{\@secondoftwo}%
\providecommand \href [0]{\begingroup \@sanitize@url \@href}%
\providecommand \@href[1]{\@@startlink{#1}\@@href}%
\providecommand \@@href[1]{\endgroup#1\@@endlink}%
\providecommand \@sanitize@url [0]{\catcode `\\12\catcode `\$12\catcode
  `\&12\catcode `\#12\catcode `\^12\catcode `\_12\catcode `\%12\relax}%
\providecommand \@@startlink[1]{}%
\providecommand \@@endlink[0]{}%
\providecommand \url  [0]{\begingroup\@sanitize@url \@url }%
\providecommand \@url [1]{\endgroup\@href {#1}{\urlprefix }}%
\providecommand \urlprefix  [0]{URL }%
\providecommand \Eprint [0]{\href }%
\providecommand \doibase [0]{https://doi.org/}%
\providecommand \selectlanguage [0]{\@gobble}%
\providecommand \bibinfo  [0]{\@secondoftwo}%
\providecommand \bibfield  [0]{\@secondoftwo}%
\providecommand \translation [1]{[#1]}%
\providecommand \BibitemOpen [0]{}%
\providecommand \bibitemStop [0]{}%
\providecommand \bibitemNoStop [0]{.\EOS\space}%
\providecommand \EOS [0]{\spacefactor3000\relax}%
\providecommand \BibitemShut  [1]{\csname bibitem#1\endcsname}%
\let\auto@bib@innerbib\@empty
\bibitem [{\citenamefont {Einstein}(1916)}]{Einstein:1916cc}%
  \BibitemOpen
  \bibfield  {author} {\bibinfo {author} {\bibfnamefont {A.}~\bibnamefont
  {Einstein}},\ }\bibfield  {title} {\bibinfo {title} {{Approximative
  Integration of the Field Equations of Gravitation}},\ }\href@noop {}
  {\bibfield  {journal} {\bibinfo  {journal} {Sitzungsber. Preuss. Akad. Wiss.
  Berlin (Math. Phys. )}\ }\textbf {\bibinfo {volume} {1916}},\ \bibinfo
  {pages} {688} (\bibinfo {year} {1916})}\BibitemShut {NoStop}%
\bibitem [{\citenamefont {Einstein}(1918)}]{Einstein:1918btx}%
  \BibitemOpen
  \bibfield  {author} {\bibinfo {author} {\bibfnamefont {A.}~\bibnamefont
  {Einstein}},\ }\bibfield  {title} {\bibinfo {title} {{\"Uber
  Gravitationswellen}},\ }\href@noop {} {\bibfield  {journal} {\bibinfo
  {journal} {Sitzungsber. Preuss. Akad. Wiss. Berlin (Math. Phys. )}\ }\textbf
  {\bibinfo {volume} {1918}},\ \bibinfo {pages} {154} (\bibinfo {year}
  {1918})}\BibitemShut {NoStop}%
\bibitem [{\citenamefont {Abbott}\ \emph
  {et~al.}(2016{\natexlab{a}})\citenamefont {Abbott} \emph
  {et~al.}}]{LIGOScientific:2016aoc}%
  \BibitemOpen
  \bibfield  {author} {\bibinfo {author} {\bibfnamefont {B.~P.}\ \bibnamefont
  {Abbott}} \emph {et~al.} (\bibinfo {collaboration} {LIGO Scientific,
  Virgo}),\ }\bibfield  {title} {\bibinfo {title} {{Observation of
  Gravitational Waves from a Binary Black Hole Merger}},\ }\href
  {https://doi.org/10.1103/PhysRevLett.116.061102} {\bibfield  {journal}
  {\bibinfo  {journal} {Phys. Rev. Lett.}\ }\textbf {\bibinfo {volume} {116}},\
  \bibinfo {pages} {061102} (\bibinfo {year} {2016}{\natexlab{a}})},\ \Eprint
  {https://arxiv.org/abs/1602.03837} {arXiv:1602.03837 [gr-qc]} \BibitemShut
  {NoStop}%
\bibitem [{\citenamefont {Aasi}\ \emph {et~al.}(2015)\citenamefont {Aasi} \emph
  {et~al.}}]{LIGOScientific:2014pky}%
  \BibitemOpen
  \bibfield  {author} {\bibinfo {author} {\bibfnamefont {J.}~\bibnamefont
  {Aasi}} \emph {et~al.} (\bibinfo {collaboration} {LIGO Scientific}),\
  }\bibfield  {title} {\bibinfo {title} {{Advanced LIGO}},\ }\href
  {https://doi.org/10.1088/0264-9381/32/7/074001} {\bibfield  {journal}
  {\bibinfo  {journal} {Class. Quant. Grav.}\ }\textbf {\bibinfo {volume}
  {32}},\ \bibinfo {pages} {074001} (\bibinfo {year} {2015})},\ \Eprint
  {https://arxiv.org/abs/1411.4547} {arXiv:1411.4547 [gr-qc]} \BibitemShut
  {NoStop}%
\bibitem [{\citenamefont {Acernese}\ \emph {et~al.}(2015)\citenamefont
  {Acernese} \emph {et~al.}}]{VIRGO:2014yos}%
  \BibitemOpen
  \bibfield  {author} {\bibinfo {author} {\bibfnamefont {F.}~\bibnamefont
  {Acernese}} \emph {et~al.} (\bibinfo {collaboration} {VIRGO}),\ }\bibfield
  {title} {\bibinfo {title} {{Advanced Virgo: a second-generation
  interferometric gravitational wave detector}},\ }\href
  {https://doi.org/10.1088/0264-9381/32/2/024001} {\bibfield  {journal}
  {\bibinfo  {journal} {Class. Quant. Grav.}\ }\textbf {\bibinfo {volume}
  {32}},\ \bibinfo {pages} {024001} (\bibinfo {year} {2015})},\ \Eprint
  {https://arxiv.org/abs/1408.3978} {arXiv:1408.3978 [gr-qc]} \BibitemShut
  {NoStop}%
\bibitem [{\citenamefont {Abbott}\ \emph
  {et~al.}(2019{\natexlab{a}})\citenamefont {Abbott} \emph
  {et~al.}}]{LIGOScientific:2018mvr}%
  \BibitemOpen
  \bibfield  {author} {\bibinfo {author} {\bibfnamefont {B.~P.}\ \bibnamefont
  {Abbott}} \emph {et~al.} (\bibinfo {collaboration} {LIGO Scientific,
  Virgo}),\ }\bibfield  {title} {\bibinfo {title} {{GWTC-1: A
  Gravitational-Wave Transient Catalog of Compact Binary Mergers Observed by
  LIGO and Virgo during the First and Second Observing Runs}},\ }\href
  {https://doi.org/10.1103/PhysRevX.9.031040} {\bibfield  {journal} {\bibinfo
  {journal} {Phys. Rev. X}\ }\textbf {\bibinfo {volume} {9}},\ \bibinfo {pages}
  {031040} (\bibinfo {year} {2019}{\natexlab{a}})},\ \Eprint
  {https://arxiv.org/abs/1811.12907} {arXiv:1811.12907 [astro-ph.HE]}
  \BibitemShut {NoStop}%
\bibitem [{\citenamefont {Abbott}\ \emph
  {et~al.}(2021{\natexlab{a}})\citenamefont {Abbott} \emph
  {et~al.}}]{LIGOScientific:2020ibl}%
  \BibitemOpen
  \bibfield  {author} {\bibinfo {author} {\bibfnamefont {R.}~\bibnamefont
  {Abbott}} \emph {et~al.} (\bibinfo {collaboration} {LIGO Scientific,
  Virgo}),\ }\bibfield  {title} {\bibinfo {title} {{GWTC-2: Compact Binary
  Coalescences Observed by LIGO and Virgo During the First Half of the Third
  Observing Run}},\ }\href {https://doi.org/10.1103/PhysRevX.11.021053}
  {\bibfield  {journal} {\bibinfo  {journal} {Phys. Rev. X}\ }\textbf {\bibinfo
  {volume} {11}},\ \bibinfo {pages} {021053} (\bibinfo {year}
  {2021}{\natexlab{a}})},\ \Eprint {https://arxiv.org/abs/2010.14527}
  {arXiv:2010.14527 [gr-qc]} \BibitemShut {NoStop}%
\bibitem [{\citenamefont {Abbott}\ \emph {et~al.}(2024)\citenamefont {Abbott}
  \emph {et~al.}}]{LIGOScientific:2021usb}%
  \BibitemOpen
  \bibfield  {author} {\bibinfo {author} {\bibfnamefont {R.}~\bibnamefont
  {Abbott}} \emph {et~al.} (\bibinfo {collaboration} {LIGO Scientific,
  VIRGO}),\ }\bibfield  {title} {\bibinfo {title} {{GWTC-2.1: Deep extended
  catalog of compact binary coalescences observed by LIGO and Virgo during the
  first half of the third observing run}},\ }\href
  {https://doi.org/10.1103/PhysRevD.109.022001} {\bibfield  {journal} {\bibinfo
   {journal} {Phys. Rev. D}\ }\textbf {\bibinfo {volume} {109}},\ \bibinfo
  {pages} {022001} (\bibinfo {year} {2024})},\ \Eprint
  {https://arxiv.org/abs/2108.01045} {arXiv:2108.01045 [gr-qc]} \BibitemShut
  {NoStop}%
\bibitem [{\citenamefont {Abbott}\ \emph
  {et~al.}(2023{\natexlab{a}})\citenamefont {Abbott} \emph
  {et~al.}}]{KAGRA:2021vkt}%
  \BibitemOpen
  \bibfield  {author} {\bibinfo {author} {\bibfnamefont {R.}~\bibnamefont
  {Abbott}} \emph {et~al.} (\bibinfo {collaboration} {KAGRA, VIRGO, LIGO
  Scientific}),\ }\bibfield  {title} {\bibinfo {title} {{GWTC-3: Compact Binary
  Coalescences Observed by LIGO and Virgo during the Second Part of the Third
  Observing Run}},\ }\href {https://doi.org/10.1103/PhysRevX.13.041039}
  {\bibfield  {journal} {\bibinfo  {journal} {Phys. Rev. X}\ }\textbf {\bibinfo
  {volume} {13}},\ \bibinfo {pages} {041039} (\bibinfo {year}
  {2023}{\natexlab{a}})},\ \Eprint {https://arxiv.org/abs/2111.03606}
  {arXiv:2111.03606 [gr-qc]} \BibitemShut {NoStop}%
\bibitem [{\citenamefont {Abbott}\ \emph
  {et~al.}(2016{\natexlab{b}})\citenamefont {Abbott} \emph
  {et~al.}}]{KAGRA:2013rdx}%
  \BibitemOpen
  \bibfield  {author} {\bibinfo {author} {\bibfnamefont {B.~P.}\ \bibnamefont
  {Abbott}} \emph {et~al.} (\bibinfo {collaboration} {KAGRA, LIGO Scientific,
  Virgo}),\ }\bibfield  {title} {\bibinfo {title} {{Prospects for observing and
  localizing gravitational-wave transients with Advanced LIGO, Advanced Virgo
  and KAGRA}},\ }\href {https://doi.org/10.1007/s41114-020-00026-9} {\bibfield
  {journal} {\bibinfo  {journal} {Living Rev. Rel.}\ }\textbf {\bibinfo
  {volume} {19}},\ \bibinfo {pages} {1} (\bibinfo {year}
  {2016}{\natexlab{b}})},\ \Eprint {https://arxiv.org/abs/1304.0670}
  {arXiv:1304.0670 [gr-qc]} \BibitemShut {NoStop}%
\bibitem [{\citenamefont {Abbott}\ \emph
  {et~al.}(2016{\natexlab{c}})\citenamefont {Abbott} \emph
  {et~al.}}]{LIGOScientific:2016lio}%
  \BibitemOpen
  \bibfield  {author} {\bibinfo {author} {\bibfnamefont {B.~P.}\ \bibnamefont
  {Abbott}} \emph {et~al.} (\bibinfo {collaboration} {LIGO Scientific,
  Virgo}),\ }\bibfield  {title} {\bibinfo {title} {{Tests of general relativity
  with GW150914}},\ }\href {https://doi.org/10.1103/PhysRevLett.116.221101}
  {\bibfield  {journal} {\bibinfo  {journal} {Phys. Rev. Lett.}\ }\textbf
  {\bibinfo {volume} {116}},\ \bibinfo {pages} {221101} (\bibinfo {year}
  {2016}{\natexlab{c}})},\ \bibinfo {note} {[Erratum: Phys.Rev.Lett. 121,
  129902 (2018)]},\ \Eprint {https://arxiv.org/abs/1602.03841}
  {arXiv:1602.03841 [gr-qc]} \BibitemShut {NoStop}%
\bibitem [{\citenamefont {Abbott}\ \emph
  {et~al.}(2019{\natexlab{b}})\citenamefont {Abbott} \emph
  {et~al.}}]{LIGOScientific:2018dkp}%
  \BibitemOpen
  \bibfield  {author} {\bibinfo {author} {\bibfnamefont {B.~P.}\ \bibnamefont
  {Abbott}} \emph {et~al.} (\bibinfo {collaboration} {LIGO Scientific,
  Virgo}),\ }\bibfield  {title} {\bibinfo {title} {{Tests of General Relativity
  with GW170817}},\ }\href {https://doi.org/10.1103/PhysRevLett.123.011102}
  {\bibfield  {journal} {\bibinfo  {journal} {Phys. Rev. Lett.}\ }\textbf
  {\bibinfo {volume} {123}},\ \bibinfo {pages} {011102} (\bibinfo {year}
  {2019}{\natexlab{b}})},\ \Eprint {https://arxiv.org/abs/1811.00364}
  {arXiv:1811.00364 [gr-qc]} \BibitemShut {NoStop}%
\bibitem [{\citenamefont {Abbott}\ \emph
  {et~al.}(2019{\natexlab{c}})\citenamefont {Abbott} \emph
  {et~al.}}]{LIGOScientific:2019fpa}%
  \BibitemOpen
  \bibfield  {author} {\bibinfo {author} {\bibfnamefont {B.~P.}\ \bibnamefont
  {Abbott}} \emph {et~al.} (\bibinfo {collaboration} {LIGO Scientific,
  Virgo}),\ }\bibfield  {title} {\bibinfo {title} {{Tests of General Relativity
  with the Binary Black Hole Signals from the LIGO-Virgo Catalog GWTC-1}},\
  }\href {https://doi.org/10.1103/PhysRevD.100.104036} {\bibfield  {journal}
  {\bibinfo  {journal} {Phys. Rev. D}\ }\textbf {\bibinfo {volume} {100}},\
  \bibinfo {pages} {104036} (\bibinfo {year} {2019}{\natexlab{c}})},\ \Eprint
  {https://arxiv.org/abs/1903.04467} {arXiv:1903.04467 [gr-qc]} \BibitemShut
  {NoStop}%
\bibitem [{\citenamefont {Abbott}\ \emph
  {et~al.}(2021{\natexlab{b}})\citenamefont {Abbott} \emph
  {et~al.}}]{LIGOScientific:2020tif}%
  \BibitemOpen
  \bibfield  {author} {\bibinfo {author} {\bibfnamefont {R.}~\bibnamefont
  {Abbott}} \emph {et~al.} (\bibinfo {collaboration} {LIGO Scientific,
  Virgo}),\ }\bibfield  {title} {\bibinfo {title} {{Tests of general relativity
  with binary black holes from the second LIGO-Virgo gravitational-wave
  transient catalog}},\ }\href {https://doi.org/10.1103/PhysRevD.103.122002}
  {\bibfield  {journal} {\bibinfo  {journal} {Phys. Rev. D}\ }\textbf {\bibinfo
  {volume} {103}},\ \bibinfo {pages} {122002} (\bibinfo {year}
  {2021}{\natexlab{b}})},\ \Eprint {https://arxiv.org/abs/2010.14529}
  {arXiv:2010.14529 [gr-qc]} \BibitemShut {NoStop}%
\bibitem [{\citenamefont {Abbott}\ \emph
  {et~al.}(2017{\natexlab{a}})\citenamefont {Abbott} \emph
  {et~al.}}]{LIGOScientific:2017adf}%
  \BibitemOpen
  \bibfield  {author} {\bibinfo {author} {\bibfnamefont {B.~P.}\ \bibnamefont
  {Abbott}} \emph {et~al.} (\bibinfo {collaboration} {LIGO Scientific, Virgo,
  1M2H, Dark Energy Camera GW-E, DES, DLT40, Las Cumbres Observatory, VINROUGE,
  MASTER}),\ }\bibfield  {title} {\bibinfo {title} {{A gravitational-wave
  standard siren measurement of the Hubble constant}},\ }\href
  {https://doi.org/10.1038/nature24471} {\bibfield  {journal} {\bibinfo
  {journal} {Nature}\ }\textbf {\bibinfo {volume} {551}},\ \bibinfo {pages}
  {85} (\bibinfo {year} {2017}{\natexlab{a}})},\ \Eprint
  {https://arxiv.org/abs/1710.05835} {arXiv:1710.05835 [astro-ph.CO]}
  \BibitemShut {NoStop}%
\bibitem [{\citenamefont {Abbott}\ \emph
  {et~al.}(2021{\natexlab{c}})\citenamefont {Abbott} \emph
  {et~al.}}]{LIGOScientific:2019zcs}%
  \BibitemOpen
  \bibfield  {author} {\bibinfo {author} {\bibfnamefont {B.~P.}\ \bibnamefont
  {Abbott}} \emph {et~al.} (\bibinfo {collaboration} {LIGO Scientific, Virgo,
  VIRGO}),\ }\bibfield  {title} {\bibinfo {title} {{A Gravitational-wave
  Measurement of the Hubble Constant Following the Second Observing Run of
  Advanced LIGO and Virgo}},\ }\href {https://doi.org/10.3847/1538-4357/abdcb7}
  {\bibfield  {journal} {\bibinfo  {journal} {Astrophys. J.}\ }\textbf
  {\bibinfo {volume} {909}},\ \bibinfo {pages} {218} (\bibinfo {year}
  {2021}{\natexlab{c}})},\ \Eprint {https://arxiv.org/abs/1908.06060}
  {arXiv:1908.06060 [astro-ph.CO]} \BibitemShut {NoStop}%
\bibitem [{\citenamefont {Abbott}\ \emph
  {et~al.}(2023{\natexlab{b}})\citenamefont {Abbott} \emph
  {et~al.}}]{LIGOScientific:2021aug}%
  \BibitemOpen
  \bibfield  {author} {\bibinfo {author} {\bibfnamefont {R.}~\bibnamefont
  {Abbott}} \emph {et~al.} (\bibinfo {collaboration} {LIGO Scientific, Virgo,
  KAGRA}),\ }\bibfield  {title} {\bibinfo {title} {{Constraints on the Cosmic
  Expansion History from GWTC\textendash{}3}},\ }\href
  {https://doi.org/10.3847/1538-4357/ac74bb} {\bibfield  {journal} {\bibinfo
  {journal} {Astrophys. J.}\ }\textbf {\bibinfo {volume} {949}},\ \bibinfo
  {pages} {76} (\bibinfo {year} {2023}{\natexlab{b}})},\ \Eprint
  {https://arxiv.org/abs/2111.03604} {arXiv:2111.03604 [astro-ph.CO]}
  \BibitemShut {NoStop}%
\bibitem [{\citenamefont {Abbott}\ \emph {et~al.}(2018)\citenamefont {Abbott}
  \emph {et~al.}}]{LIGOScientific:2018cki}%
  \BibitemOpen
  \bibfield  {author} {\bibinfo {author} {\bibfnamefont {B.~P.}\ \bibnamefont
  {Abbott}} \emph {et~al.} (\bibinfo {collaboration} {LIGO Scientific,
  Virgo}),\ }\bibfield  {title} {\bibinfo {title} {{GW170817: Measurements of
  neutron star radii and equation of state}},\ }\href
  {https://doi.org/10.1103/PhysRevLett.121.161101} {\bibfield  {journal}
  {\bibinfo  {journal} {Phys. Rev. Lett.}\ }\textbf {\bibinfo {volume} {121}},\
  \bibinfo {pages} {161101} (\bibinfo {year} {2018})},\ \Eprint
  {https://arxiv.org/abs/1805.11581} {arXiv:1805.11581 [gr-qc]} \BibitemShut
  {NoStop}%
\bibitem [{\citenamefont {Barack}\ \emph {et~al.}(2019)\citenamefont {Barack}
  \emph {et~al.}}]{Barack:2018yly}%
  \BibitemOpen
  \bibfield  {author} {\bibinfo {author} {\bibfnamefont {L.}~\bibnamefont
  {Barack}} \emph {et~al.},\ }\bibfield  {title} {\bibinfo {title} {{Black
  holes, gravitational waves and fundamental physics: a roadmap}},\ }\href
  {https://doi.org/10.1088/1361-6382/ab0587} {\bibfield  {journal} {\bibinfo
  {journal} {Class. Quant. Grav.}\ }\textbf {\bibinfo {volume} {36}},\ \bibinfo
  {pages} {143001} (\bibinfo {year} {2019})},\ \Eprint
  {https://arxiv.org/abs/1806.05195} {arXiv:1806.05195 [gr-qc]} \BibitemShut
  {NoStop}%
\bibitem [{\citenamefont {Maggiore}\ \emph {et~al.}(2020)\citenamefont
  {Maggiore} \emph {et~al.}}]{Maggiore:2019uih}%
  \BibitemOpen
  \bibfield  {author} {\bibinfo {author} {\bibfnamefont {M.}~\bibnamefont
  {Maggiore}} \emph {et~al.},\ }\bibfield  {title} {\bibinfo {title} {{Science
  Case for the Einstein Telescope}},\ }\href
  {https://doi.org/10.1088/1475-7516/2020/03/050} {\bibfield  {journal}
  {\bibinfo  {journal} {JCAP}\ }\textbf {\bibinfo {volume} {03}},\ \bibinfo
  {pages} {050}},\ \Eprint {https://arxiv.org/abs/1912.02622} {arXiv:1912.02622
  [astro-ph.CO]} \BibitemShut {NoStop}%
\bibitem [{\citenamefont {Evans}\ \emph {et~al.}(2021)\citenamefont {Evans}
  \emph {et~al.}}]{Evans:2021gyd}%
  \BibitemOpen
  \bibfield  {author} {\bibinfo {author} {\bibfnamefont {M.}~\bibnamefont
  {Evans}} \emph {et~al.},\ }\bibfield  {title} {\bibinfo {title} {{A Horizon
  Study for Cosmic Explorer: Science, Observatories, and Community}},\ }\href
  {https://doi.org/10.48550/arXiv.2109.09882} {\bibfield  {journal} {\bibinfo
  {journal} {arXiv e-prints}\ ,\ \bibinfo {eid} {arXiv:2109.09882}} (\bibinfo
  {year} {2021})},\ \Eprint {https://arxiv.org/abs/2109.09882}
  {arXiv:2109.09882 [astro-ph.IM]} \BibitemShut {NoStop}%
\bibitem [{\citenamefont {Colpi}\ \emph {et~al.}(2024)\citenamefont {Colpi}
  \emph {et~al.}}]{Colpi:2024xhw}%
  \BibitemOpen
  \bibfield  {author} {\bibinfo {author} {\bibfnamefont {M.}~\bibnamefont
  {Colpi}} \emph {et~al.},\ }\bibfield  {title} {\bibinfo {title} {{LISA
  Definition Study Report}},\ }\href
  {https://doi.org/10.48550/arXiv.2402.07571} {\bibfield  {journal} {\bibinfo
  {journal} {arXiv e-prints}\ ,\ \bibinfo {eid} {arXiv:2402.07571}} (\bibinfo
  {year} {2024})},\ \Eprint {https://arxiv.org/abs/2402.07571}
  {arXiv:2402.07571 [astro-ph.CO]} \BibitemShut {NoStop}%
\bibitem [{\citenamefont {Mandel}\ and\ \citenamefont
  {Broekgaarden}(2022)}]{Mandel:2021smh}%
  \BibitemOpen
  \bibfield  {author} {\bibinfo {author} {\bibfnamefont {I.}~\bibnamefont
  {Mandel}}\ and\ \bibinfo {author} {\bibfnamefont {F.~S.}\ \bibnamefont
  {Broekgaarden}},\ }\bibfield  {title} {\bibinfo {title} {{Rates of compact
  object coalescences}},\ }\href {https://doi.org/10.1007/s41114-021-00034-3}
  {\bibfield  {journal} {\bibinfo  {journal} {Living Rev. Rel.}\ }\textbf
  {\bibinfo {volume} {25}},\ \bibinfo {pages} {1} (\bibinfo {year} {2022})},\
  \Eprint {https://arxiv.org/abs/2107.14239} {arXiv:2107.14239 [astro-ph.HE]}
  \BibitemShut {NoStop}%
\bibitem [{\citenamefont {Mandel}\ and\ \citenamefont
  {Farmer}(2022)}]{Mandel:2018hfr}%
  \BibitemOpen
  \bibfield  {author} {\bibinfo {author} {\bibfnamefont {I.}~\bibnamefont
  {Mandel}}\ and\ \bibinfo {author} {\bibfnamefont {A.}~\bibnamefont
  {Farmer}},\ }\bibfield  {title} {\bibinfo {title} {{Merging stellar-mass
  binary black holes}},\ }\href {https://doi.org/10.1016/j.physrep.2022.01.003}
  {\bibfield  {journal} {\bibinfo  {journal} {Phys. Rept.}\ }\textbf {\bibinfo
  {volume} {955}},\ \bibinfo {pages} {1} (\bibinfo {year} {2022})},\ \Eprint
  {https://arxiv.org/abs/1806.05820} {arXiv:1806.05820 [astro-ph.HE]}
  \BibitemShut {NoStop}%
\bibitem [{\citenamefont {Abbott}\ \emph
  {et~al.}(2023{\natexlab{c}})\citenamefont {Abbott} \emph
  {et~al.}}]{KAGRA:2021duu}%
  \BibitemOpen
  \bibfield  {author} {\bibinfo {author} {\bibfnamefont {R.}~\bibnamefont
  {Abbott}} \emph {et~al.} (\bibinfo {collaboration} {KAGRA, VIRGO, LIGO
  Scientific}),\ }\bibfield  {title} {\bibinfo {title} {{Population of Merging
  Compact Binaries Inferred Using Gravitational Waves through GWTC-3}},\ }\href
  {https://doi.org/10.1103/PhysRevX.13.011048} {\bibfield  {journal} {\bibinfo
  {journal} {Phys. Rev. X}\ }\textbf {\bibinfo {volume} {13}},\ \bibinfo
  {pages} {011048} (\bibinfo {year} {2023}{\natexlab{c}})},\ \Eprint
  {https://arxiv.org/abs/2111.03634} {arXiv:2111.03634 [astro-ph.HE]}
  \BibitemShut {NoStop}%
\bibitem [{\citenamefont {Woosley}(2016)}]{Woosley:2016nnw}%
  \BibitemOpen
  \bibfield  {author} {\bibinfo {author} {\bibfnamefont {S.~E.}\ \bibnamefont
  {Woosley}},\ }\bibfield  {title} {\bibinfo {title} {{The Progenitor of
  Gw150914}},\ }\href {https://doi.org/10.3847/2041-8205/824/1/L10} {\bibfield
  {journal} {\bibinfo  {journal} {Astrophys. J. Lett.}\ }\textbf {\bibinfo
  {volume} {824}},\ \bibinfo {pages} {L10} (\bibinfo {year} {2016})},\ \Eprint
  {https://arxiv.org/abs/1603.00511} {arXiv:1603.00511 [astro-ph.HE]}
  \BibitemShut {NoStop}%
\bibitem [{\citenamefont {Fedrow}\ \emph {et~al.}(2017)\citenamefont {Fedrow},
  \citenamefont {Ott}, \citenamefont {Sperhake}, \citenamefont {Blackman},
  \citenamefont {Haas}, \citenamefont {Reisswig},\ and\ \citenamefont
  {De~Felice}}]{Fedrow:2017dpk}%
  \BibitemOpen
  \bibfield  {author} {\bibinfo {author} {\bibfnamefont {J.~M.}\ \bibnamefont
  {Fedrow}}, \bibinfo {author} {\bibfnamefont {C.~D.}\ \bibnamefont {Ott}},
  \bibinfo {author} {\bibfnamefont {U.}~\bibnamefont {Sperhake}}, \bibinfo
  {author} {\bibfnamefont {J.}~\bibnamefont {Blackman}}, \bibinfo {author}
  {\bibfnamefont {R.}~\bibnamefont {Haas}}, \bibinfo {author} {\bibfnamefont
  {C.}~\bibnamefont {Reisswig}},\ and\ \bibinfo {author} {\bibfnamefont
  {A.}~\bibnamefont {De~Felice}},\ }\bibfield  {title} {\bibinfo {title}
  {{Gravitational Waves from Binary Black Hole Mergers inside Stars}},\ }\href
  {https://doi.org/10.1103/PhysRevLett.119.171103} {\bibfield  {journal}
  {\bibinfo  {journal} {Phys. Rev. Lett.}\ }\textbf {\bibinfo {volume} {119}},\
  \bibinfo {pages} {171103} (\bibinfo {year} {2017})},\ \Eprint
  {https://arxiv.org/abs/1704.07383} {arXiv:1704.07383 [astro-ph.HE]}
  \BibitemShut {NoStop}%
\bibitem [{\citenamefont {Caneva~Santoro}\ \emph {et~al.}(2024)\citenamefont
  {Caneva~Santoro}, \citenamefont {Roy}, \citenamefont {Vicente}, \citenamefont
  {Haney}, \citenamefont {Piccinni}, \citenamefont {Del~Pozzo},\ and\
  \citenamefont {Martinez}}]{CanevaSantoro:2023aol}%
  \BibitemOpen
  \bibfield  {author} {\bibinfo {author} {\bibfnamefont {G.}~\bibnamefont
  {Caneva~Santoro}}, \bibinfo {author} {\bibfnamefont {S.}~\bibnamefont {Roy}},
  \bibinfo {author} {\bibfnamefont {R.}~\bibnamefont {Vicente}}, \bibinfo
  {author} {\bibfnamefont {M.}~\bibnamefont {Haney}}, \bibinfo {author}
  {\bibfnamefont {O.~J.}\ \bibnamefont {Piccinni}}, \bibinfo {author}
  {\bibfnamefont {W.}~\bibnamefont {Del~Pozzo}},\ and\ \bibinfo {author}
  {\bibfnamefont {M.}~\bibnamefont {Martinez}},\ }\bibfield  {title} {\bibinfo
  {title} {{First Constraints on Compact Binary Environments from LIGO-Virgo
  Data}},\ }\href {https://doi.org/10.1103/PhysRevLett.132.251401} {\bibfield
  {journal} {\bibinfo  {journal} {Phys. Rev. Lett.}\ }\textbf {\bibinfo
  {volume} {132}},\ \bibinfo {pages} {251401} (\bibinfo {year} {2024})},\
  \Eprint {https://arxiv.org/abs/2309.05061} {arXiv:2309.05061 [gr-qc]}
  \BibitemShut {NoStop}%
\bibitem [{\citenamefont {Farris}\ \emph {et~al.}(2011)\citenamefont {Farris},
  \citenamefont {Liu},\ and\ \citenamefont {Shapiro}}]{Farris:2011vx}%
  \BibitemOpen
  \bibfield  {author} {\bibinfo {author} {\bibfnamefont {B.~D.}\ \bibnamefont
  {Farris}}, \bibinfo {author} {\bibfnamefont {Y.~T.}\ \bibnamefont {Liu}},\
  and\ \bibinfo {author} {\bibfnamefont {S.~L.}\ \bibnamefont {Shapiro}},\
  }\bibfield  {title} {\bibinfo {title} {{Binary black hole mergers in gaseous
  disks: Simulations in general relativity}},\ }\href
  {https://doi.org/10.1103/PhysRevD.84.024024} {\bibfield  {journal} {\bibinfo
  {journal} {Phys. Rev. D}\ }\textbf {\bibinfo {volume} {84}},\ \bibinfo
  {pages} {024024} (\bibinfo {year} {2011})},\ \Eprint
  {https://arxiv.org/abs/1105.2821} {arXiv:1105.2821 [astro-ph.HE]}
  \BibitemShut {NoStop}%
\bibitem [{\citenamefont {Farris}\ \emph {et~al.}(2012)\citenamefont {Farris},
  \citenamefont {Gold}, \citenamefont {Paschalidis}, \citenamefont {Etienne},\
  and\ \citenamefont {Shapiro}}]{Farris:2012ux}%
  \BibitemOpen
  \bibfield  {author} {\bibinfo {author} {\bibfnamefont {B.~D.}\ \bibnamefont
  {Farris}}, \bibinfo {author} {\bibfnamefont {R.}~\bibnamefont {Gold}},
  \bibinfo {author} {\bibfnamefont {V.}~\bibnamefont {Paschalidis}}, \bibinfo
  {author} {\bibfnamefont {Z.~B.}\ \bibnamefont {Etienne}},\ and\ \bibinfo
  {author} {\bibfnamefont {S.~L.}\ \bibnamefont {Shapiro}},\ }\bibfield
  {title} {\bibinfo {title} {{Binary black hole mergers in magnetized disks:
  simulations in full general relativity}},\ }\href
  {https://doi.org/10.1103/PhysRevLett.109.221102} {\bibfield  {journal}
  {\bibinfo  {journal} {Phys. Rev. Lett.}\ }\textbf {\bibinfo {volume} {109}},\
  \bibinfo {pages} {221102} (\bibinfo {year} {2012})},\ \Eprint
  {https://arxiv.org/abs/1207.3354} {arXiv:1207.3354 [astro-ph.HE]}
  \BibitemShut {NoStop}%
\bibitem [{\citenamefont {Paschalidis}\ \emph {et~al.}(2021)\citenamefont
  {Paschalidis}, \citenamefont {Bright}, \citenamefont {Ruiz},\ and\
  \citenamefont {Gold}}]{Paschalidis:2021ntt}%
  \BibitemOpen
  \bibfield  {author} {\bibinfo {author} {\bibfnamefont {V.}~\bibnamefont
  {Paschalidis}}, \bibinfo {author} {\bibfnamefont {J.}~\bibnamefont {Bright}},
  \bibinfo {author} {\bibfnamefont {M.}~\bibnamefont {Ruiz}},\ and\ \bibinfo
  {author} {\bibfnamefont {R.}~\bibnamefont {Gold}},\ }\bibfield  {title}
  {\bibinfo {title} {{Minidisk dynamics in accreting, spinning black hole
  binaries: Simulations in full general relativity}},\ }\href
  {https://doi.org/10.3847/2041-8213/abee21} {\bibfield  {journal} {\bibinfo
  {journal} {Astrophys. J. Lett.}\ }\textbf {\bibinfo {volume} {910}},\
  \bibinfo {pages} {L26} (\bibinfo {year} {2021})},\ \Eprint
  {https://arxiv.org/abs/2102.06712} {arXiv:2102.06712 [astro-ph.HE]}
  \BibitemShut {NoStop}%
\bibitem [{\citenamefont {Ruiz}\ \emph {et~al.}(2023)\citenamefont {Ruiz},
  \citenamefont {Tsokaros},\ and\ \citenamefont {Shapiro}}]{Ruiz:2023hit}%
  \BibitemOpen
  \bibfield  {author} {\bibinfo {author} {\bibfnamefont {M.}~\bibnamefont
  {Ruiz}}, \bibinfo {author} {\bibfnamefont {A.}~\bibnamefont {Tsokaros}},\
  and\ \bibinfo {author} {\bibfnamefont {S.~L.}\ \bibnamefont {Shapiro}},\
  }\bibfield  {title} {\bibinfo {title} {{General relativistic
  magnetohydrodynamic simulations of accretion disks around tilted binary black
  holes of unequal mass}},\ }\href
  {https://doi.org/10.1103/PhysRevD.108.124043} {\bibfield  {journal} {\bibinfo
   {journal} {Phys. Rev. D}\ }\textbf {\bibinfo {volume} {108}},\ \bibinfo
  {pages} {124043} (\bibinfo {year} {2023})},\ \Eprint
  {https://arxiv.org/abs/2302.09083} {arXiv:2302.09083 [astro-ph.HE]}
  \BibitemShut {NoStop}%
\bibitem [{\citenamefont {Choudhary}\ \emph {et~al.}(2021)\citenamefont
  {Choudhary}, \citenamefont {Sanchis-Gual}, \citenamefont {Gupta},
  \citenamefont {Degollado}, \citenamefont {Bose},\ and\ \citenamefont
  {Font}}]{Choudhary:2020pxy}%
  \BibitemOpen
  \bibfield  {author} {\bibinfo {author} {\bibfnamefont {S.}~\bibnamefont
  {Choudhary}}, \bibinfo {author} {\bibfnamefont {N.}~\bibnamefont
  {Sanchis-Gual}}, \bibinfo {author} {\bibfnamefont {A.}~\bibnamefont {Gupta}},
  \bibinfo {author} {\bibfnamefont {J.~C.}\ \bibnamefont {Degollado}}, \bibinfo
  {author} {\bibfnamefont {S.}~\bibnamefont {Bose}},\ and\ \bibinfo {author}
  {\bibfnamefont {J.~A.}\ \bibnamefont {Font}},\ }\bibfield  {title} {\bibinfo
  {title} {{Gravitational waves from binary black hole mergers surrounded by
  scalar field clouds: Numerical simulations and observational implications}},\
  }\href {https://doi.org/10.1103/PhysRevD.103.044032} {\bibfield  {journal}
  {\bibinfo  {journal} {Phys. Rev. D}\ }\textbf {\bibinfo {volume} {103}},\
  \bibinfo {pages} {044032} (\bibinfo {year} {2021})},\ \Eprint
  {https://arxiv.org/abs/2010.00935} {arXiv:2010.00935 [gr-qc]} \BibitemShut
  {NoStop}%
\bibitem [{\citenamefont {Zhang}\ \emph {et~al.}(2023)\citenamefont {Zhang},
  \citenamefont {Gracia-Linares}, \citenamefont {Laguna}, \citenamefont
  {Shoemaker},\ and\ \citenamefont {Liu}}]{Zhang:2022rex}%
  \BibitemOpen
  \bibfield  {author} {\bibinfo {author} {\bibfnamefont {Y.-P.}\ \bibnamefont
  {Zhang}}, \bibinfo {author} {\bibfnamefont {M.}~\bibnamefont
  {Gracia-Linares}}, \bibinfo {author} {\bibfnamefont {P.}~\bibnamefont
  {Laguna}}, \bibinfo {author} {\bibfnamefont {D.}~\bibnamefont {Shoemaker}},\
  and\ \bibinfo {author} {\bibfnamefont {Y.-X.}\ \bibnamefont {Liu}},\
  }\bibfield  {title} {\bibinfo {title} {{Gravitational recoil from binary
  black hole mergers in scalar field clouds}},\ }\href
  {https://doi.org/10.1103/PhysRevD.107.044039} {\bibfield  {journal} {\bibinfo
   {journal} {Phys. Rev. D}\ }\textbf {\bibinfo {volume} {107}},\ \bibinfo
  {pages} {044039} (\bibinfo {year} {2023})},\ \Eprint
  {https://arxiv.org/abs/2209.11814} {arXiv:2209.11814 [gr-qc]} \BibitemShut
  {NoStop}%
\bibitem [{\citenamefont {Aurrekoetxea}\ \emph {et~al.}(2024)\citenamefont
  {Aurrekoetxea}, \citenamefont {Clough}, \citenamefont {Bamber},\ and\
  \citenamefont {Ferreira}}]{Aurrekoetxea:2023jwk}%
  \BibitemOpen
  \bibfield  {author} {\bibinfo {author} {\bibfnamefont {J.~C.}\ \bibnamefont
  {Aurrekoetxea}}, \bibinfo {author} {\bibfnamefont {K.}~\bibnamefont
  {Clough}}, \bibinfo {author} {\bibfnamefont {J.}~\bibnamefont {Bamber}},\
  and\ \bibinfo {author} {\bibfnamefont {P.~G.}\ \bibnamefont {Ferreira}},\
  }\bibfield  {title} {\bibinfo {title} {{Effect of Wave Dark Matter on Equal
  Mass Black Hole Mergers}},\ }\href
  {https://doi.org/10.1103/PhysRevLett.132.211401} {\bibfield  {journal}
  {\bibinfo  {journal} {Phys. Rev. Lett.}\ }\textbf {\bibinfo {volume} {132}},\
  \bibinfo {pages} {211401} (\bibinfo {year} {2024})},\ \Eprint
  {https://arxiv.org/abs/2311.18156} {arXiv:2311.18156 [gr-qc]} \BibitemShut
  {NoStop}%
\bibitem [{\citenamefont {Loeb}(2016)}]{Loeb:2016fzn}%
  \BibitemOpen
  \bibfield  {author} {\bibinfo {author} {\bibfnamefont {A.}~\bibnamefont
  {Loeb}},\ }\bibfield  {title} {\bibinfo {title} {{Electromagnetic
  Counterparts to Black Hole Mergers Detected by LIGO}},\ }\href
  {https://doi.org/10.3847/2041-8205/819/2/L21} {\bibfield  {journal} {\bibinfo
   {journal} {Astrophys. J. Lett.}\ }\textbf {\bibinfo {volume} {819}},\
  \bibinfo {pages} {L21} (\bibinfo {year} {2016})},\ \Eprint
  {https://arxiv.org/abs/1602.04735} {arXiv:1602.04735 [astro-ph.HE]}
  \BibitemShut {NoStop}%
\bibitem [{\citenamefont {Reisswig}\ \emph {et~al.}(2013)\citenamefont
  {Reisswig}, \citenamefont {Ott}, \citenamefont {Abdikamalov}, \citenamefont
  {Haas}, \citenamefont {Moesta},\ and\ \citenamefont
  {Schnetter}}]{Reisswig:2013sqa}%
  \BibitemOpen
  \bibfield  {author} {\bibinfo {author} {\bibfnamefont {C.}~\bibnamefont
  {Reisswig}}, \bibinfo {author} {\bibfnamefont {C.~D.}\ \bibnamefont {Ott}},
  \bibinfo {author} {\bibfnamefont {E.}~\bibnamefont {Abdikamalov}}, \bibinfo
  {author} {\bibfnamefont {R.}~\bibnamefont {Haas}}, \bibinfo {author}
  {\bibfnamefont {P.}~\bibnamefont {Moesta}},\ and\ \bibinfo {author}
  {\bibfnamefont {E.}~\bibnamefont {Schnetter}},\ }\bibfield  {title} {\bibinfo
  {title} {{Formation and Coalescence of Cosmological Supermassive Black Hole
  Binaries in Supermassive Star Collapse}},\ }\href
  {https://doi.org/10.1103/PhysRevLett.111.151101} {\bibfield  {journal}
  {\bibinfo  {journal} {Phys. Rev. Lett.}\ }\textbf {\bibinfo {volume} {111}},\
  \bibinfo {pages} {151101} (\bibinfo {year} {2013})},\ \Eprint
  {https://arxiv.org/abs/1304.7787} {arXiv:1304.7787 [astro-ph.CO]}
  \BibitemShut {NoStop}%
\bibitem [{\citenamefont {Dai}\ \emph {et~al.}(2017)\citenamefont {Dai},
  \citenamefont {McKinney},\ and\ \citenamefont {Miller}}]{Dai:2016ejj}%
  \BibitemOpen
  \bibfield  {author} {\bibinfo {author} {\bibfnamefont {L.}~\bibnamefont
  {Dai}}, \bibinfo {author} {\bibfnamefont {J.~C.}\ \bibnamefont {McKinney}},\
  and\ \bibinfo {author} {\bibfnamefont {M.~C.}\ \bibnamefont {Miller}},\
  }\bibfield  {title} {\bibinfo {title} {{Energetic constraints on
  electromagnetic signals from double black hole mergers}},\ }\href
  {https://doi.org/10.1093/mnrasl/slx086} {\bibfield  {journal} {\bibinfo
  {journal} {Mon. Not. Roy. Astron. Soc.}\ }\textbf {\bibinfo {volume} {470}},\
  \bibinfo {pages} {L92} (\bibinfo {year} {2017})},\ \Eprint
  {https://arxiv.org/abs/1611.00764} {arXiv:1611.00764 [astro-ph.HE]}
  \BibitemShut {NoStop}%
\bibitem [{\citenamefont {Leong}\ \emph {et~al.}(2023)\citenamefont {Leong},
  \citenamefont {Calder\'on~Bustillo}, \citenamefont {Gracia-Linares},\ and\
  \citenamefont {Laguna}}]{Leong:2023nuk}%
  \BibitemOpen
  \bibfield  {author} {\bibinfo {author} {\bibfnamefont {S.~H.~W.}\
  \bibnamefont {Leong}}, \bibinfo {author} {\bibfnamefont {J.}~\bibnamefont
  {Calder\'on~Bustillo}}, \bibinfo {author} {\bibfnamefont {M.}~\bibnamefont
  {Gracia-Linares}},\ and\ \bibinfo {author} {\bibfnamefont {P.}~\bibnamefont
  {Laguna}},\ }\bibfield  {title} {\bibinfo {title} {{Detectability of
  dense-environment effects on black-hole mergers: The scalar field case,
  higher-order ringdown modes, and parameter biases}},\ }\href
  {https://doi.org/10.1103/PhysRevD.108.124079} {\bibfield  {journal} {\bibinfo
   {journal} {Phys. Rev. D}\ }\textbf {\bibinfo {volume} {108}},\ \bibinfo
  {pages} {124079} (\bibinfo {year} {2023})},\ \Eprint
  {https://arxiv.org/abs/2308.03250} {arXiv:2308.03250 [gr-qc]} \BibitemShut
  {NoStop}%
\bibitem [{\citenamefont {Brandt}\ \emph {et~al.}(2024)\citenamefont {Brandt},
  \citenamefont {Haas}, \citenamefont {Diener}, \citenamefont {Ennoggi},
  \citenamefont {Ferguson} \emph {et~al.}}]{EinsteinToolkit:2024_05}%
  \BibitemOpen
  \bibfield  {author} {\bibinfo {author} {\bibfnamefont {S.~R.}\ \bibnamefont
  {Brandt}}, \bibinfo {author} {\bibfnamefont {R.}~\bibnamefont {Haas}},
  \bibinfo {author} {\bibfnamefont {P.}~\bibnamefont {Diener}}, \bibinfo
  {author} {\bibfnamefont {L.}~\bibnamefont {Ennoggi}}, \bibinfo {author}
  {\bibfnamefont {D.}~\bibnamefont {Ferguson}}, \emph {et~al.},\ }\href
  {https://doi.org/10.5281/zenodo.12588764} {\bibinfo {title} {The einstein
  toolkit}} (\bibinfo {year} {2024})\BibitemShut {NoStop}%
\bibitem [{\citenamefont {L{\"{o}}ffler}\ \emph {et~al.}(2012)\citenamefont
  {L{\"{o}}ffler}, \citenamefont {Faber}, \citenamefont {Bentivegna},
  \citenamefont {Bode}, \citenamefont {Diener} \emph
  {et~al.}}]{Loffler:2011ay}%
  \BibitemOpen
  \bibfield  {author} {\bibinfo {author} {\bibfnamefont {F.}~\bibnamefont
  {L{\"{o}}ffler}}, \bibinfo {author} {\bibfnamefont {J.}~\bibnamefont
  {Faber}}, \bibinfo {author} {\bibfnamefont {E.}~\bibnamefont {Bentivegna}},
  \bibinfo {author} {\bibfnamefont {T.}~\bibnamefont {Bode}}, \bibinfo {author}
  {\bibfnamefont {P.}~\bibnamefont {Diener}}, \emph {et~al.},\ }\bibfield
  {title} {\bibinfo {title} {{{T}he {E}instein {T}oolkit: {A} {C}ommunity
  {C}omputational {I}nfrastructure for {R}elativistic {A}strophysics}},\ }\href
  {https://doi.org/doi:10.1088/0264-9381/29/11/115001} {\bibfield  {journal}
  {\bibinfo  {journal} {Class. Quantum Grav.}\ }\textbf {\bibinfo {volume}
  {29}},\ \bibinfo {pages} {115001} (\bibinfo {year} {2012})},\ \Eprint
  {https://arxiv.org/abs/arXiv:1111.3344 [gr-qc]} {arXiv:1111.3344 [gr-qc]}
  \BibitemShut {NoStop}%
\bibitem [{\citenamefont {Shibata}\ and\ \citenamefont
  {Nakamura}(1995)}]{Shibata:1995we}%
  \BibitemOpen
  \bibfield  {author} {\bibinfo {author} {\bibfnamefont {M.}~\bibnamefont
  {Shibata}}\ and\ \bibinfo {author} {\bibfnamefont {T.}~\bibnamefont
  {Nakamura}},\ }\bibfield  {title} {\bibinfo {title} {{Evolution of
  three-dimensional gravitational waves: Harmonic slicing case}},\ }\href
  {https://doi.org/10.1103/PhysRevD.52.5428} {\bibfield  {journal} {\bibinfo
  {journal} {Phys. Rev. D}\ }\textbf {\bibinfo {volume} {52}},\ \bibinfo
  {pages} {5428} (\bibinfo {year} {1995})}\BibitemShut {NoStop}%
\bibitem [{\citenamefont {Baumgarte}\ and\ \citenamefont
  {Shapiro}(1998)}]{Baumgarte:1998te}%
  \BibitemOpen
  \bibfield  {author} {\bibinfo {author} {\bibfnamefont {T.~W.}\ \bibnamefont
  {Baumgarte}}\ and\ \bibinfo {author} {\bibfnamefont {S.~L.}\ \bibnamefont
  {Shapiro}},\ }\bibfield  {title} {\bibinfo {title} {{On the numerical
  integration of Einstein's field equations}},\ }\href
  {https://doi.org/10.1103/PhysRevD.59.024007} {\bibfield  {journal} {\bibinfo
  {journal} {Phys. Rev. D}\ }\textbf {\bibinfo {volume} {59}},\ \bibinfo
  {pages} {024007} (\bibinfo {year} {1998})},\ \Eprint
  {https://arxiv.org/abs/gr-qc/9810065} {arXiv:gr-qc/9810065} \BibitemShut
  {NoStop}%
\bibitem [{\citenamefont {Shapiro}\ and\ \citenamefont
  {Teukolsky}(1983)}]{Shapiro:1983du}%
  \BibitemOpen
  \bibfield  {author} {\bibinfo {author} {\bibfnamefont {S.~L.}\ \bibnamefont
  {Shapiro}}\ and\ \bibinfo {author} {\bibfnamefont {S.~A.}\ \bibnamefont
  {Teukolsky}},\ }\href {https://doi.org/10.1002/9783527617661} {\emph
  {\bibinfo {title} {{Black holes, white dwarfs, and neutron stars: The physics
  of compact objects}}}}\ (\bibinfo {year} {1983})\BibitemShut {NoStop}%
\bibitem [{\citenamefont {{Hoyle}}\ and\ \citenamefont
  {{Lyttleton}}(1941)}]{Hoyle1941}%
  \BibitemOpen
  \bibfield  {author} {\bibinfo {author} {\bibfnamefont {F.}~\bibnamefont
  {{Hoyle}}}\ and\ \bibinfo {author} {\bibfnamefont {R.~A.}\ \bibnamefont
  {{Lyttleton}}},\ }\bibfield  {title} {\bibinfo {title} {{On the accretion
  theory of stellar evolution}},\ }\href
  {https://doi.org/10.1093/mnras/101.4.227} {\bibfield  {journal} {\bibinfo
  {journal} {{Mon. Not. Roy. Astron. Soc.}}\ }\textbf {\bibinfo {volume}
  {101}},\ \bibinfo {pages} {227} (\bibinfo {year} {1941})}\BibitemShut
  {NoStop}%
\bibitem [{\citenamefont {Edgar}(2004)}]{Edgar:2004mk}%
  \BibitemOpen
  \bibfield  {author} {\bibinfo {author} {\bibfnamefont {R.~G.}\ \bibnamefont
  {Edgar}},\ }\bibfield  {title} {\bibinfo {title} {{A Review of
  Bondi-Hoyle-Lyttleton accretion}},\ }\href
  {https://doi.org/10.1016/j.newar.2004.06.001} {\bibfield  {journal} {\bibinfo
   {journal} {New Astron. Rev.}\ }\textbf {\bibinfo {volume} {48}},\ \bibinfo
  {pages} {843} (\bibinfo {year} {2004})},\ \Eprint
  {https://arxiv.org/abs/astro-ph/0406166} {arXiv:astro-ph/0406166}
  \BibitemShut {NoStop}%
\bibitem [{\citenamefont {Cutler}\ and\ \citenamefont
  {Flanagan}(1994)}]{Cutler:1994ys}%
  \BibitemOpen
  \bibfield  {author} {\bibinfo {author} {\bibfnamefont {C.}~\bibnamefont
  {Cutler}}\ and\ \bibinfo {author} {\bibfnamefont {E.~E.}\ \bibnamefont
  {Flanagan}},\ }\bibfield  {title} {\bibinfo {title} {{Gravitational waves
  from merging compact binaries: How accurately can one extract the binary's
  parameters from the inspiral wave form?}},\ }\href
  {https://doi.org/10.1103/PhysRevD.49.2658} {\bibfield  {journal} {\bibinfo
  {journal} {Phys. Rev. D}\ }\textbf {\bibinfo {volume} {49}},\ \bibinfo
  {pages} {2658} (\bibinfo {year} {1994})},\ \Eprint
  {https://arxiv.org/abs/gr-qc/9402014} {arXiv:gr-qc/9402014} \BibitemShut
  {NoStop}%
\bibitem [{\citenamefont {Droz}\ \emph {et~al.}(1999)\citenamefont {Droz},
  \citenamefont {Knapp}, \citenamefont {Poisson},\ and\ \citenamefont
  {Owen}}]{Droz:1999qx}%
  \BibitemOpen
  \bibfield  {author} {\bibinfo {author} {\bibfnamefont {S.}~\bibnamefont
  {Droz}}, \bibinfo {author} {\bibfnamefont {D.~J.}\ \bibnamefont {Knapp}},
  \bibinfo {author} {\bibfnamefont {E.}~\bibnamefont {Poisson}},\ and\ \bibinfo
  {author} {\bibfnamefont {B.~J.}\ \bibnamefont {Owen}},\ }\bibfield  {title}
  {\bibinfo {title} {{Gravitational waves from inspiraling compact binaries:
  Validity of the stationary phase approximation to the Fourier transform}},\
  }\href {https://doi.org/10.1103/PhysRevD.59.124016} {\bibfield  {journal}
  {\bibinfo  {journal} {Phys. Rev. D}\ }\textbf {\bibinfo {volume} {59}},\
  \bibinfo {pages} {124016} (\bibinfo {year} {1999})},\ \Eprint
  {https://arxiv.org/abs/gr-qc/9901076} {arXiv:gr-qc/9901076} \BibitemShut
  {NoStop}%
\bibitem [{\citenamefont {Pratten}\ \emph {et~al.}(2021)\citenamefont {Pratten}
  \emph {et~al.}}]{Pratten:2020ceb}%
  \BibitemOpen
  \bibfield  {author} {\bibinfo {author} {\bibfnamefont {G.}~\bibnamefont
  {Pratten}} \emph {et~al.},\ }\bibfield  {title} {\bibinfo {title}
  {{Computationally efficient models for the dominant and subdominant harmonic
  modes of precessing binary black holes}},\ }\href
  {https://doi.org/10.1103/PhysRevD.103.104056} {\bibfield  {journal} {\bibinfo
   {journal} {Phys. Rev. D}\ }\textbf {\bibinfo {volume} {103}},\ \bibinfo
  {pages} {104056} (\bibinfo {year} {2021})},\ \Eprint
  {https://arxiv.org/abs/2004.06503} {arXiv:2004.06503 [gr-qc]} \BibitemShut
  {NoStop}%
\bibitem [{\citenamefont {Dietrich}\ \emph {et~al.}(2019)\citenamefont
  {Dietrich}, \citenamefont {Samajdar}, \citenamefont {Khan}, \citenamefont
  {Johnson-McDaniel}, \citenamefont {Dudi},\ and\ \citenamefont
  {Tichy}}]{Dietrich:2019kaq}%
  \BibitemOpen
  \bibfield  {author} {\bibinfo {author} {\bibfnamefont {T.}~\bibnamefont
  {Dietrich}}, \bibinfo {author} {\bibfnamefont {A.}~\bibnamefont {Samajdar}},
  \bibinfo {author} {\bibfnamefont {S.}~\bibnamefont {Khan}}, \bibinfo {author}
  {\bibfnamefont {N.~K.}\ \bibnamefont {Johnson-McDaniel}}, \bibinfo {author}
  {\bibfnamefont {R.}~\bibnamefont {Dudi}},\ and\ \bibinfo {author}
  {\bibfnamefont {W.}~\bibnamefont {Tichy}},\ }\bibfield  {title} {\bibinfo
  {title} {{Improving the NRTidal model for binary neutron star systems}},\
  }\href {https://doi.org/10.1103/PhysRevD.100.044003} {\bibfield  {journal}
  {\bibinfo  {journal} {Phys. Rev. D}\ }\textbf {\bibinfo {volume} {100}},\
  \bibinfo {pages} {044003} (\bibinfo {year} {2019})},\ \Eprint
  {https://arxiv.org/abs/1905.06011} {arXiv:1905.06011 [gr-qc]} \BibitemShut
  {NoStop}%
\bibitem [{\citenamefont {Barausse}\ \emph {et~al.}(2014)\citenamefont
  {Barausse}, \citenamefont {Cardoso},\ and\ \citenamefont
  {Pani}}]{Barausse:2014tra}%
  \BibitemOpen
  \bibfield  {author} {\bibinfo {author} {\bibfnamefont {E.}~\bibnamefont
  {Barausse}}, \bibinfo {author} {\bibfnamefont {V.}~\bibnamefont {Cardoso}},\
  and\ \bibinfo {author} {\bibfnamefont {P.}~\bibnamefont {Pani}},\ }\bibfield
  {title} {\bibinfo {title} {{Can environmental effects spoil precision
  gravitational-wave astrophysics?}},\ }\href
  {https://doi.org/10.1103/PhysRevD.89.104059} {\bibfield  {journal} {\bibinfo
  {journal} {Phys. Rev. D}\ }\textbf {\bibinfo {volume} {89}},\ \bibinfo
  {pages} {104059} (\bibinfo {year} {2014})},\ \Eprint
  {https://arxiv.org/abs/1404.7149} {arXiv:1404.7149 [gr-qc]} \BibitemShut
  {NoStop}%
\bibitem [{\citenamefont {Cardoso}\ and\ \citenamefont
  {Maselli}(2020)}]{Cardoso:2019rou}%
  \BibitemOpen
  \bibfield  {author} {\bibinfo {author} {\bibfnamefont {V.}~\bibnamefont
  {Cardoso}}\ and\ \bibinfo {author} {\bibfnamefont {A.}~\bibnamefont
  {Maselli}},\ }\bibfield  {title} {\bibinfo {title} {{Constraints on the
  astrophysical environment of binaries with gravitational-wave
  observations}},\ }\href {https://doi.org/10.1051/0004-6361/202037654}
  {\bibfield  {journal} {\bibinfo  {journal} {Astron. Astrophys.}\ }\textbf
  {\bibinfo {volume} {644}},\ \bibinfo {pages} {A147} (\bibinfo {year}
  {2020})},\ \Eprint {https://arxiv.org/abs/1909.05870} {arXiv:1909.05870
  [astro-ph.HE]} \BibitemShut {NoStop}%
\bibitem [{\citenamefont {{LIGO Scientific Collaboration}}\ \emph
  {et~al.}(2018)\citenamefont {{LIGO Scientific Collaboration}}, \citenamefont
  {{Virgo Collaboration}},\ and\ \citenamefont {{KAGRA
  Collaboration}}}]{lalsuite}%
  \BibitemOpen
  \bibfield  {author} {\bibinfo {author} {\bibnamefont {{LIGO Scientific
  Collaboration}}}, \bibinfo {author} {\bibnamefont {{Virgo Collaboration}}},\
  and\ \bibinfo {author} {\bibnamefont {{KAGRA Collaboration}}},\ }\href
  {https://doi.org/10.7935/GT1W-FZ16} {\bibinfo {title} {{LVK} {A}lgorithm
  {L}ibrary - {LALS}uite}},\ \bibinfo {howpublished} {Free software (GPL)}
  (\bibinfo {year} {2018})\BibitemShut {NoStop}%
\bibitem [{\citenamefont {Ashton}\ \emph {et~al.}(2019)\citenamefont {Ashton}
  \emph {et~al.}}]{Ashton:2018jfp}%
  \BibitemOpen
  \bibfield  {author} {\bibinfo {author} {\bibfnamefont {G.}~\bibnamefont
  {Ashton}} \emph {et~al.},\ }\bibfield  {title} {\bibinfo {title} {{BILBY: A
  user-friendly Bayesian inference library for gravitational-wave astronomy}},\
  }\href {https://doi.org/10.3847/1538-4365/ab06fc} {\bibfield  {journal}
  {\bibinfo  {journal} {Astrophys. J. Suppl.}\ }\textbf {\bibinfo {volume}
  {241}},\ \bibinfo {pages} {27} (\bibinfo {year} {2019})},\ \Eprint
  {https://arxiv.org/abs/1811.02042} {arXiv:1811.02042 [astro-ph.IM]}
  \BibitemShut {NoStop}%
\bibitem [{\citenamefont {Speagle}(2020)}]{Speagle:2019ivv}%
  \BibitemOpen
  \bibfield  {author} {\bibinfo {author} {\bibfnamefont {J.~S.}\ \bibnamefont
  {Speagle}},\ }\bibfield  {title} {\bibinfo {title} {{dynesty: a dynamic
  nested sampling package for estimating Bayesian posteriors and evidences}},\
  }\href {https://doi.org/10.1093/mnras/staa278} {\bibfield  {journal}
  {\bibinfo  {journal} {Mon. Not. Roy. Astron. Soc.}\ }\textbf {\bibinfo
  {volume} {493}},\ \bibinfo {pages} {3132} (\bibinfo {year} {2020})},\ \Eprint
  {https://arxiv.org/abs/1904.02180} {arXiv:1904.02180 [astro-ph.IM]}
  \BibitemShut {NoStop}%
\bibitem [{\citenamefont {B\'ecsy}\ \emph {et~al.}(2017)\citenamefont
  {B\'ecsy}, \citenamefont {Raffai}, \citenamefont {Cornish}, \citenamefont
  {Essick}, \citenamefont {Kanner}, \citenamefont {Katsavounidis},
  \citenamefont {Littenberg}, \citenamefont {Millhouse},\ and\ \citenamefont
  {Vitale}}]{Becsy:2016ofp}%
  \BibitemOpen
  \bibfield  {author} {\bibinfo {author} {\bibfnamefont {B.}~\bibnamefont
  {B\'ecsy}}, \bibinfo {author} {\bibfnamefont {P.}~\bibnamefont {Raffai}},
  \bibinfo {author} {\bibfnamefont {N.~J.}\ \bibnamefont {Cornish}}, \bibinfo
  {author} {\bibfnamefont {R.}~\bibnamefont {Essick}}, \bibinfo {author}
  {\bibfnamefont {J.}~\bibnamefont {Kanner}}, \bibinfo {author} {\bibfnamefont
  {E.}~\bibnamefont {Katsavounidis}}, \bibinfo {author} {\bibfnamefont {T.~B.}\
  \bibnamefont {Littenberg}}, \bibinfo {author} {\bibfnamefont
  {M.}~\bibnamefont {Millhouse}},\ and\ \bibinfo {author} {\bibfnamefont
  {S.}~\bibnamefont {Vitale}},\ }\bibfield  {title} {\bibinfo {title}
  {{Parameter estimation for gravitational-wave bursts with the BayesWave
  pipeline}},\ }\href {https://doi.org/10.3847/1538-4357/aa63ef} {\bibfield
  {journal} {\bibinfo  {journal} {Astrophys. J.}\ }\textbf {\bibinfo {volume}
  {839}},\ \bibinfo {pages} {15} (\bibinfo {year} {2017})},\ \Eprint
  {https://arxiv.org/abs/1612.02003} {arXiv:1612.02003 [astro-ph.HE]}
  \BibitemShut {NoStop}%
\bibitem [{\citenamefont {Roy}(2022)}]{Roy:2022teu}%
  \BibitemOpen
  \bibfield  {author} {\bibinfo {author} {\bibfnamefont {S.}~\bibnamefont
  {Roy}},\ }\bibfield  {title} {\bibinfo {title} {{Nonorthogonal wavelet
  transformation for reconstructing gravitational wave signals}},\ }\href
  {https://doi.org/10.1103/PhysRevResearch.4.033078} {\bibfield  {journal}
  {\bibinfo  {journal} {Phys. Rev. Res.}\ }\textbf {\bibinfo {volume} {4}},\
  \bibinfo {pages} {033078} (\bibinfo {year} {2022})},\ \Eprint
  {https://arxiv.org/abs/2201.01526} {arXiv:2201.01526 [gr-qc]} \BibitemShut
  {NoStop}%
\bibitem [{\citenamefont {Hannam}\ \emph {et~al.}(2014)\citenamefont {Hannam},
  \citenamefont {Schmidt}, \citenamefont {Boh\'e}, \citenamefont {Haegel},
  \citenamefont {Husa}, \citenamefont {Ohme}, \citenamefont {Pratten},\ and\
  \citenamefont {P\"urrer}}]{Hannam:2013oca}%
  \BibitemOpen
  \bibfield  {author} {\bibinfo {author} {\bibfnamefont {M.}~\bibnamefont
  {Hannam}}, \bibinfo {author} {\bibfnamefont {P.}~\bibnamefont {Schmidt}},
  \bibinfo {author} {\bibfnamefont {A.}~\bibnamefont {Boh\'e}}, \bibinfo
  {author} {\bibfnamefont {L.}~\bibnamefont {Haegel}}, \bibinfo {author}
  {\bibfnamefont {S.}~\bibnamefont {Husa}}, \bibinfo {author} {\bibfnamefont
  {F.}~\bibnamefont {Ohme}}, \bibinfo {author} {\bibfnamefont {G.}~\bibnamefont
  {Pratten}},\ and\ \bibinfo {author} {\bibfnamefont {M.}~\bibnamefont
  {P\"urrer}},\ }\bibfield  {title} {\bibinfo {title} {{Simple Model of
  Complete Precessing Black-Hole-Binary Gravitational Waveforms}},\ }\href
  {https://doi.org/10.1103/PhysRevLett.113.151101} {\bibfield  {journal}
  {\bibinfo  {journal} {Phys. Rev. Lett.}\ }\textbf {\bibinfo {volume} {113}},\
  \bibinfo {pages} {151101} (\bibinfo {year} {2014})},\ \Eprint
  {https://arxiv.org/abs/1308.3271} {arXiv:1308.3271 [gr-qc]} \BibitemShut
  {NoStop}%
\bibitem [{\citenamefont {Schmidt}\ \emph {et~al.}(2015)\citenamefont
  {Schmidt}, \citenamefont {Ohme},\ and\ \citenamefont
  {Hannam}}]{Schmidt:2014iyl}%
  \BibitemOpen
  \bibfield  {author} {\bibinfo {author} {\bibfnamefont {P.}~\bibnamefont
  {Schmidt}}, \bibinfo {author} {\bibfnamefont {F.}~\bibnamefont {Ohme}},\ and\
  \bibinfo {author} {\bibfnamefont {M.}~\bibnamefont {Hannam}},\ }\bibfield
  {title} {\bibinfo {title} {{Towards models of gravitational waveforms from
  generic binaries II: Modelling precession effects with a single effective
  precession parameter}},\ }\href {https://doi.org/10.1103/PhysRevD.91.024043}
  {\bibfield  {journal} {\bibinfo  {journal} {Phys. Rev. D}\ }\textbf {\bibinfo
  {volume} {91}},\ \bibinfo {pages} {024043} (\bibinfo {year} {2015})},\
  \Eprint {https://arxiv.org/abs/1408.1810} {arXiv:1408.1810 [gr-qc]}
  \BibitemShut {NoStop}%
\bibitem [{\citenamefont {Abbott}\ \emph
  {et~al.}(2016{\natexlab{d}})\citenamefont {Abbott} \emph
  {et~al.}}]{LIGOScientific:2016vbw}%
  \BibitemOpen
  \bibfield  {author} {\bibinfo {author} {\bibfnamefont {B.~P.}\ \bibnamefont
  {Abbott}} \emph {et~al.} (\bibinfo {collaboration} {LIGO Scientific,
  Virgo}),\ }\bibfield  {title} {\bibinfo {title} {{GW150914: First results
  from the search for binary black hole coalescence with Advanced LIGO}},\
  }\href {https://doi.org/10.1103/PhysRevD.93.122003} {\bibfield  {journal}
  {\bibinfo  {journal} {Phys. Rev. D}\ }\textbf {\bibinfo {volume} {93}},\
  \bibinfo {pages} {122003} (\bibinfo {year} {2016}{\natexlab{d}})},\ \Eprint
  {https://arxiv.org/abs/1602.03839} {arXiv:1602.03839 [gr-qc]} \BibitemShut
  {NoStop}%
\bibitem [{\citenamefont {Capano}\ \emph {et~al.}(2016)\citenamefont {Capano},
  \citenamefont {Harry}, \citenamefont {Privitera},\ and\ \citenamefont
  {Buonanno}}]{Capano:2016dsf}%
  \BibitemOpen
  \bibfield  {author} {\bibinfo {author} {\bibfnamefont {C.}~\bibnamefont
  {Capano}}, \bibinfo {author} {\bibfnamefont {I.}~\bibnamefont {Harry}},
  \bibinfo {author} {\bibfnamefont {S.}~\bibnamefont {Privitera}},\ and\
  \bibinfo {author} {\bibfnamefont {A.}~\bibnamefont {Buonanno}},\ }\bibfield
  {title} {\bibinfo {title} {{Implementing a search for gravitational waves
  from binary black holes with nonprecessing spin}},\ }\href
  {https://doi.org/10.1103/PhysRevD.93.124007} {\bibfield  {journal} {\bibinfo
  {journal} {Phys. Rev. D}\ }\textbf {\bibinfo {volume} {93}},\ \bibinfo
  {pages} {124007} (\bibinfo {year} {2016})},\ \Eprint
  {https://arxiv.org/abs/1602.03509} {arXiv:1602.03509 [gr-qc]} \BibitemShut
  {NoStop}%
\bibitem [{\citenamefont {{PyCBC Project}}()}]{pycbc-config}%
  \BibitemOpen
  \bibfield  {author} {\bibinfo {author} {\bibnamefont {{PyCBC Project}}},\
  }\href@noop {} {\bibinfo {title} {{PyCBC configuration for O1}}},\ \bibinfo
  {howpublished}
  {\url{https://github.com/gwastro/pycbc-config/tree/master/O1}},\ \bibinfo
  {note} {accessed: 2023-12-09}\BibitemShut {NoStop}%
\bibitem [{\citenamefont {Taracchini}\ \emph {et~al.}(2014)\citenamefont
  {Taracchini} \emph {et~al.}}]{Taracchini:2013rva}%
  \BibitemOpen
  \bibfield  {author} {\bibinfo {author} {\bibfnamefont {A.}~\bibnamefont
  {Taracchini}} \emph {et~al.},\ }\bibfield  {title} {\bibinfo {title}
  {{Effective-one-body model for black-hole binaries with generic mass ratios
  and spins}},\ }\href {https://doi.org/10.1103/PhysRevD.89.061502} {\bibfield
  {journal} {\bibinfo  {journal} {Phys. Rev. D}\ }\textbf {\bibinfo {volume}
  {89}},\ \bibinfo {pages} {061502} (\bibinfo {year} {2014})},\ \Eprint
  {https://arxiv.org/abs/1311.2544} {arXiv:1311.2544 [gr-qc]} \BibitemShut
  {NoStop}%
\bibitem [{\citenamefont {Usman}\ \emph {et~al.}(2016)\citenamefont {Usman}
  \emph {et~al.}}]{Usman:2015kfa}%
  \BibitemOpen
  \bibfield  {author} {\bibinfo {author} {\bibfnamefont {S.~A.}\ \bibnamefont
  {Usman}} \emph {et~al.},\ }\bibfield  {title} {\bibinfo {title} {{The PyCBC
  search for gravitational waves from compact binary coalescence}},\ }\href
  {https://doi.org/10.1088/0264-9381/33/21/215004} {\bibfield  {journal}
  {\bibinfo  {journal} {Class. Quant. Grav.}\ }\textbf {\bibinfo {volume}
  {33}},\ \bibinfo {pages} {215004} (\bibinfo {year} {2016})},\ \Eprint
  {https://arxiv.org/abs/1508.02357} {arXiv:1508.02357 [gr-qc]} \BibitemShut
  {NoStop}%
\bibitem [{\citenamefont {Allen}\ \emph {et~al.}(2012)\citenamefont {Allen},
  \citenamefont {Anderson}, \citenamefont {Brady}, \citenamefont {Brown},\ and\
  \citenamefont {Creighton}}]{Allen:2005fk}%
  \BibitemOpen
  \bibfield  {author} {\bibinfo {author} {\bibfnamefont {B.}~\bibnamefont
  {Allen}}, \bibinfo {author} {\bibfnamefont {W.~G.}\ \bibnamefont {Anderson}},
  \bibinfo {author} {\bibfnamefont {P.~R.}\ \bibnamefont {Brady}}, \bibinfo
  {author} {\bibfnamefont {D.~A.}\ \bibnamefont {Brown}},\ and\ \bibinfo
  {author} {\bibfnamefont {J.~D.~E.}\ \bibnamefont {Creighton}},\ }\bibfield
  {title} {\bibinfo {title} {{FINDCHIRP: An Algorithm for detection of
  gravitational waves from inspiraling compact binaries}},\ }\href
  {https://doi.org/10.1103/PhysRevD.85.122006} {\bibfield  {journal} {\bibinfo
  {journal} {Phys. Rev. D}\ }\textbf {\bibinfo {volume} {85}},\ \bibinfo
  {pages} {122006} (\bibinfo {year} {2012})},\ \Eprint
  {https://arxiv.org/abs/gr-qc/0509116} {arXiv:gr-qc/0509116} \BibitemShut
  {NoStop}%
\bibitem [{\citenamefont {Allen}(2005)}]{Allen:2004gu}%
  \BibitemOpen
  \bibfield  {author} {\bibinfo {author} {\bibfnamefont {B.}~\bibnamefont
  {Allen}},\ }\bibfield  {title} {\bibinfo {title} {{${\chi}^{2}$
  time-frequency discriminator for gravitational wave detection}},\ }\href
  {https://doi.org/10.1103/PhysRevD.71.062001} {\bibfield  {journal} {\bibinfo
  {journal} {Phys. Rev. D}\ }\textbf {\bibinfo {volume} {71}},\ \bibinfo
  {pages} {062001} (\bibinfo {year} {2005})},\ \Eprint
  {https://arxiv.org/abs/gr-qc/0405045} {arXiv:gr-qc/0405045} \BibitemShut
  {NoStop}%
\bibitem [{\citenamefont {Nitz}\ \emph {et~al.}(2017)\citenamefont {Nitz},
  \citenamefont {Dent}, \citenamefont {Dal~Canton}, \citenamefont {Fairhurst},\
  and\ \citenamefont {Brown}}]{Nitz:2017svb}%
  \BibitemOpen
  \bibfield  {author} {\bibinfo {author} {\bibfnamefont {A.~H.}\ \bibnamefont
  {Nitz}}, \bibinfo {author} {\bibfnamefont {T.}~\bibnamefont {Dent}}, \bibinfo
  {author} {\bibfnamefont {T.}~\bibnamefont {Dal~Canton}}, \bibinfo {author}
  {\bibfnamefont {S.}~\bibnamefont {Fairhurst}},\ and\ \bibinfo {author}
  {\bibfnamefont {D.~A.}\ \bibnamefont {Brown}},\ }\bibfield  {title} {\bibinfo
  {title} {{Detecting binary compact-object mergers with gravitational waves:
  Understanding and Improving the sensitivity of the PyCBC search}},\ }\href
  {https://doi.org/10.3847/1538-4357/aa8f50} {\bibfield  {journal} {\bibinfo
  {journal} {Astrophys. J.}\ }\textbf {\bibinfo {volume} {849}},\ \bibinfo
  {pages} {118} (\bibinfo {year} {2017})},\ \Eprint
  {https://arxiv.org/abs/1705.01513} {arXiv:1705.01513 [gr-qc]} \BibitemShut
  {NoStop}%
\bibitem [{\citenamefont {Dal~Canton}\ \emph {et~al.}(2014)\citenamefont
  {Dal~Canton} \emph {et~al.}}]{DalCanton:2014hxh}%
  \BibitemOpen
  \bibfield  {author} {\bibinfo {author} {\bibfnamefont {T.}~\bibnamefont
  {Dal~Canton}} \emph {et~al.},\ }\bibfield  {title} {\bibinfo {title}
  {{Implementing a search for aligned-spin neutron star-black hole systems with
  advanced ground based gravitational wave detectors}},\ }\href
  {https://doi.org/10.1103/PhysRevD.90.082004} {\bibfield  {journal} {\bibinfo
  {journal} {Phys. Rev. D}\ }\textbf {\bibinfo {volume} {90}},\ \bibinfo
  {pages} {082004} (\bibinfo {year} {2014})},\ \Eprint
  {https://arxiv.org/abs/1405.6731} {arXiv:1405.6731 [gr-qc]} \BibitemShut
  {NoStop}%
\bibitem [{\citenamefont {Apostolatos}(1995)}]{Apostolatos-1995}%
  \BibitemOpen
  \bibfield  {author} {\bibinfo {author} {\bibfnamefont {T.~A.}\ \bibnamefont
  {Apostolatos}},\ }\bibfield  {title} {\bibinfo {title} {Search templates for
  gravitational waves from precessing, inspiraling binaries},\ }\href
  {https://doi.org/10.1103/PhysRevD.52.605} {\bibfield  {journal} {\bibinfo
  {journal} {Phys. Rev. D}\ }\textbf {\bibinfo {volume} {52}},\ \bibinfo
  {pages} {605} (\bibinfo {year} {1995})}\BibitemShut {NoStop}%
\bibitem [{\citenamefont {Narola}\ \emph {et~al.}(2023)\citenamefont {Narola},
  \citenamefont {Roy},\ and\ \citenamefont {Sengupta}}]{Narola:2022aob}%
  \BibitemOpen
  \bibfield  {author} {\bibinfo {author} {\bibfnamefont {H.}~\bibnamefont
  {Narola}}, \bibinfo {author} {\bibfnamefont {S.}~\bibnamefont {Roy}},\ and\
  \bibinfo {author} {\bibfnamefont {A.~S.}\ \bibnamefont {Sengupta}},\
  }\bibfield  {title} {\bibinfo {title} {{Beyond general relativity: Designing
  a template-based search for exotic gravitational wave signals}},\ }\href
  {https://doi.org/10.1103/PhysRevD.107.024017} {\bibfield  {journal} {\bibinfo
   {journal} {Phys. Rev. D}\ }\textbf {\bibinfo {volume} {107}},\ \bibinfo
  {pages} {024017} (\bibinfo {year} {2023})},\ \Eprint
  {https://arxiv.org/abs/2207.10410} {arXiv:2207.10410 [gr-qc]} \BibitemShut
  {NoStop}%
\bibitem [{\citenamefont {Sharma}\ \emph {et~al.}(2024)\citenamefont {Sharma},
  \citenamefont {Roy},\ and\ \citenamefont {Sengupta}}]{Sharma:2023djw}%
  \BibitemOpen
  \bibfield  {author} {\bibinfo {author} {\bibfnamefont {A.}~\bibnamefont
  {Sharma}}, \bibinfo {author} {\bibfnamefont {S.}~\bibnamefont {Roy}},\ and\
  \bibinfo {author} {\bibfnamefont {A.~S.}\ \bibnamefont {Sengupta}},\
  }\bibfield  {title} {\bibinfo {title} {{Template bank to search for exotic
  gravitational wave signals from astrophysical compact binaries}},\ }\href
  {https://doi.org/10.1103/PhysRevD.109.124049} {\bibfield  {journal} {\bibinfo
   {journal} {Phys. Rev. D}\ }\textbf {\bibinfo {volume} {109}},\ \bibinfo
  {pages} {124049} (\bibinfo {year} {2024})},\ \Eprint
  {https://arxiv.org/abs/2311.03274} {arXiv:2311.03274 [gr-qc]} \BibitemShut
  {NoStop}%
\bibitem [{\citenamefont {Ghosh}\ \emph {et~al.}(2016)\citenamefont {Ghosh}
  \emph {et~al.}}]{Ghosh:2016qgn}%
  \BibitemOpen
  \bibfield  {author} {\bibinfo {author} {\bibfnamefont {A.}~\bibnamefont
  {Ghosh}} \emph {et~al.},\ }\bibfield  {title} {\bibinfo {title} {{Testing
  general relativity using golden black-hole binaries}},\ }\href
  {https://doi.org/10.1103/PhysRevD.94.021101} {\bibfield  {journal} {\bibinfo
  {journal} {Phys. Rev. D}\ }\textbf {\bibinfo {volume} {94}},\ \bibinfo
  {pages} {021101} (\bibinfo {year} {2016})},\ \Eprint
  {https://arxiv.org/abs/1602.02453} {arXiv:1602.02453 [gr-qc]} \BibitemShut
  {NoStop}%
\bibitem [{\citenamefont {Ghosh}\ \emph {et~al.}(2018)\citenamefont {Ghosh},
  \citenamefont {Johnson-Mcdaniel}, \citenamefont {Ghosh}, \citenamefont
  {Mishra}, \citenamefont {Ajith}, \citenamefont {Del~Pozzo}, \citenamefont
  {Berry}, \citenamefont {Nielsen},\ and\ \citenamefont
  {London}}]{Ghosh:2017gfp}%
  \BibitemOpen
  \bibfield  {author} {\bibinfo {author} {\bibfnamefont {A.}~\bibnamefont
  {Ghosh}}, \bibinfo {author} {\bibfnamefont {N.~K.}\ \bibnamefont
  {Johnson-Mcdaniel}}, \bibinfo {author} {\bibfnamefont {A.}~\bibnamefont
  {Ghosh}}, \bibinfo {author} {\bibfnamefont {C.~K.}\ \bibnamefont {Mishra}},
  \bibinfo {author} {\bibfnamefont {P.}~\bibnamefont {Ajith}}, \bibinfo
  {author} {\bibfnamefont {W.}~\bibnamefont {Del~Pozzo}}, \bibinfo {author}
  {\bibfnamefont {C.~P.~L.}\ \bibnamefont {Berry}}, \bibinfo {author}
  {\bibfnamefont {A.~B.}\ \bibnamefont {Nielsen}},\ and\ \bibinfo {author}
  {\bibfnamefont {L.}~\bibnamefont {London}},\ }\bibfield  {title} {\bibinfo
  {title} {{Testing general relativity using gravitational wave signals from
  the inspiral, merger and ringdown of binary black holes}},\ }\href
  {https://doi.org/10.1088/1361-6382/aa972e} {\bibfield  {journal} {\bibinfo
  {journal} {Class. Quant. Grav.}\ }\textbf {\bibinfo {volume} {35}},\ \bibinfo
  {pages} {014002} (\bibinfo {year} {2018})},\ \Eprint
  {https://arxiv.org/abs/1704.06784} {arXiv:1704.06784 [gr-qc]} \BibitemShut
  {NoStop}%
\bibitem [{\citenamefont {Li}\ \emph {et~al.}(2012)\citenamefont {Li},
  \citenamefont {Del~Pozzo}, \citenamefont {Vitale}, \citenamefont {Van
  Den~Broeck}, \citenamefont {Agathos}, \citenamefont {Veitch}, \citenamefont
  {Grover}, \citenamefont {Sidery}, \citenamefont {Sturani},\ and\
  \citenamefont {Vecchio}}]{Li:2011cg}%
  \BibitemOpen
  \bibfield  {author} {\bibinfo {author} {\bibfnamefont {T.~G.~F.}\
  \bibnamefont {Li}}, \bibinfo {author} {\bibfnamefont {W.}~\bibnamefont
  {Del~Pozzo}}, \bibinfo {author} {\bibfnamefont {S.}~\bibnamefont {Vitale}},
  \bibinfo {author} {\bibfnamefont {C.}~\bibnamefont {Van Den~Broeck}},
  \bibinfo {author} {\bibfnamefont {M.}~\bibnamefont {Agathos}}, \bibinfo
  {author} {\bibfnamefont {J.}~\bibnamefont {Veitch}}, \bibinfo {author}
  {\bibfnamefont {K.}~\bibnamefont {Grover}}, \bibinfo {author} {\bibfnamefont
  {T.}~\bibnamefont {Sidery}}, \bibinfo {author} {\bibfnamefont
  {R.}~\bibnamefont {Sturani}},\ and\ \bibinfo {author} {\bibfnamefont
  {A.}~\bibnamefont {Vecchio}},\ }\bibfield  {title} {\bibinfo {title}
  {{Towards a generic test of the strong field dynamics of general relativity
  using compact binary coalescence}},\ }\href
  {https://doi.org/10.1103/PhysRevD.85.082003} {\bibfield  {journal} {\bibinfo
  {journal} {Phys. Rev. D}\ }\textbf {\bibinfo {volume} {85}},\ \bibinfo
  {pages} {082003} (\bibinfo {year} {2012})},\ \Eprint
  {https://arxiv.org/abs/1110.0530} {arXiv:1110.0530 [gr-qc]} \BibitemShut
  {NoStop}%
\bibitem [{\citenamefont {Agathos}\ \emph {et~al.}(2014)\citenamefont
  {Agathos}, \citenamefont {Del~Pozzo}, \citenamefont {Li}, \citenamefont {Van
  Den~Broeck}, \citenamefont {Veitch},\ and\ \citenamefont
  {Vitale}}]{Agathos:2013upa}%
  \BibitemOpen
  \bibfield  {author} {\bibinfo {author} {\bibfnamefont {M.}~\bibnamefont
  {Agathos}}, \bibinfo {author} {\bibfnamefont {W.}~\bibnamefont {Del~Pozzo}},
  \bibinfo {author} {\bibfnamefont {T.~G.~F.}\ \bibnamefont {Li}}, \bibinfo
  {author} {\bibfnamefont {C.}~\bibnamefont {Van Den~Broeck}}, \bibinfo
  {author} {\bibfnamefont {J.}~\bibnamefont {Veitch}},\ and\ \bibinfo {author}
  {\bibfnamefont {S.}~\bibnamefont {Vitale}},\ }\bibfield  {title} {\bibinfo
  {title} {{TIGER: A data analysis pipeline for testing the strong-field
  dynamics of general relativity with gravitational wave signals from
  coalescing compact binaries}},\ }\href
  {https://doi.org/10.1103/PhysRevD.89.082001} {\bibfield  {journal} {\bibinfo
  {journal} {Phys. Rev. D}\ }\textbf {\bibinfo {volume} {89}},\ \bibinfo
  {pages} {082001} (\bibinfo {year} {2014})},\ \Eprint
  {https://arxiv.org/abs/1311.0420} {arXiv:1311.0420 [gr-qc]} \BibitemShut
  {NoStop}%
\bibitem [{\citenamefont {Meidam}\ \emph {et~al.}(2018)\citenamefont {Meidam}
  \emph {et~al.}}]{Meidam:2017dgf}%
  \BibitemOpen
  \bibfield  {author} {\bibinfo {author} {\bibfnamefont {J.}~\bibnamefont
  {Meidam}} \emph {et~al.},\ }\bibfield  {title} {\bibinfo {title}
  {{Parametrized tests of the strong-field dynamics of general relativity using
  gravitational wave signals from coalescing binary black holes: Fast
  likelihood calculations and sensitivity of the method}},\ }\href
  {https://doi.org/10.1103/PhysRevD.97.044033} {\bibfield  {journal} {\bibinfo
  {journal} {Phys. Rev. D}\ }\textbf {\bibinfo {volume} {97}},\ \bibinfo
  {pages} {044033} (\bibinfo {year} {2018})},\ \Eprint
  {https://arxiv.org/abs/1712.08772} {arXiv:1712.08772 [gr-qc]} \BibitemShut
  {NoStop}%
\bibitem [{\citenamefont {Roy}\ \emph {et~al.}(2024)\citenamefont {Roy},
  \citenamefont {Haney}, \citenamefont {Pratten}, \citenamefont {Van
  Den~Broeck} \emph {et~al.}}]{roy:2024tiger}%
  \BibitemOpen
  \bibfield  {author} {\bibinfo {author} {\bibfnamefont {S.}~\bibnamefont
  {Roy}}, \bibinfo {author} {\bibfnamefont {M.}~\bibnamefont {Haney}}, \bibinfo
  {author} {\bibfnamefont {G.}~\bibnamefont {Pratten}}, \bibinfo {author}
  {\bibfnamefont {C.}~\bibnamefont {Van Den~Broeck}}, \emph {et~al.},\
  }\bibfield  {title} {\bibinfo {title} {{An improved parametrized test of
  general relativity using the IMRPhenomX waveform family: Including
  higher-harmonics and precession}}} (\bibinfo {year} {2024}),\ \bibinfo {note}
  {in preparation}\BibitemShut {NoStop}%
\bibitem [{\citenamefont {Hoy}\ and\ \citenamefont
  {Raymond}(2021)}]{Hoy:2020vys}%
  \BibitemOpen
  \bibfield  {author} {\bibinfo {author} {\bibfnamefont {C.}~\bibnamefont
  {Hoy}}\ and\ \bibinfo {author} {\bibfnamefont {V.}~\bibnamefont {Raymond}},\
  }\bibfield  {title} {\bibinfo {title} {{PESummary: the code agnostic
  Parameter Estimation Summary page builder}},\ }\href
  {https://doi.org/10.1016/j.softx.2021.100765} {\bibfield  {journal} {\bibinfo
   {journal} {SoftwareX}\ }\textbf {\bibinfo {volume} {15}},\ \bibinfo {pages}
  {100765} (\bibinfo {year} {2021})},\ \Eprint
  {https://arxiv.org/abs/2006.06639} {arXiv:2006.06639 [astro-ph.IM]}
  \BibitemShut {NoStop}%
\bibitem [{\citenamefont {Mehta}\ \emph {et~al.}(2023)\citenamefont {Mehta},
  \citenamefont {Buonanno}, \citenamefont {Cotesta}, \citenamefont {Ghosh},
  \citenamefont {Sennett},\ and\ \citenamefont {Steinhoff}}]{Mehta:2022pcn}%
  \BibitemOpen
  \bibfield  {author} {\bibinfo {author} {\bibfnamefont {A.~K.}\ \bibnamefont
  {Mehta}}, \bibinfo {author} {\bibfnamefont {A.}~\bibnamefont {Buonanno}},
  \bibinfo {author} {\bibfnamefont {R.}~\bibnamefont {Cotesta}}, \bibinfo
  {author} {\bibfnamefont {A.}~\bibnamefont {Ghosh}}, \bibinfo {author}
  {\bibfnamefont {N.}~\bibnamefont {Sennett}},\ and\ \bibinfo {author}
  {\bibfnamefont {J.}~\bibnamefont {Steinhoff}},\ }\bibfield  {title} {\bibinfo
  {title} {{Tests of general relativity with gravitational-wave observations
  using a flexible theory-independent method}},\ }\href
  {https://doi.org/10.1103/PhysRevD.107.044020} {\bibfield  {journal} {\bibinfo
   {journal} {Phys. Rev. D}\ }\textbf {\bibinfo {volume} {107}},\ \bibinfo
  {pages} {044020} (\bibinfo {year} {2023})},\ \Eprint
  {https://arxiv.org/abs/2203.13937} {arXiv:2203.13937 [gr-qc]} \BibitemShut
  {NoStop}%
\bibitem [{\citenamefont {{S{\"a}nger}}\ \emph {et~al.}(2024)\citenamefont
  {{S{\"a}nger}}, \citenamefont {{Roy}}, \citenamefont {{Agathos}},
  \citenamefont {{Birnholtz}}, \citenamefont {{Buonanno}}, \citenamefont
  {{Dietrich}}, \citenamefont {{Haney}}, \citenamefont {{Juli{\'e}}},
  \citenamefont {{Pratten}}, \citenamefont {{Steinhoff}}, \citenamefont {{Van
  Den Broeck}}, \citenamefont {{Biscoveanu}}, \citenamefont {{Char}},
  \citenamefont {{Heffernan}}, \citenamefont {{Joshi}}, \citenamefont
  {{Kedia}}, \citenamefont {{Schofield}}, \citenamefont {{Trevor}},\ and\
  \citenamefont {{Zevin}}}]{Sanger:2024axs}%
  \BibitemOpen
  \bibfield  {author} {\bibinfo {author} {\bibfnamefont {E.~M.}\ \bibnamefont
  {{S{\"a}nger}}}, \bibinfo {author} {\bibfnamefont {S.}~\bibnamefont {{Roy}}},
  \bibinfo {author} {\bibfnamefont {M.}~\bibnamefont {{Agathos}}}, \bibinfo
  {author} {\bibfnamefont {O.}~\bibnamefont {{Birnholtz}}}, \bibinfo {author}
  {\bibfnamefont {A.}~\bibnamefont {{Buonanno}}}, \bibinfo {author}
  {\bibfnamefont {T.}~\bibnamefont {{Dietrich}}}, \bibinfo {author}
  {\bibfnamefont {M.}~\bibnamefont {{Haney}}}, \bibinfo {author} {\bibfnamefont
  {F.-L.}\ \bibnamefont {{Juli{\'e}}}}, \bibinfo {author} {\bibfnamefont
  {G.}~\bibnamefont {{Pratten}}}, \bibinfo {author} {\bibfnamefont
  {J.}~\bibnamefont {{Steinhoff}}}, \bibinfo {author} {\bibfnamefont
  {C.}~\bibnamefont {{Van Den Broeck}}}, \bibinfo {author} {\bibfnamefont
  {S.}~\bibnamefont {{Biscoveanu}}}, \bibinfo {author} {\bibfnamefont
  {P.}~\bibnamefont {{Char}}}, \bibinfo {author} {\bibfnamefont
  {A.}~\bibnamefont {{Heffernan}}}, \bibinfo {author} {\bibfnamefont
  {P.}~\bibnamefont {{Joshi}}}, \bibinfo {author} {\bibfnamefont
  {A.}~\bibnamefont {{Kedia}}}, \bibinfo {author} {\bibfnamefont {R.~M.~S.}\
  \bibnamefont {{Schofield}}}, \bibinfo {author} {\bibfnamefont
  {M.}~\bibnamefont {{Trevor}}},\ and\ \bibinfo {author} {\bibfnamefont
  {M.}~\bibnamefont {{Zevin}}},\ }\bibfield  {title} {\bibinfo {title} {{Tests
  of General Relativity with GW230529: a neutron star merging with a lower
  mass-gap compact object}},\ }\href
  {https://doi.org/10.48550/arXiv.2406.03568} {\bibfield  {journal} {\bibinfo
  {journal} {arXiv e-prints}\ ,\ \bibinfo {eid} {arXiv:2406.03568}} (\bibinfo
  {year} {2024})},\ \Eprint {https://arxiv.org/abs/2406.03568}
  {arXiv:2406.03568 [gr-qc]} \BibitemShut {NoStop}%
\bibitem [{\citenamefont {{Ashton}}\ \emph {et~al.}(2024)\citenamefont
  {{Ashton}}, \citenamefont {{Talbot}}, \citenamefont {{Roy}}, \citenamefont
  {{Pratten}}, \citenamefont {{Pang}}, \citenamefont {{Agathos}}, \citenamefont
  {{Baka}}, \citenamefont {{S{\"a}nger}}, \citenamefont {{Mehta}},
  \citenamefont {{Steinhoff}}, \citenamefont {{Maggio}}, \citenamefont
  {{Ghosh}},\ and\ \citenamefont {{Vijaykumar}}}]{bilby-tgr}%
  \BibitemOpen
  \bibfield  {author} {\bibinfo {author} {\bibfnamefont {G.}~\bibnamefont
  {{Ashton}}}, \bibinfo {author} {\bibfnamefont {C.}~\bibnamefont {{Talbot}}},
  \bibinfo {author} {\bibfnamefont {S.}~\bibnamefont {{Roy}}}, \bibinfo
  {author} {\bibfnamefont {G.}~\bibnamefont {{Pratten}}}, \bibinfo {author}
  {\bibfnamefont {T.-H.}\ \bibnamefont {{Pang}}}, \bibinfo {author}
  {\bibfnamefont {M.}~\bibnamefont {{Agathos}}}, \bibinfo {author}
  {\bibfnamefont {T.}~\bibnamefont {{Baka}}}, \bibinfo {author} {\bibfnamefont
  {E.}~\bibnamefont {{S{\"a}nger}}}, \bibinfo {author} {\bibfnamefont
  {A.}~\bibnamefont {{Mehta}}}, \bibinfo {author} {\bibfnamefont
  {J.}~\bibnamefont {{Steinhoff}}}, \bibinfo {author} {\bibfnamefont
  {E.}~\bibnamefont {{Maggio}}}, \bibinfo {author} {\bibfnamefont
  {A.}~\bibnamefont {{Ghosh}}},\ and\ \bibinfo {author} {\bibfnamefont
  {A.}~\bibnamefont {{Vijaykumar}}},\ }\href
  {https://doi.org/10.5281/zenodo.10940210} {\bibinfo {title} {{Bilby TGR}}}
  (\bibinfo {year} {2024})\BibitemShut {NoStop}%
\bibitem [{\citenamefont {Pratten}\ \emph {et~al.}(2020)\citenamefont
  {Pratten}, \citenamefont {Husa}, \citenamefont {Garcia-Quiros}, \citenamefont
  {Colleoni}, \citenamefont {Ramos-Buades}, \citenamefont {Estelles},\ and\
  \citenamefont {Jaume}}]{Pratten:2020fqn}%
  \BibitemOpen
  \bibfield  {author} {\bibinfo {author} {\bibfnamefont {G.}~\bibnamefont
  {Pratten}}, \bibinfo {author} {\bibfnamefont {S.}~\bibnamefont {Husa}},
  \bibinfo {author} {\bibfnamefont {C.}~\bibnamefont {Garcia-Quiros}}, \bibinfo
  {author} {\bibfnamefont {M.}~\bibnamefont {Colleoni}}, \bibinfo {author}
  {\bibfnamefont {A.}~\bibnamefont {Ramos-Buades}}, \bibinfo {author}
  {\bibfnamefont {H.}~\bibnamefont {Estelles}},\ and\ \bibinfo {author}
  {\bibfnamefont {R.}~\bibnamefont {Jaume}},\ }\bibfield  {title} {\bibinfo
  {title} {{Setting the cornerstone for a family of models for gravitational
  waves from compact binaries: The dominant harmonic for nonprecessing
  quasicircular black holes}},\ }\href
  {https://doi.org/10.1103/PhysRevD.102.064001} {\bibfield  {journal} {\bibinfo
   {journal} {Phys. Rev. D}\ }\textbf {\bibinfo {volume} {102}},\ \bibinfo
  {pages} {064001} (\bibinfo {year} {2020})},\ \Eprint
  {https://arxiv.org/abs/2001.11412} {arXiv:2001.11412 [gr-qc]} \BibitemShut
  {NoStop}%
\bibitem [{\citenamefont {Gerosa}\ \emph {et~al.}(2018)\citenamefont {Gerosa},
  \citenamefont {Berti}, \citenamefont {O'Shaughnessy}, \citenamefont
  {Belczynski}, \citenamefont {Kesden}, \citenamefont {Wysocki},\ and\
  \citenamefont {Gladysz}}]{Gerosa:2018wbw}%
  \BibitemOpen
  \bibfield  {author} {\bibinfo {author} {\bibfnamefont {D.}~\bibnamefont
  {Gerosa}}, \bibinfo {author} {\bibfnamefont {E.}~\bibnamefont {Berti}},
  \bibinfo {author} {\bibfnamefont {R.}~\bibnamefont {O'Shaughnessy}}, \bibinfo
  {author} {\bibfnamefont {K.}~\bibnamefont {Belczynski}}, \bibinfo {author}
  {\bibfnamefont {M.}~\bibnamefont {Kesden}}, \bibinfo {author} {\bibfnamefont
  {D.}~\bibnamefont {Wysocki}},\ and\ \bibinfo {author} {\bibfnamefont
  {W.}~\bibnamefont {Gladysz}},\ }\bibfield  {title} {\bibinfo {title} {{Spin
  orientations of merging black holes formed from the evolution of stellar
  binaries}},\ }\href {https://doi.org/10.1103/PhysRevD.98.084036} {\bibfield
  {journal} {\bibinfo  {journal} {Phys. Rev. D}\ }\textbf {\bibinfo {volume}
  {98}},\ \bibinfo {pages} {084036} (\bibinfo {year} {2018})},\ \Eprint
  {https://arxiv.org/abs/1808.02491} {arXiv:1808.02491 [astro-ph.HE]}
  \BibitemShut {NoStop}%
\bibitem [{\citenamefont {{Banerjee}}\ \emph {et~al.}(2010)\citenamefont
  {{Banerjee}}, \citenamefont {{Baumgardt}},\ and\ \citenamefont
  {{Kroupa}}}]{Banerjee:2010}%
  \BibitemOpen
  \bibfield  {author} {\bibinfo {author} {\bibfnamefont {S.}~\bibnamefont
  {{Banerjee}}}, \bibinfo {author} {\bibfnamefont {H.}~\bibnamefont
  {{Baumgardt}}},\ and\ \bibinfo {author} {\bibfnamefont {P.}~\bibnamefont
  {{Kroupa}}},\ }\bibfield  {title} {\bibinfo {title} {{Stellar-mass black
  holes in star clusters: implications for gravitational wave radiation}},\
  }\href {https://doi.org/10.1111/j.1365-2966.2009.15880.x} {\bibfield
  {journal} {\bibinfo  {journal} {{Mon. Not. Roy. Astron. Soc.}}\ }\textbf
  {\bibinfo {volume} {402}},\ \bibinfo {pages} {371} (\bibinfo {year}
  {2010})},\ \Eprint {https://arxiv.org/abs/0910.3954} {arXiv:0910.3954
  [astro-ph.SR]} \BibitemShut {NoStop}%
\bibitem [{\citenamefont {{Chatterjee}}\ \emph {et~al.}(2017)\citenamefont
  {{Chatterjee}}, \citenamefont {{Rodriguez}},\ and\ \citenamefont
  {{Rasio}}}]{Chatterjee:2017}%
  \BibitemOpen
  \bibfield  {author} {\bibinfo {author} {\bibfnamefont {S.}~\bibnamefont
  {{Chatterjee}}}, \bibinfo {author} {\bibfnamefont {C.~L.}\ \bibnamefont
  {{Rodriguez}}},\ and\ \bibinfo {author} {\bibfnamefont {F.~A.}\ \bibnamefont
  {{Rasio}}},\ }\bibfield  {title} {\bibinfo {title} {{Binary Black Holes in
  Dense Star Clusters: Exploring the Theoretical Uncertainties}},\ }\href
  {https://doi.org/10.3847/1538-4357/834/1/68} {\bibfield  {journal} {\bibinfo
  {journal} {\apj}\ }\textbf {\bibinfo {volume} {834}},\ \bibinfo {eid} {68}
  (\bibinfo {year} {2017})},\ \Eprint {https://arxiv.org/abs/1603.00884}
  {arXiv:1603.00884 [astro-ph.GA]} \BibitemShut {NoStop}%
\bibitem [{\citenamefont {Ziosi}\ \emph {et~al.}(2014)\citenamefont {Ziosi},
  \citenamefont {Mapelli}, \citenamefont {Branchesi},\ and\ \citenamefont
  {Tormen}}]{Ziosi:2014sra}%
  \BibitemOpen
  \bibfield  {author} {\bibinfo {author} {\bibfnamefont {B.~M.}\ \bibnamefont
  {Ziosi}}, \bibinfo {author} {\bibfnamefont {M.}~\bibnamefont {Mapelli}},
  \bibinfo {author} {\bibfnamefont {M.}~\bibnamefont {Branchesi}},\ and\
  \bibinfo {author} {\bibfnamefont {G.}~\bibnamefont {Tormen}},\ }\bibfield
  {title} {\bibinfo {title} {{Dynamics of stellar black holes in young star
  clusters with different metallicities \textendash{} II. Black
  hole\textendash{}black hole binaries}},\ }\href
  {https://doi.org/10.1093/mnras/stu824} {\bibfield  {journal} {\bibinfo
  {journal} {Mon. Not. Roy. Astron. Soc.}\ }\textbf {\bibinfo {volume} {441}},\
  \bibinfo {pages} {3703} (\bibinfo {year} {2014})},\ \Eprint
  {https://arxiv.org/abs/1404.7147} {arXiv:1404.7147 [astro-ph.GA]}
  \BibitemShut {NoStop}%
\bibitem [{\citenamefont {Rodriguez}\ \emph {et~al.}(2015)\citenamefont
  {Rodriguez}, \citenamefont {Morscher}, \citenamefont {Pattabiraman},
  \citenamefont {Chatterjee}, \citenamefont {Haster},\ and\ \citenamefont
  {Rasio}}]{Rodriguez:2015oxa}%
  \BibitemOpen
  \bibfield  {author} {\bibinfo {author} {\bibfnamefont {C.~L.}\ \bibnamefont
  {Rodriguez}}, \bibinfo {author} {\bibfnamefont {M.}~\bibnamefont {Morscher}},
  \bibinfo {author} {\bibfnamefont {B.}~\bibnamefont {Pattabiraman}}, \bibinfo
  {author} {\bibfnamefont {S.}~\bibnamefont {Chatterjee}}, \bibinfo {author}
  {\bibfnamefont {C.-J.}\ \bibnamefont {Haster}},\ and\ \bibinfo {author}
  {\bibfnamefont {F.~A.}\ \bibnamefont {Rasio}},\ }\bibfield  {title} {\bibinfo
  {title} {{Binary Black Hole Mergers from Globular Clusters: Implications for
  Advanced LIGO}},\ }\href {https://doi.org/10.1103/PhysRevLett.115.051101}
  {\bibfield  {journal} {\bibinfo  {journal} {Phys. Rev. Lett.}\ }\textbf
  {\bibinfo {volume} {115}},\ \bibinfo {pages} {051101} (\bibinfo {year}
  {2015})},\ \bibinfo {note} {[Erratum: Phys.Rev.Lett. 116, 029901 (2016)]},\
  \Eprint {https://arxiv.org/abs/1505.00792} {arXiv:1505.00792 [astro-ph.HE]}
  \BibitemShut {NoStop}%
\bibitem [{\citenamefont {Mapelli}\ \emph {et~al.}(2022)\citenamefont
  {Mapelli}, \citenamefont {Bouffanais}, \citenamefont {Santoliquido},
  \citenamefont {Sedda},\ and\ \citenamefont {Artale}}]{Mapelli:2021gyv}%
  \BibitemOpen
  \bibfield  {author} {\bibinfo {author} {\bibfnamefont {M.}~\bibnamefont
  {Mapelli}}, \bibinfo {author} {\bibfnamefont {Y.}~\bibnamefont {Bouffanais}},
  \bibinfo {author} {\bibfnamefont {F.}~\bibnamefont {Santoliquido}}, \bibinfo
  {author} {\bibfnamefont {M.~A.}\ \bibnamefont {Sedda}},\ and\ \bibinfo
  {author} {\bibfnamefont {M.~C.}\ \bibnamefont {Artale}},\ }\bibfield  {title}
  {\bibinfo {title} {{The cosmic evolution of binary black holes in young,
  globular, and nuclear star clusters: rates, masses, spins, and mixing
  fractions}},\ }\href {https://doi.org/10.1093/mnras/stac422} {\bibfield
  {journal} {\bibinfo  {journal} {Mon. Not. Roy. Astron. Soc.}\ }\textbf
  {\bibinfo {volume} {511}},\ \bibinfo {pages} {5797} (\bibinfo {year}
  {2022})},\ \Eprint {https://arxiv.org/abs/2109.06222} {arXiv:2109.06222
  [astro-ph.HE]} \BibitemShut {NoStop}%
\bibitem [{\citenamefont {Tagawa}\ \emph {et~al.}(2020)\citenamefont {Tagawa},
  \citenamefont {Haiman},\ and\ \citenamefont {Kocsis}}]{Tagawa:2019osr}%
  \BibitemOpen
  \bibfield  {author} {\bibinfo {author} {\bibfnamefont {H.}~\bibnamefont
  {Tagawa}}, \bibinfo {author} {\bibfnamefont {Z.}~\bibnamefont {Haiman}},\
  and\ \bibinfo {author} {\bibfnamefont {B.}~\bibnamefont {Kocsis}},\
  }\bibfield  {title} {\bibinfo {title} {{Formation and Evolution of Compact
  Object Binaries in AGN Disks}},\ }\href
  {https://doi.org/10.3847/1538-4357/ab9b8c} {\bibfield  {journal} {\bibinfo
  {journal} {Astrophys. J.}\ }\textbf {\bibinfo {volume} {898}},\ \bibinfo
  {pages} {25} (\bibinfo {year} {2020})},\ \Eprint
  {https://arxiv.org/abs/1912.08218} {arXiv:1912.08218 [astro-ph.GA]}
  \BibitemShut {NoStop}%
\bibitem [{\citenamefont {Sedda}\ \emph {et~al.}(2023)\citenamefont {Sedda},
  \citenamefont {Naoz},\ and\ \citenamefont {Kocsis}}]{Sedda:2023big}%
  \BibitemOpen
  \bibfield  {author} {\bibinfo {author} {\bibfnamefont {M.~A.}\ \bibnamefont
  {Sedda}}, \bibinfo {author} {\bibfnamefont {S.}~\bibnamefont {Naoz}},\ and\
  \bibinfo {author} {\bibfnamefont {B.}~\bibnamefont {Kocsis}},\ }\bibfield
  {title} {\bibinfo {title} {{Quiescent and Active Galactic Nuclei as Factories
  of Merging Compact Objects in the Era of Gravitational Wave Astronomy}},\
  }\href {https://doi.org/10.3390/universe9030138} {\bibfield  {journal}
  {\bibinfo  {journal} {Universe}\ }\textbf {\bibinfo {volume} {9}},\ \bibinfo
  {pages} {138} (\bibinfo {year} {2023})},\ \Eprint
  {https://arxiv.org/abs/2302.14071} {arXiv:2302.14071 [astro-ph.GA]}
  \BibitemShut {NoStop}%
\bibitem [{\citenamefont {{Rowan}}\ \emph {et~al.}(2023)\citenamefont
  {{Rowan}}, \citenamefont {{Boekholt}}, \citenamefont {{Kocsis}},\ and\
  \citenamefont {{Haiman}}}]{Rowan:2023}%
  \BibitemOpen
  \bibfield  {author} {\bibinfo {author} {\bibfnamefont {C.}~\bibnamefont
  {{Rowan}}}, \bibinfo {author} {\bibfnamefont {T.}~\bibnamefont {{Boekholt}}},
  \bibinfo {author} {\bibfnamefont {B.}~\bibnamefont {{Kocsis}}},\ and\
  \bibinfo {author} {\bibfnamefont {Z.}~\bibnamefont {{Haiman}}},\ }\bibfield
  {title} {\bibinfo {title} {{Black hole binary formation in AGN discs: from
  isolation to merger}},\ }\href {https://doi.org/10.1093/mnras/stad1926}
  {\bibfield  {journal} {\bibinfo  {journal} {Mon. Not. Roy. Astron. Soc.}\
  }\textbf {\bibinfo {volume} {524}},\ \bibinfo {pages} {2770} (\bibinfo {year}
  {2023})},\ \Eprint {https://arxiv.org/abs/2212.06133} {arXiv:2212.06133
  [astro-ph.GA]} \BibitemShut {NoStop}%
\bibitem [{\citenamefont {Ishibashi}\ and\ \citenamefont
  {Gr\"obner}(2020)}]{Ishibashi:2020zzy}%
  \BibitemOpen
  \bibfield  {author} {\bibinfo {author} {\bibfnamefont {W.}~\bibnamefont
  {Ishibashi}}\ and\ \bibinfo {author} {\bibfnamefont {M.}~\bibnamefont
  {Gr\"obner}},\ }\bibfield  {title} {\bibinfo {title} {{Evolution of binary
  black holes in AGN accretion discs: Disc-binary interaction and gravitational
  wave emission}},\ }\href {https://doi.org/10.1051/0004-6361/202037799}
  {\bibfield  {journal} {\bibinfo  {journal} {Astron. Astrophys.}\ }\textbf
  {\bibinfo {volume} {639}},\ \bibinfo {pages} {A108} (\bibinfo {year}
  {2020})},\ \Eprint {https://arxiv.org/abs/2006.07407} {arXiv:2006.07407
  [astro-ph.GA]} \BibitemShut {NoStop}%
\bibitem [{\citenamefont {Barausse}\ and\ \citenamefont
  {Rezzolla}(2008)}]{Barausse:2007dy}%
  \BibitemOpen
  \bibfield  {author} {\bibinfo {author} {\bibfnamefont {E.}~\bibnamefont
  {Barausse}}\ and\ \bibinfo {author} {\bibfnamefont {L.}~\bibnamefont
  {Rezzolla}},\ }\bibfield  {title} {\bibinfo {title} {{The Influence of the
  hydrodynamic drag from an accretion torus on extreme mass-ratio inspirals}},\
  }\href {https://doi.org/10.1103/PhysRevD.77.104027} {\bibfield  {journal}
  {\bibinfo  {journal} {Phys. Rev. D}\ }\textbf {\bibinfo {volume} {77}},\
  \bibinfo {pages} {104027} (\bibinfo {year} {2008})},\ \Eprint
  {https://arxiv.org/abs/0711.4558} {arXiv:0711.4558 [gr-qc]} \BibitemShut
  {NoStop}%
\bibitem [{\citenamefont {Kocsis}\ \emph {et~al.}(2011)\citenamefont {Kocsis},
  \citenamefont {Yunes},\ and\ \citenamefont {Loeb}}]{Kocsis:2011dr}%
  \BibitemOpen
  \bibfield  {author} {\bibinfo {author} {\bibfnamefont {B.}~\bibnamefont
  {Kocsis}}, \bibinfo {author} {\bibfnamefont {N.}~\bibnamefont {Yunes}},\ and\
  \bibinfo {author} {\bibfnamefont {A.}~\bibnamefont {Loeb}},\ }\bibfield
  {title} {\bibinfo {title} {{Observable Signatures of EMRI Black Hole Binaries
  Embedded in Thin Accretion Disks}},\ }\href
  {https://doi.org/10.1103/PhysRevD.86.049907} {\bibfield  {journal} {\bibinfo
  {journal} {Phys. Rev. D}\ }\textbf {\bibinfo {volume} {84}},\ \bibinfo
  {pages} {024032} (\bibinfo {year} {2011})},\ \Eprint
  {https://arxiv.org/abs/1104.2322} {arXiv:1104.2322 [astro-ph.GA]}
  \BibitemShut {NoStop}%
\bibitem [{\citenamefont {Yunes}\ \emph {et~al.}(2011)\citenamefont {Yunes},
  \citenamefont {Kocsis}, \citenamefont {Loeb},\ and\ \citenamefont
  {Haiman}}]{Yunes:2011ws}%
  \BibitemOpen
  \bibfield  {author} {\bibinfo {author} {\bibfnamefont {N.}~\bibnamefont
  {Yunes}}, \bibinfo {author} {\bibfnamefont {B.}~\bibnamefont {Kocsis}},
  \bibinfo {author} {\bibfnamefont {A.}~\bibnamefont {Loeb}},\ and\ \bibinfo
  {author} {\bibfnamefont {Z.}~\bibnamefont {Haiman}},\ }\bibfield  {title}
  {\bibinfo {title} {{Imprint of Accretion Disk-Induced Migration on
  Gravitational Waves from Extreme Mass Ratio Inspirals}},\ }\href
  {https://doi.org/10.1103/PhysRevLett.107.171103} {\bibfield  {journal}
  {\bibinfo  {journal} {Phys. Rev. Lett.}\ }\textbf {\bibinfo {volume} {107}},\
  \bibinfo {pages} {171103} (\bibinfo {year} {2011})},\ \Eprint
  {https://arxiv.org/abs/1103.4609} {arXiv:1103.4609 [astro-ph.CO]}
  \BibitemShut {NoStop}%
\bibitem [{\citenamefont {Tamanini}\ \emph {et~al.}(2020)\citenamefont
  {Tamanini}, \citenamefont {Klein}, \citenamefont {Bonvin}, \citenamefont
  {Barausse},\ and\ \citenamefont {Caprini}}]{Tamanini:2019usx}%
  \BibitemOpen
  \bibfield  {author} {\bibinfo {author} {\bibfnamefont {N.}~\bibnamefont
  {Tamanini}}, \bibinfo {author} {\bibfnamefont {A.}~\bibnamefont {Klein}},
  \bibinfo {author} {\bibfnamefont {C.}~\bibnamefont {Bonvin}}, \bibinfo
  {author} {\bibfnamefont {E.}~\bibnamefont {Barausse}},\ and\ \bibinfo
  {author} {\bibfnamefont {C.}~\bibnamefont {Caprini}},\ }\bibfield  {title}
  {\bibinfo {title} {{Peculiar acceleration of stellar-origin black hole
  binaries: Measurement and biases with LISA}},\ }\href
  {https://doi.org/10.1103/PhysRevD.101.063002} {\bibfield  {journal} {\bibinfo
   {journal} {Phys. Rev. D}\ }\textbf {\bibinfo {volume} {101}},\ \bibinfo
  {pages} {063002} (\bibinfo {year} {2020})},\ \Eprint
  {https://arxiv.org/abs/1907.02018} {arXiv:1907.02018 [astro-ph.IM]}
  \BibitemShut {NoStop}%
\bibitem [{\citenamefont {Toubiana}\ \emph {et~al.}(2021)\citenamefont
  {Toubiana} \emph {et~al.}}]{Toubiana:2020drf}%
  \BibitemOpen
  \bibfield  {author} {\bibinfo {author} {\bibfnamefont {A.}~\bibnamefont
  {Toubiana}} \emph {et~al.},\ }\bibfield  {title} {\bibinfo {title}
  {{Detectable environmental effects in GW190521-like black-hole binaries with
  LISA}},\ }\href {https://doi.org/10.1103/PhysRevLett.126.101105} {\bibfield
  {journal} {\bibinfo  {journal} {Phys. Rev. Lett.}\ }\textbf {\bibinfo
  {volume} {126}},\ \bibinfo {pages} {101105} (\bibinfo {year} {2021})},\
  \Eprint {https://arxiv.org/abs/2010.06056} {arXiv:2010.06056 [astro-ph.HE]}
  \BibitemShut {NoStop}%
\bibitem [{\citenamefont {Derdzinski}\ \emph {et~al.}(2021)\citenamefont
  {Derdzinski}, \citenamefont {D'Orazio}, \citenamefont {Duffell},
  \citenamefont {Haiman},\ and\ \citenamefont
  {MacFadyen}}]{Derdzinski:2020wlw}%
  \BibitemOpen
  \bibfield  {author} {\bibinfo {author} {\bibfnamefont {A.}~\bibnamefont
  {Derdzinski}}, \bibinfo {author} {\bibfnamefont {D.}~\bibnamefont
  {D'Orazio}}, \bibinfo {author} {\bibfnamefont {P.}~\bibnamefont {Duffell}},
  \bibinfo {author} {\bibfnamefont {Z.}~\bibnamefont {Haiman}},\ and\ \bibinfo
  {author} {\bibfnamefont {A.}~\bibnamefont {MacFadyen}},\ }\bibfield  {title}
  {\bibinfo {title} {{Evolution of gas disc\textendash{}embedded intermediate
  mass ratio inspirals in the $LISA$ band}},\ }\href
  {https://doi.org/10.1093/mnras/staa3976} {\bibfield  {journal} {\bibinfo
  {journal} {Mon. Not. Roy. Astron. Soc.}\ }\textbf {\bibinfo {volume} {501}},\
  \bibinfo {pages} {3540} (\bibinfo {year} {2021})},\ \Eprint
  {https://arxiv.org/abs/2005.11333} {arXiv:2005.11333 [astro-ph.HE]}
  \BibitemShut {NoStop}%
\bibitem [{\citenamefont {Zwick}\ \emph {et~al.}(2022)\citenamefont {Zwick},
  \citenamefont {Derdzinski}, \citenamefont {Garg}, \citenamefont {Capelo},\
  and\ \citenamefont {Mayer}}]{Zwick:2021dlg}%
  \BibitemOpen
  \bibfield  {author} {\bibinfo {author} {\bibfnamefont {L.}~\bibnamefont
  {Zwick}}, \bibinfo {author} {\bibfnamefont {A.}~\bibnamefont {Derdzinski}},
  \bibinfo {author} {\bibfnamefont {M.}~\bibnamefont {Garg}}, \bibinfo {author}
  {\bibfnamefont {P.~R.}\ \bibnamefont {Capelo}},\ and\ \bibinfo {author}
  {\bibfnamefont {L.}~\bibnamefont {Mayer}},\ }\bibfield  {title} {\bibinfo
  {title} {{Dirty waveforms: multiband harmonic content of gas-embedded
  gravitational wave sources}},\ }\href {https://doi.org/10.1093/mnras/stac299}
  {\bibfield  {journal} {\bibinfo  {journal} {Mon. Not. Roy. Astron. Soc.}\
  }\textbf {\bibinfo {volume} {511}},\ \bibinfo {pages} {6143} (\bibinfo {year}
  {2022})},\ \Eprint {https://arxiv.org/abs/2110.09097} {arXiv:2110.09097
  [astro-ph.HE]} \BibitemShut {NoStop}%
\bibitem [{\citenamefont {Sberna}\ \emph {et~al.}(2022)\citenamefont {Sberna}
  \emph {et~al.}}]{Sberna:2022qbn}%
  \BibitemOpen
  \bibfield  {author} {\bibinfo {author} {\bibfnamefont {L.}~\bibnamefont
  {Sberna}} \emph {et~al.},\ }\bibfield  {title} {\bibinfo {title} {{Observing
  GW190521-like binary black holes and their environment with LISA}},\ }\href
  {https://doi.org/10.1103/PhysRevD.106.064056} {\bibfield  {journal} {\bibinfo
   {journal} {Phys. Rev. D}\ }\textbf {\bibinfo {volume} {106}},\ \bibinfo
  {pages} {064056} (\bibinfo {year} {2022})},\ \Eprint
  {https://arxiv.org/abs/2205.08550} {arXiv:2205.08550 [gr-qc]} \BibitemShut
  {NoStop}%
\bibitem [{\citenamefont {Zwick}\ \emph {et~al.}(2023)\citenamefont {Zwick},
  \citenamefont {Capelo},\ and\ \citenamefont {Mayer}}]{Zwick:2022dih}%
  \BibitemOpen
  \bibfield  {author} {\bibinfo {author} {\bibfnamefont {L.}~\bibnamefont
  {Zwick}}, \bibinfo {author} {\bibfnamefont {P.~R.}\ \bibnamefont {Capelo}},\
  and\ \bibinfo {author} {\bibfnamefont {L.}~\bibnamefont {Mayer}},\ }\bibfield
   {title} {\bibinfo {title} {{Priorities in gravitational waveforms for future
  space-borne detectors: vacuum accuracy or environment?}},\ }\href
  {https://doi.org/10.1093/mnras/stad707} {\bibfield  {journal} {\bibinfo
  {journal} {Mon. Not. Roy. Astron. Soc.}\ }\textbf {\bibinfo {volume} {521}},\
  \bibinfo {pages} {4645} (\bibinfo {year} {2023})},\ \Eprint
  {https://arxiv.org/abs/2209.04060} {arXiv:2209.04060 [gr-qc]} \BibitemShut
  {NoStop}%
\bibitem [{\citenamefont {Vijaykumar}\ \emph {et~al.}(2023)\citenamefont
  {Vijaykumar}, \citenamefont {Tiwari}, \citenamefont {Kapadia}, \citenamefont
  {Arun},\ and\ \citenamefont {Ajith}}]{Vijaykumar:2023tjg}%
  \BibitemOpen
  \bibfield  {author} {\bibinfo {author} {\bibfnamefont {A.}~\bibnamefont
  {Vijaykumar}}, \bibinfo {author} {\bibfnamefont {A.}~\bibnamefont {Tiwari}},
  \bibinfo {author} {\bibfnamefont {S.~J.}\ \bibnamefont {Kapadia}}, \bibinfo
  {author} {\bibfnamefont {K.~G.}\ \bibnamefont {Arun}},\ and\ \bibinfo
  {author} {\bibfnamefont {P.}~\bibnamefont {Ajith}},\ }\bibfield  {title}
  {\bibinfo {title} {{Waltzing Binaries: Probing the Line-of-sight Acceleration
  of Merging Compact Objects with Gravitational Waves}},\ }\href
  {https://doi.org/10.3847/1538-4357/acd77d} {\bibfield  {journal} {\bibinfo
  {journal} {Astrophys. J.}\ }\textbf {\bibinfo {volume} {954}},\ \bibinfo
  {pages} {105} (\bibinfo {year} {2023})},\ \Eprint
  {https://arxiv.org/abs/2302.09651} {arXiv:2302.09651 [astro-ph.HE]}
  \BibitemShut {NoStop}%
\bibitem [{\citenamefont {Annulli}\ \emph {et~al.}(2020)\citenamefont
  {Annulli}, \citenamefont {Cardoso},\ and\ \citenamefont
  {Vicente}}]{Annulli:2020lyc}%
  \BibitemOpen
  \bibfield  {author} {\bibinfo {author} {\bibfnamefont {L.}~\bibnamefont
  {Annulli}}, \bibinfo {author} {\bibfnamefont {V.}~\bibnamefont {Cardoso}},\
  and\ \bibinfo {author} {\bibfnamefont {R.}~\bibnamefont {Vicente}},\
  }\bibfield  {title} {\bibinfo {title} {{Response of ultralight dark matter to
  supermassive black holes and binaries}},\ }\href
  {https://doi.org/10.1103/PhysRevD.102.063022} {\bibfield  {journal} {\bibinfo
   {journal} {Phys. Rev. D}\ }\textbf {\bibinfo {volume} {102}},\ \bibinfo
  {pages} {063022} (\bibinfo {year} {2020})},\ \Eprint
  {https://arxiv.org/abs/2009.00012} {arXiv:2009.00012 [gr-qc]} \BibitemShut
  {NoStop}%
\bibitem [{\citenamefont {Kavanagh}\ \emph {et~al.}(2020)\citenamefont
  {Kavanagh}, \citenamefont {Nichols}, \citenamefont {Bertone},\ and\
  \citenamefont {Gaggero}}]{Kavanagh:2020cfn}%
  \BibitemOpen
  \bibfield  {author} {\bibinfo {author} {\bibfnamefont {B.~J.}\ \bibnamefont
  {Kavanagh}}, \bibinfo {author} {\bibfnamefont {D.~A.}\ \bibnamefont
  {Nichols}}, \bibinfo {author} {\bibfnamefont {G.}~\bibnamefont {Bertone}},\
  and\ \bibinfo {author} {\bibfnamefont {D.}~\bibnamefont {Gaggero}},\
  }\bibfield  {title} {\bibinfo {title} {{Detecting dark matter around black
  holes with gravitational waves: Effects of dark-matter dynamics on the
  gravitational waveform}},\ }\href
  {https://doi.org/10.1103/PhysRevD.102.083006} {\bibfield  {journal} {\bibinfo
   {journal} {Phys. Rev. D}\ }\textbf {\bibinfo {volume} {102}},\ \bibinfo
  {pages} {083006} (\bibinfo {year} {2020})},\ \Eprint
  {https://arxiv.org/abs/2002.12811} {arXiv:2002.12811 [gr-qc]} \BibitemShut
  {NoStop}%
\bibitem [{\citenamefont {Baumann}\ \emph {et~al.}(2022)\citenamefont
  {Baumann}, \citenamefont {Bertone}, \citenamefont {Stout},\ and\
  \citenamefont {Tomaselli}}]{Baumann:2021fkf}%
  \BibitemOpen
  \bibfield  {author} {\bibinfo {author} {\bibfnamefont {D.}~\bibnamefont
  {Baumann}}, \bibinfo {author} {\bibfnamefont {G.}~\bibnamefont {Bertone}},
  \bibinfo {author} {\bibfnamefont {J.}~\bibnamefont {Stout}},\ and\ \bibinfo
  {author} {\bibfnamefont {G.~M.}\ \bibnamefont {Tomaselli}},\ }\bibfield
  {title} {\bibinfo {title} {{Ionization of gravitational atoms}},\ }\href
  {https://doi.org/10.1103/PhysRevD.105.115036} {\bibfield  {journal} {\bibinfo
   {journal} {Phys. Rev. D}\ }\textbf {\bibinfo {volume} {105}},\ \bibinfo
  {pages} {115036} (\bibinfo {year} {2022})},\ \Eprint
  {https://arxiv.org/abs/2112.14777} {arXiv:2112.14777 [gr-qc]} \BibitemShut
  {NoStop}%
\bibitem [{\citenamefont {Vicente}\ and\ \citenamefont
  {Cardoso}(2022)}]{Vicente:2022ivh}%
  \BibitemOpen
  \bibfield  {author} {\bibinfo {author} {\bibfnamefont {R.}~\bibnamefont
  {Vicente}}\ and\ \bibinfo {author} {\bibfnamefont {V.}~\bibnamefont
  {Cardoso}},\ }\bibfield  {title} {\bibinfo {title} {{Dynamical friction of
  black holes in ultralight dark matter}},\ }\href
  {https://doi.org/10.1103/PhysRevD.105.083008} {\bibfield  {journal} {\bibinfo
   {journal} {Phys. Rev. D}\ }\textbf {\bibinfo {volume} {105}},\ \bibinfo
  {pages} {083008} (\bibinfo {year} {2022})},\ \Eprint
  {https://arxiv.org/abs/2201.08854} {arXiv:2201.08854 [gr-qc]} \BibitemShut
  {NoStop}%
\bibitem [{\citenamefont {Traykova}\ \emph {et~al.}(2023)\citenamefont
  {Traykova}, \citenamefont {Vicente}, \citenamefont {Clough}, \citenamefont
  {Helfer}, \citenamefont {Berti}, \citenamefont {Ferreira},\ and\
  \citenamefont {Hui}}]{Traykova:2023qyv}%
  \BibitemOpen
  \bibfield  {author} {\bibinfo {author} {\bibfnamefont {D.}~\bibnamefont
  {Traykova}}, \bibinfo {author} {\bibfnamefont {R.}~\bibnamefont {Vicente}},
  \bibinfo {author} {\bibfnamefont {K.}~\bibnamefont {Clough}}, \bibinfo
  {author} {\bibfnamefont {T.}~\bibnamefont {Helfer}}, \bibinfo {author}
  {\bibfnamefont {E.}~\bibnamefont {Berti}}, \bibinfo {author} {\bibfnamefont
  {P.~G.}\ \bibnamefont {Ferreira}},\ and\ \bibinfo {author} {\bibfnamefont
  {L.}~\bibnamefont {Hui}},\ }\bibfield  {title} {\bibinfo {title}
  {{Relativistic drag forces on black holes from scalar dark matter clouds of
  all sizes}},\ }\href {https://doi.org/10.48550/arXiv.2305.10492} {\bibfield
  {journal} {\bibinfo  {journal} {arXiv e-prints}\ ,\ \bibinfo {eid}
  {arXiv:2305.10492}} (\bibinfo {year} {2023})},\ \Eprint
  {https://arxiv.org/abs/2305.10492} {arXiv:2305.10492 [gr-qc]} \BibitemShut
  {NoStop}%
\bibitem [{\citenamefont {Duque}\ \emph {et~al.}(2024)\citenamefont {Duque},
  \citenamefont {Macedo}, \citenamefont {Vicente},\ and\ \citenamefont
  {Cardoso}}]{Duque:2023seg}%
  \BibitemOpen
  \bibfield  {author} {\bibinfo {author} {\bibfnamefont {F.}~\bibnamefont
  {Duque}}, \bibinfo {author} {\bibfnamefont {C.~F.~B.}\ \bibnamefont
  {Macedo}}, \bibinfo {author} {\bibfnamefont {R.}~\bibnamefont {Vicente}},\
  and\ \bibinfo {author} {\bibfnamefont {V.}~\bibnamefont {Cardoso}},\
  }\bibfield  {title} {\bibinfo {title} {{Extreme-Mass-Ratio Inspirals in
  Ultralight Dark Matter}},\ }\href
  {https://doi.org/10.1103/PhysRevLett.133.121404} {\bibfield  {journal}
  {\bibinfo  {journal} {Phys. Rev. Lett.}\ }\textbf {\bibinfo {volume} {133}},\
  \bibinfo {pages} {121404} (\bibinfo {year} {2024})},\ \Eprint
  {https://arxiv.org/abs/2312.06767} {arXiv:2312.06767 [gr-qc]} \BibitemShut
  {NoStop}%
\bibitem [{\citenamefont {Cardoso}\ \emph {et~al.}(2022)\citenamefont
  {Cardoso}, \citenamefont {Destounis}, \citenamefont {Duque}, \citenamefont
  {Panosso~Macedo},\ and\ \citenamefont {Maselli}}]{Cardoso:2022whc}%
  \BibitemOpen
  \bibfield  {author} {\bibinfo {author} {\bibfnamefont {V.}~\bibnamefont
  {Cardoso}}, \bibinfo {author} {\bibfnamefont {K.}~\bibnamefont {Destounis}},
  \bibinfo {author} {\bibfnamefont {F.}~\bibnamefont {Duque}}, \bibinfo
  {author} {\bibfnamefont {R.}~\bibnamefont {Panosso~Macedo}},\ and\ \bibinfo
  {author} {\bibfnamefont {A.}~\bibnamefont {Maselli}},\ }\bibfield  {title}
  {\bibinfo {title} {{Gravitational Waves from Extreme-Mass-Ratio Systems in
  Astrophysical Environments}},\ }\href
  {https://doi.org/10.1103/PhysRevLett.129.241103} {\bibfield  {journal}
  {\bibinfo  {journal} {Phys. Rev. Lett.}\ }\textbf {\bibinfo {volume} {129}},\
  \bibinfo {pages} {241103} (\bibinfo {year} {2022})},\ \Eprint
  {https://arxiv.org/abs/2210.01133} {arXiv:2210.01133 [gr-qc]} \BibitemShut
  {NoStop}%
\bibitem [{\citenamefont {Cole}\ \emph {et~al.}(2023)\citenamefont {Cole},
  \citenamefont {Bertone}, \citenamefont {Coogan}, \citenamefont {Gaggero},
  \citenamefont {Karydas}, \citenamefont {Kavanagh}, \citenamefont {Spieksma},\
  and\ \citenamefont {Tomaselli}}]{Cole:2022yzw}%
  \BibitemOpen
  \bibfield  {author} {\bibinfo {author} {\bibfnamefont {P.~S.}\ \bibnamefont
  {Cole}}, \bibinfo {author} {\bibfnamefont {G.}~\bibnamefont {Bertone}},
  \bibinfo {author} {\bibfnamefont {A.}~\bibnamefont {Coogan}}, \bibinfo
  {author} {\bibfnamefont {D.}~\bibnamefont {Gaggero}}, \bibinfo {author}
  {\bibfnamefont {T.}~\bibnamefont {Karydas}}, \bibinfo {author} {\bibfnamefont
  {B.~J.}\ \bibnamefont {Kavanagh}}, \bibinfo {author} {\bibfnamefont
  {T.~F.~M.}\ \bibnamefont {Spieksma}},\ and\ \bibinfo {author} {\bibfnamefont
  {G.~M.}\ \bibnamefont {Tomaselli}},\ }\bibfield  {title} {\bibinfo {title}
  {{Distinguishing environmental effects on binary black hole gravitational
  waveforms}},\ }\href {https://doi.org/10.1038/s41550-023-01990-2} {\bibfield
  {journal} {\bibinfo  {journal} {Nature Astron.}\ }\textbf {\bibinfo {volume}
  {7}},\ \bibinfo {pages} {943} (\bibinfo {year} {2023})},\ \Eprint
  {https://arxiv.org/abs/2211.01362} {arXiv:2211.01362 [gr-qc]} \BibitemShut
  {NoStop}%
\bibitem [{\citenamefont {Tomaselli}\ \emph {et~al.}(2023)\citenamefont
  {Tomaselli}, \citenamefont {Spieksma},\ and\ \citenamefont
  {Bertone}}]{Tomaselli:2023ysb}%
  \BibitemOpen
  \bibfield  {author} {\bibinfo {author} {\bibfnamefont {G.~M.}\ \bibnamefont
  {Tomaselli}}, \bibinfo {author} {\bibfnamefont {T.~F.~M.}\ \bibnamefont
  {Spieksma}},\ and\ \bibinfo {author} {\bibfnamefont {G.}~\bibnamefont
  {Bertone}},\ }\bibfield  {title} {\bibinfo {title} {{Dynamical friction in
  gravitational atoms}},\ }\href
  {https://doi.org/10.1088/1475-7516/2023/07/070} {\bibfield  {journal}
  {\bibinfo  {journal} {JCAP}\ }\textbf {\bibinfo {volume} {07}},\ \bibinfo
  {pages} {070}},\ \Eprint {https://arxiv.org/abs/2305.15460} {arXiv:2305.15460
  [gr-qc]} \BibitemShut {NoStop}%
\bibitem [{\citenamefont {Rahman}\ \emph {et~al.}(2024)\citenamefont {Rahman},
  \citenamefont {Kumar},\ and\ \citenamefont {Bhattacharyya}}]{Rahman:2023sof}%
  \BibitemOpen
  \bibfield  {author} {\bibinfo {author} {\bibfnamefont {M.}~\bibnamefont
  {Rahman}}, \bibinfo {author} {\bibfnamefont {S.}~\bibnamefont {Kumar}},\ and\
  \bibinfo {author} {\bibfnamefont {A.}~\bibnamefont {Bhattacharyya}},\
  }\bibfield  {title} {\bibinfo {title} {{Probing astrophysical environment
  with eccentric extreme mass-ratio inspirals}},\ }\href
  {https://doi.org/10.1088/1475-7516/2024/01/035} {\bibfield  {journal}
  {\bibinfo  {journal} {JCAP}\ }\textbf {\bibinfo {volume} {01}},\ \bibinfo
  {pages} {035}},\ \Eprint {https://arxiv.org/abs/2306.14971} {arXiv:2306.14971
  [gr-qc]} \BibitemShut {NoStop}%
\bibitem [{\citenamefont {{Karydas}}\ \emph {et~al.}(2024)\citenamefont
  {{Karydas}}, \citenamefont {{Kavanagh}},\ and\ \citenamefont
  {{Bertone}}}]{Karydas:2024fcn}%
  \BibitemOpen
  \bibfield  {author} {\bibinfo {author} {\bibfnamefont {T.~K.}\ \bibnamefont
  {{Karydas}}}, \bibinfo {author} {\bibfnamefont {B.~J.}\ \bibnamefont
  {{Kavanagh}}},\ and\ \bibinfo {author} {\bibfnamefont {G.}~\bibnamefont
  {{Bertone}}},\ }\bibfield  {title} {\bibinfo {title} {{Sharpening the dark
  matter signature in gravitational waveforms I: Accretion and eccentricity
  evolution}},\ }\href {https://doi.org/10.48550/arXiv.2402.13053} {\bibfield
  {journal} {\bibinfo  {journal} {arXiv e-prints}\ ,\ \bibinfo {eid}
  {arXiv:2402.13053}} (\bibinfo {year} {2024})},\ \Eprint
  {https://arxiv.org/abs/2402.13053} {arXiv:2402.13053 [gr-qc]} \BibitemShut
  {NoStop}%
\bibitem [{\citenamefont {{Kavanagh}}\ \emph {et~al.}(2024)\citenamefont
  {{Kavanagh}}, \citenamefont {{Karydas}}, \citenamefont {{Bertone}},
  \citenamefont {{Di Cintio}},\ and\ \citenamefont
  {{Pasquato}}}]{Kavanagh:2024lgq}%
  \BibitemOpen
  \bibfield  {author} {\bibinfo {author} {\bibfnamefont {B.~J.}\ \bibnamefont
  {{Kavanagh}}}, \bibinfo {author} {\bibfnamefont {T.~K.}\ \bibnamefont
  {{Karydas}}}, \bibinfo {author} {\bibfnamefont {G.}~\bibnamefont
  {{Bertone}}}, \bibinfo {author} {\bibfnamefont {P.}~\bibnamefont {{Di
  Cintio}}},\ and\ \bibinfo {author} {\bibfnamefont {M.}~\bibnamefont
  {{Pasquato}}},\ }\bibfield  {title} {\bibinfo {title} {{Sharpening the dark
  matter signature in gravitational waveforms II: Numerical simulations with
  the NbodyIMRI code}},\ }\href {https://doi.org/10.48550/arXiv.2402.13762}
  {\bibfield  {journal} {\bibinfo  {journal} {arXiv e-prints}\ ,\ \bibinfo
  {eid} {arXiv:2402.13762}} (\bibinfo {year} {2024})},\ \Eprint
  {https://arxiv.org/abs/2402.13762} {arXiv:2402.13762 [gr-qc]} \BibitemShut
  {NoStop}%
\bibitem [{\citenamefont {{Spieksma}}\ \emph {et~al.}(2024)\citenamefont
  {{Spieksma}}, \citenamefont {{Cardoso}}, \citenamefont {{Carullo}},
  \citenamefont {{Della Rocca}},\ and\ \citenamefont
  {{Duque}}}]{Spieksma:2024voy}%
  \BibitemOpen
  \bibfield  {author} {\bibinfo {author} {\bibfnamefont {T.~F.~M.}\
  \bibnamefont {{Spieksma}}}, \bibinfo {author} {\bibfnamefont
  {V.}~\bibnamefont {{Cardoso}}}, \bibinfo {author} {\bibfnamefont
  {G.}~\bibnamefont {{Carullo}}}, \bibinfo {author} {\bibfnamefont
  {M.}~\bibnamefont {{Della Rocca}}},\ and\ \bibinfo {author} {\bibfnamefont
  {F.}~\bibnamefont {{Duque}}},\ }\bibfield  {title} {\bibinfo {title} {{Black
  hole spectroscopy in environments: detectability prospects}},\ }\href
  {https://doi.org/10.48550/arXiv.2409.05950} {\bibfield  {journal} {\bibinfo
  {journal} {arXiv e-prints}\ ,\ \bibinfo {eid} {arXiv:2409.05950}} (\bibinfo
  {year} {2024})},\ \Eprint {https://arxiv.org/abs/2409.05950}
  {arXiv:2409.05950 [gr-qc]} \BibitemShut {NoStop}%
\bibitem [{\citenamefont {Connaughton}\ \emph {et~al.}(2016)\citenamefont
  {Connaughton} \emph {et~al.}}]{Connaughton:2016umz}%
  \BibitemOpen
  \bibfield  {author} {\bibinfo {author} {\bibfnamefont {V.}~\bibnamefont
  {Connaughton}} \emph {et~al.},\ }\bibfield  {title} {\bibinfo {title} {{Fermi
  GBM Observations of LIGO Gravitational Wave event GW150914}},\ }\href
  {https://doi.org/10.3847/2041-8205/826/1/L6} {\bibfield  {journal} {\bibinfo
  {journal} {Astrophys. J. Lett.}\ }\textbf {\bibinfo {volume} {826}},\
  \bibinfo {pages} {L6} (\bibinfo {year} {2016})},\ \Eprint
  {https://arxiv.org/abs/1602.03920} {arXiv:1602.03920 [astro-ph.HE]}
  \BibitemShut {NoStop}%
\bibitem [{\citenamefont {{Garg}}\ \emph {et~al.}(2024)\citenamefont {{Garg}},
  \citenamefont {{Sberna}}, \citenamefont {{Speri}}, \citenamefont {{Duque}},\
  and\ \citenamefont {{Gair}}}]{Garg:2024qxq}%
  \BibitemOpen
  \bibfield  {author} {\bibinfo {author} {\bibfnamefont {M.}~\bibnamefont
  {{Garg}}}, \bibinfo {author} {\bibfnamefont {L.}~\bibnamefont {{Sberna}}},
  \bibinfo {author} {\bibfnamefont {L.}~\bibnamefont {{Speri}}}, \bibinfo
  {author} {\bibfnamefont {F.}~\bibnamefont {{Duque}}},\ and\ \bibinfo {author}
  {\bibfnamefont {J.}~\bibnamefont {{Gair}}},\ }\bibfield  {title} {\bibinfo
  {title} {{Systematics in tests of general relativity using LISA massive black
  hole binaries}},\ }\href@noop {} {\bibfield  {journal} {\bibinfo  {journal}
  {arXiv e-prints}\ ,\ \bibinfo {eid} {arXiv:2410.02910}} (\bibinfo {year}
  {2024})},\ \Eprint {https://arxiv.org/abs/2410.02910} {arXiv:2410.02910
  [astro-ph.GA]} \BibitemShut {NoStop}%
\bibitem [{\citenamefont {Cardoso}\ \emph {et~al.}(2021)\citenamefont
  {Cardoso}, \citenamefont {Macedo},\ and\ \citenamefont
  {Vicente}}]{Cardoso:2020iji}%
  \BibitemOpen
  \bibfield  {author} {\bibinfo {author} {\bibfnamefont {V.}~\bibnamefont
  {Cardoso}}, \bibinfo {author} {\bibfnamefont {C.~F.~B.}\ \bibnamefont
  {Macedo}},\ and\ \bibinfo {author} {\bibfnamefont {R.}~\bibnamefont
  {Vicente}},\ }\bibfield  {title} {\bibinfo {title} {{Eccentricity evolution
  of compact binaries and applications to gravitational-wave physics}},\ }\href
  {https://doi.org/10.1103/PhysRevD.103.023015} {\bibfield  {journal} {\bibinfo
   {journal} {Phys. Rev. D}\ }\textbf {\bibinfo {volume} {103}},\ \bibinfo
  {pages} {023015} (\bibinfo {year} {2021})},\ \Eprint
  {https://arxiv.org/abs/2010.15151} {arXiv:2010.15151 [gr-qc]} \BibitemShut
  {NoStop}%
\bibitem [{\citenamefont {Gayathri}\ \emph {et~al.}(2022)\citenamefont
  {Gayathri}, \citenamefont {Healy}, \citenamefont {Lange}, \citenamefont
  {O'Brien}, \citenamefont {Szczepanczyk}, \citenamefont {Bartos},
  \citenamefont {Campanelli}, \citenamefont {Klimenko}, \citenamefont
  {Lousto},\ and\ \citenamefont {O'Shaughnessy}}]{Gayathri:2020coq}%
  \BibitemOpen
  \bibfield  {author} {\bibinfo {author} {\bibfnamefont {V.}~\bibnamefont
  {Gayathri}}, \bibinfo {author} {\bibfnamefont {J.}~\bibnamefont {Healy}},
  \bibinfo {author} {\bibfnamefont {J.}~\bibnamefont {Lange}}, \bibinfo
  {author} {\bibfnamefont {B.}~\bibnamefont {O'Brien}}, \bibinfo {author}
  {\bibfnamefont {M.}~\bibnamefont {Szczepanczyk}}, \bibinfo {author}
  {\bibfnamefont {I.}~\bibnamefont {Bartos}}, \bibinfo {author} {\bibfnamefont
  {M.}~\bibnamefont {Campanelli}}, \bibinfo {author} {\bibfnamefont
  {S.}~\bibnamefont {Klimenko}}, \bibinfo {author} {\bibfnamefont {C.~O.}\
  \bibnamefont {Lousto}},\ and\ \bibinfo {author} {\bibfnamefont
  {R.}~\bibnamefont {O'Shaughnessy}},\ }\bibfield  {title} {\bibinfo {title}
  {{Eccentricity estimate for black hole mergers with numerical relativity
  simulations}},\ }\href {https://doi.org/10.1038/s41550-021-01568-w}
  {\bibfield  {journal} {\bibinfo  {journal} {Nature Astron.}\ }\textbf
  {\bibinfo {volume} {6}},\ \bibinfo {pages} {344} (\bibinfo {year} {2022})},\
  \Eprint {https://arxiv.org/abs/2009.05461} {arXiv:2009.05461 [astro-ph.HE]}
  \BibitemShut {NoStop}%
\bibitem [{\citenamefont {Calder\'on~Bustillo}\ \emph
  {et~al.}(2021)\citenamefont {Calder\'on~Bustillo}, \citenamefont
  {Sanchis-Gual}, \citenamefont {Torres-Forn\'e},\ and\ \citenamefont
  {Font}}]{CalderonBustillo:2020xms}%
  \BibitemOpen
  \bibfield  {author} {\bibinfo {author} {\bibfnamefont {J.}~\bibnamefont
  {Calder\'on~Bustillo}}, \bibinfo {author} {\bibfnamefont {N.}~\bibnamefont
  {Sanchis-Gual}}, \bibinfo {author} {\bibfnamefont {A.}~\bibnamefont
  {Torres-Forn\'e}},\ and\ \bibinfo {author} {\bibfnamefont {J.~A.}\
  \bibnamefont {Font}},\ }\bibfield  {title} {\bibinfo {title} {{Confusing
  Head-On Collisions with Precessing Intermediate-Mass Binary Black Hole
  Mergers}},\ }\href {https://doi.org/10.1103/PhysRevLett.126.201101}
  {\bibfield  {journal} {\bibinfo  {journal} {Phys. Rev. Lett.}\ }\textbf
  {\bibinfo {volume} {126}},\ \bibinfo {pages} {201101} (\bibinfo {year}
  {2021})},\ \Eprint {https://arxiv.org/abs/2009.01066} {arXiv:2009.01066
  [gr-qc]} \BibitemShut {NoStop}%
\bibitem [{\citenamefont {Favata}\ \emph {et~al.}(2022)\citenamefont {Favata},
  \citenamefont {Kim}, \citenamefont {Arun}, \citenamefont {Kim},\ and\
  \citenamefont {Lee}}]{Favata:2021vhw}%
  \BibitemOpen
  \bibfield  {author} {\bibinfo {author} {\bibfnamefont {M.}~\bibnamefont
  {Favata}}, \bibinfo {author} {\bibfnamefont {C.}~\bibnamefont {Kim}},
  \bibinfo {author} {\bibfnamefont {K.~G.}\ \bibnamefont {Arun}}, \bibinfo
  {author} {\bibfnamefont {J.}~\bibnamefont {Kim}},\ and\ \bibinfo {author}
  {\bibfnamefont {H.~W.}\ \bibnamefont {Lee}},\ }\bibfield  {title} {\bibinfo
  {title} {{Constraining the orbital eccentricity of inspiralling compact
  binary systems with Advanced LIGO}},\ }\href
  {https://doi.org/10.1103/PhysRevD.105.023003} {\bibfield  {journal} {\bibinfo
   {journal} {Phys. Rev. D}\ }\textbf {\bibinfo {volume} {105}},\ \bibinfo
  {pages} {023003} (\bibinfo {year} {2022})},\ \Eprint
  {https://arxiv.org/abs/2108.05861} {arXiv:2108.05861 [gr-qc]} \BibitemShut
  {NoStop}%
\bibitem [{\citenamefont {Divyajyoti}\ \emph {et~al.}(2024)\citenamefont
  {Divyajyoti}, \citenamefont {Kumar}, \citenamefont {Tibrewal}, \citenamefont
  {Romero-Shaw},\ and\ \citenamefont {Mishra}}]{Divyajyoti:2023rht}%
  \BibitemOpen
  \bibfield  {author} {\bibinfo {author} {\bibnamefont {Divyajyoti}}, \bibinfo
  {author} {\bibfnamefont {S.}~\bibnamefont {Kumar}}, \bibinfo {author}
  {\bibfnamefont {S.}~\bibnamefont {Tibrewal}}, \bibinfo {author}
  {\bibfnamefont {I.~M.}\ \bibnamefont {Romero-Shaw}},\ and\ \bibinfo {author}
  {\bibfnamefont {C.~K.}\ \bibnamefont {Mishra}},\ }\bibfield  {title}
  {\bibinfo {title} {{Blind spots and biases: The dangers of ignoring
  eccentricity in gravitational-wave signals from binary black holes}},\ }\href
  {https://doi.org/10.1103/PhysRevD.109.043037} {\bibfield  {journal} {\bibinfo
   {journal} {Phys. Rev. D}\ }\textbf {\bibinfo {volume} {109}},\ \bibinfo
  {pages} {043037} (\bibinfo {year} {2024})},\ \Eprint
  {https://arxiv.org/abs/2309.16638} {arXiv:2309.16638 [gr-qc]} \BibitemShut
  {NoStop}%
\bibitem [{\citenamefont {Garg}\ \emph {et~al.}(2024)\citenamefont {Garg},
  \citenamefont {Derdzinski}, \citenamefont {Tiwari}, \citenamefont {Gair},\
  and\ \citenamefont {Mayer}}]{Garg:2024oeu}%
  \BibitemOpen
  \bibfield  {author} {\bibinfo {author} {\bibfnamefont {M.}~\bibnamefont
  {Garg}}, \bibinfo {author} {\bibfnamefont {A.}~\bibnamefont {Derdzinski}},
  \bibinfo {author} {\bibfnamefont {S.}~\bibnamefont {Tiwari}}, \bibinfo
  {author} {\bibfnamefont {J.}~\bibnamefont {Gair}},\ and\ \bibinfo {author}
  {\bibfnamefont {L.}~\bibnamefont {Mayer}},\ }\bibfield  {title} {\bibinfo
  {title} {{Measuring eccentricity and gas-induced perturbation from
  gravitational waves of LISA massive black hole binaries}},\ }\href
  {https://doi.org/10.1093/mnras/stae1764} {\bibfield  {journal} {\bibinfo
  {journal} {Mon. Not. Roy. Astron. Soc.}\ }\textbf {\bibinfo {volume} {532}},\
  \bibinfo {pages} {4060} (\bibinfo {year} {2024})},\ \Eprint
  {https://arxiv.org/abs/2402.14058} {arXiv:2402.14058 [astro-ph.GA]}
  \BibitemShut {NoStop}%
\bibitem [{\citenamefont {Abbott}\ \emph
  {et~al.}(2017{\natexlab{b}})\citenamefont {Abbott} \emph
  {et~al.}}]{LIGOScientific:2017vwq}%
  \BibitemOpen
  \bibfield  {author} {\bibinfo {author} {\bibfnamefont {B.~P.}\ \bibnamefont
  {Abbott}} \emph {et~al.} (\bibinfo {collaboration} {LIGO Scientific,
  Virgo}),\ }\bibfield  {title} {\bibinfo {title} {{GW170817: Observation of
  Gravitational Waves from a Binary Neutron Star Inspiral}},\ }\href
  {https://doi.org/10.1103/PhysRevLett.119.161101} {\bibfield  {journal}
  {\bibinfo  {journal} {Phys. Rev. Lett.}\ }\textbf {\bibinfo {volume} {119}},\
  \bibinfo {pages} {161101} (\bibinfo {year} {2017}{\natexlab{b}})},\ \Eprint
  {https://arxiv.org/abs/1710.05832} {arXiv:1710.05832 [gr-qc]} \BibitemShut
  {NoStop}%
\bibitem [{\citenamefont {Abbott}\ \emph
  {et~al.}(2019{\natexlab{d}})\citenamefont {Abbott} \emph
  {et~al.}}]{LIGOScientific:2018hze}%
  \BibitemOpen
  \bibfield  {author} {\bibinfo {author} {\bibfnamefont {B.~P.}\ \bibnamefont
  {Abbott}} \emph {et~al.} (\bibinfo {collaboration} {LIGO Scientific,
  Virgo}),\ }\bibfield  {title} {\bibinfo {title} {{Properties of the binary
  neutron star merger GW170817}},\ }\href
  {https://doi.org/10.1103/PhysRevX.9.011001} {\bibfield  {journal} {\bibinfo
  {journal} {Phys. Rev. X}\ }\textbf {\bibinfo {volume} {9}},\ \bibinfo {pages}
  {011001} (\bibinfo {year} {2019}{\natexlab{d}})},\ \Eprint
  {https://arxiv.org/abs/1805.11579} {arXiv:1805.11579 [gr-qc]} \BibitemShut
  {NoStop}%
\bibitem [{\citenamefont {Barsotti}\ \emph {et~al.}(2018)\citenamefont
  {Barsotti}, \citenamefont {Gras}, \citenamefont {Evans},\ and\ \citenamefont
  {Fritschel}}]{aLIGO_ZDHP}%
  \BibitemOpen
  \bibfield  {author} {\bibinfo {author} {\bibfnamefont {L.}~\bibnamefont
  {Barsotti}}, \bibinfo {author} {\bibfnamefont {S.}~\bibnamefont {Gras}},
  \bibinfo {author} {\bibfnamefont {M.}~\bibnamefont {Evans}},\ and\ \bibinfo
  {author} {\bibfnamefont {P.}~\bibnamefont {Fritschel}},\ }\href
  {https://dcc.ligo.org/LIGO-T1800044/public} {\emph {\bibinfo {title} {The
  updated Advanced LIGO design curve}}},\ \bibinfo {type} {LIGO Technical
  Note}\ \bibinfo {number} {T1800044-v5}\ (\bibinfo  {institution} {LIGO
  Scientific Collaboration},\ \bibinfo {year} {2018})\ \bibinfo {note} {updated
  from T0900288-v3}\BibitemShut {NoStop}%
\bibitem [{\citenamefont {{Manzotti}}\ and\ \citenamefont
  {{Dietz}}(2012)}]{2012arXiv1202.4031M}%
  \BibitemOpen
  \bibfield  {author} {\bibinfo {author} {\bibfnamefont {A.}~\bibnamefont
  {{Manzotti}}}\ and\ \bibinfo {author} {\bibfnamefont {A.}~\bibnamefont
  {{Dietz}}},\ }\bibfield  {title} {\bibinfo {title} {{Prospects for early
  localization of gravitational-wave signals from compact binary coalescences
  with advanced detectors}},\ }\href {https://doi.org/10.48550/arXiv.1202.4031}
  {\bibfield  {journal} {\bibinfo  {journal} {arXiv e-prints}\ ,\ \bibinfo
  {eid} {arXiv:1202.4031}} (\bibinfo {year} {2012})},\ \Eprint
  {https://arxiv.org/abs/1202.4031} {arXiv:1202.4031 [gr-qc]} \BibitemShut
  {NoStop}%
\bibitem [{\citenamefont {Punturo}\ \emph {et~al.}(2010)\citenamefont {Punturo}
  \emph {et~al.}}]{Punturo:2010zz}%
  \BibitemOpen
  \bibfield  {author} {\bibinfo {author} {\bibfnamefont {M.}~\bibnamefont
  {Punturo}} \emph {et~al.},\ }\bibfield  {title} {\bibinfo {title} {{The
  Einstein Telescope: A third-generation gravitational wave observatory}},\
  }\href {https://doi.org/10.1088/0264-9381/27/19/194002} {\bibfield  {journal}
  {\bibinfo  {journal} {Class. Quant. Grav.}\ }\textbf {\bibinfo {volume}
  {27}},\ \bibinfo {pages} {194002} (\bibinfo {year} {2010})}\BibitemShut
  {NoStop}%
\bibitem [{\citenamefont {Reitze}\ \emph {et~al.}(2019)\citenamefont {Reitze}
  \emph {et~al.}}]{Reitze:2019iox}%
  \BibitemOpen
  \bibfield  {author} {\bibinfo {author} {\bibfnamefont {D.}~\bibnamefont
  {Reitze}} \emph {et~al.},\ }\bibfield  {title} {\bibinfo {title} {{Cosmic
  Explorer: The U.S. Contribution to Gravitational-Wave Astronomy beyond
  LIGO}},\ }\href@noop {} {\bibfield  {journal} {\bibinfo  {journal} {Bull. Am.
  Astron. Soc.}\ }\textbf {\bibinfo {volume} {51}},\ \bibinfo {pages} {035}
  (\bibinfo {year} {2019})},\ \Eprint {https://arxiv.org/abs/1907.04833}
  {arXiv:1907.04833 [astro-ph.IM]} \BibitemShut {NoStop}%
\bibitem [{\citenamefont {Martin}\ \emph {et~al.}(2018)\citenamefont {Martin},
  \citenamefont {Nixon}, \citenamefont {Xie},\ and\ \citenamefont
  {King}}]{Martin:2018iov}%
  \BibitemOpen
  \bibfield  {author} {\bibinfo {author} {\bibfnamefont {R.~G.}\ \bibnamefont
  {Martin}}, \bibinfo {author} {\bibfnamefont {C.}~\bibnamefont {Nixon}},
  \bibinfo {author} {\bibfnamefont {F.-G.}\ \bibnamefont {Xie}},\ and\ \bibinfo
  {author} {\bibfnamefont {A.}~\bibnamefont {King}},\ }\bibfield  {title}
  {\bibinfo {title} {{Circumbinary discs around merging stellar-mass black
  holes}},\ }\href {https://doi.org/10.1093/mnras/sty2178} {\bibfield
  {journal} {\bibinfo  {journal} {Mon. Not. Roy. Astron. Soc.}\ }\textbf
  {\bibinfo {volume} {480}},\ \bibinfo {pages} {4732} (\bibinfo {year}
  {2018})},\ \Eprint {https://arxiv.org/abs/1808.06023} {arXiv:1808.06023
  [astro-ph.HE]} \BibitemShut {NoStop}%
\bibitem [{\citenamefont {Tuna}\ and\ \citenamefont
  {Metzger}(2023)}]{Tuna:2023jgw}%
  \BibitemOpen
  \bibfield  {author} {\bibinfo {author} {\bibfnamefont {S.}~\bibnamefont
  {Tuna}}\ and\ \bibinfo {author} {\bibfnamefont {B.~D.}\ \bibnamefont
  {Metzger}},\ }\bibfield  {title} {\bibinfo {title} {{Long-term Evolution of
  Massive-star Post-common-envelope Circumbinary Disks and the Environments of
  Fast Luminous Transients}},\ }\href
  {https://doi.org/10.3847/1538-4357/acef17} {\bibfield  {journal} {\bibinfo
  {journal} {Astrophys. J.}\ }\textbf {\bibinfo {volume} {955}},\ \bibinfo
  {pages} {125} (\bibinfo {year} {2023})},\ \Eprint
  {https://arxiv.org/abs/2306.10111} {arXiv:2306.10111 [astro-ph.HE]}
  \BibitemShut {NoStop}%
\bibitem [{\citenamefont {Taylor}\ \emph {et~al.}(2012)\citenamefont {Taylor},
  \citenamefont {Gair},\ and\ \citenamefont {Mandel}}]{Taylor:2011fs}%
  \BibitemOpen
  \bibfield  {author} {\bibinfo {author} {\bibfnamefont {S.~R.}\ \bibnamefont
  {Taylor}}, \bibinfo {author} {\bibfnamefont {J.~R.}\ \bibnamefont {Gair}},\
  and\ \bibinfo {author} {\bibfnamefont {I.}~\bibnamefont {Mandel}},\
  }\bibfield  {title} {\bibinfo {title} {{Hubble without the Hubble: Cosmology
  using advanced gravitational-wave detectors alone}},\ }\href
  {https://doi.org/10.1103/PhysRevD.85.023535} {\bibfield  {journal} {\bibinfo
  {journal} {Phys. Rev. D}\ }\textbf {\bibinfo {volume} {85}},\ \bibinfo
  {pages} {023535} (\bibinfo {year} {2012})},\ \Eprint
  {https://arxiv.org/abs/1108.5161} {arXiv:1108.5161 [gr-qc]} \BibitemShut
  {NoStop}%
\bibitem [{\citenamefont {Taylor}\ and\ \citenamefont
  {Gair}(2012)}]{Taylor:2012db}%
  \BibitemOpen
  \bibfield  {author} {\bibinfo {author} {\bibfnamefont {S.~R.}\ \bibnamefont
  {Taylor}}\ and\ \bibinfo {author} {\bibfnamefont {J.~R.}\ \bibnamefont
  {Gair}},\ }\bibfield  {title} {\bibinfo {title} {{Cosmology with the lights
  off: standard sirens in the Einstein Telescope era}},\ }\href
  {https://doi.org/10.1103/PhysRevD.86.023502} {\bibfield  {journal} {\bibinfo
  {journal} {Phys. Rev. D}\ }\textbf {\bibinfo {volume} {86}},\ \bibinfo
  {pages} {023502} (\bibinfo {year} {2012})},\ \Eprint
  {https://arxiv.org/abs/1204.6739} {arXiv:1204.6739 [astro-ph.CO]}
  \BibitemShut {NoStop}%
\bibitem [{\citenamefont {Harris}\ \emph {et~al.}(2020)\citenamefont {Harris}
  \emph {et~al.}}]{Harris:2020xlr}%
  \BibitemOpen
  \bibfield  {author} {\bibinfo {author} {\bibfnamefont {C.~R.}\ \bibnamefont
  {Harris}} \emph {et~al.},\ }\bibfield  {title} {\bibinfo {title} {{Array
  programming with NumPy}},\ }\href {https://doi.org/10.1038/s41586-020-2649-2}
  {\bibfield  {journal} {\bibinfo  {journal} {Nature}\ }\textbf {\bibinfo
  {volume} {585}},\ \bibinfo {pages} {357} (\bibinfo {year} {2020})},\ \Eprint
  {https://arxiv.org/abs/2006.10256} {arXiv:2006.10256 [cs.MS]} \BibitemShut
  {NoStop}%
\bibitem [{\citenamefont {Virtanen}\ \emph {et~al.}(2020)\citenamefont
  {Virtanen} \emph {et~al.}}]{Virtanen:2019joe}%
  \BibitemOpen
  \bibfield  {author} {\bibinfo {author} {\bibfnamefont {P.}~\bibnamefont
  {Virtanen}} \emph {et~al.},\ }\bibfield  {title} {\bibinfo {title} {{SciPy
  1.0--Fundamental Algorithms for Scientific Computing in Python}},\ }\href
  {https://doi.org/10.1038/s41592-019-0686-2} {\bibfield  {journal} {\bibinfo
  {journal} {Nat. Meth.}\ }\textbf {\bibinfo {volume} {17}},\ \bibinfo {pages}
  {261} (\bibinfo {year} {2020})},\ \Eprint {https://arxiv.org/abs/1907.10121}
  {arXiv:1907.10121 [cs.MS]} \BibitemShut {NoStop}%
\bibitem [{\citenamefont {Hunter}(2007)}]{Hunter:2007ouj}%
  \BibitemOpen
  \bibfield  {author} {\bibinfo {author} {\bibfnamefont {J.~D.}\ \bibnamefont
  {Hunter}},\ }\bibfield  {title} {\bibinfo {title} {{Matplotlib: A 2D Graphics
  Environment}},\ }\href {https://doi.org/10.1109/MCSE.2007.55} {\bibfield
  {journal} {\bibinfo  {journal} {Comput. Sci. Eng.}\ }\textbf {\bibinfo
  {volume} {9}},\ \bibinfo {pages} {90} (\bibinfo {year} {2007})}\BibitemShut
  {NoStop}%
\bibitem [{BBH(2017)}]{BBHGAS_StellarCollapse}%
  \BibitemOpen
  \href {https://stellarcollapse.org/bbhgas.html} {}\bibinfo {howpublished}
  {\url{https://stellarcollapse.org/bbhgas.html}} (\bibinfo {year}
  {2017})\BibitemShut {NoStop}%
\bibitem [{gw1(2019)}]{gw150914psds}%
  \BibitemOpen
  \href@noop {} {}\bibinfo {howpublished}
  {\url{https://dcc.ligo.org/LIGO-P1800370/public}} (\bibinfo {year}
  {2019})\BibitemShut {NoStop}%
\end{thebibliography}%
\end{document}